\documentclass[%
 reprint,
 amsmath,amssymb,
 aps,
]{revtex4-2}
\usepackage{graphicx}
\usepackage{dcolumn}
\usepackage{bm}
\usepackage{float}
\usepackage{color}
\usepackage{import}
\usepackage{xr} 
\usepackage{amssymb,amsmath,latexsym,upgreek,mathtools}
\usepackage{pdfpages}
\usepackage{pgffor}
\usepackage{filecontents}

\usepackage{xr}
\newcommand{\br}{\mathbf{r}}
\newcommand{\bq}{\mathbf{q}}

\newcommand{\bxi}{\boldsymbol\xi}

\newcommand{\diff}[1]{\mathop{}\!\mathrm{d}#1\;}

\newcommand{\va}{v_{\text{a}}}
\newcommand{\taua}{\tau_{\text{a}}}
\newcommand{\la}{\lambda_{\text{a}}}
\newcommand{\nL}{n_{\text{largest}}}
\newcommand{\nLA}{\langle \nL \rangle}
\newcommand{\xipart}{\langle |\bxi|\rangle_{\text{particles}}}
\newcommand{\xispace}{\langle |\bxi|\rangle_{\text{space}}}
\newcommand{\xidiff}{\Delta \langle |\bxi|\rangle}
\newcommand{\rg}{R_{\text{g}}}

\newcommand\numberthis{\addtocounter{equation}{1}\tag{\theequation}}

\makeatletter
\AtBeginDocument{\let\LS@rot\@undefined}
\makeatother

\begin{document}


\preprint{APS/123-QED}

\title{Active noise\textendash induced dynamic clustering of passive colloids}

\author{Layne B. Frechette}
\email{laynefrechette@brandeis.edu}
\author{Aparna Baskaran}
 \email{aparna@brandeis.edu}
 \author{Michael F. Hagan}%
\email{hagan@brandeis.edu}
\affiliation{%
 Martin Fisher School of Physics, Brandeis University, Waltham, Massachusetts 02453, USA
}%

\date{\today}

\begin{abstract}
Active fluids generate spontaneous, often chaotic mesoscale flows. Harnessing these flows to drive embedded soft materials into structures with controlled length scales and lifetimes is a key challenge at the interface between the fields of active matter and nonequilibrium self-assembly. Here, we present a simple and efficient computational approach to model soft materials advected by active fluids, by simulating particles moving in a spatiotemporally correlated noise field.  To illustrate our approach, we simulate the dynamical self-organization of repulsive colloids within such an active noise field in two and three dimensions. The colloids form structures whose sizes and dynamics can be tuned by the correlation time and length of the active fluid, and range from small rotating droplets to clusters with internal flows and system-spanning sizes that vastly exceed the active correlation length. Our results elucidate how the interplay between active fluid time/length scales and emergent driven assembly can be used to rationally design functional assemblies. More broadly, our approach can be used to efficiently simulate diverse active fluids and other systems with spatiotemporally correlated noise.

\end{abstract}

\maketitle


\section*{\label{sec:intro}Introduction}

Equilibrium self-assembly can generate incredibly intricate structures \cite{Whitesides2002,Boles2016,Hagan2021}, but is limited by kinetic traps that suppress yields and the fact that equilibrium structures lack functionalities exhibited by living matter. Embedding passive assembly components within an active fluid provides a potent route to overcome these limitations. By transforming microscopic energy consumption into large-scale directed motion, active materials can drive spatiotemporal organization that would be thermodynamically forbidden in a passive system. For example, in contrast to equilibrium hard spheres, purely repulsive colloids can cluster or phase-separate in an active bath, e.g., of motile bacteria \cite{Grober2023}. Furthermore, the interplay between activity and assembly energetics can greatly expand the manifold of accessible structures. Actin filaments advected by a kinesin-microtubule active fluid provide a powerful illustration: when the actin filaments are entangled but not crosslinked, activity drives the assembly of compact asters \cite{Berezney2022}. But, when the filament network is further strengthened by chemical crosslinks, the filaments instead condense into a viscoelastic membrane \cite{Berezney2024}. 

Such organization is driven by a distinctive feature common to many active materials --- spatiotemporal correlations among active forces. Clarifying how these correlations are related to the structure and dynamics of soft materials would allow systematic control of their properties, enabling a new class of materials with features currently found only in living organisms. Yet, the wide range of length scales involved, from nanoscale chemical reactions to macroscopic fluid flows \cite{Lemma2019}, makes this extremely challenging from a theoretical and computational perspective. 

\begin{figure*}[ht]
    \centering
    \includegraphics[width=\linewidth]{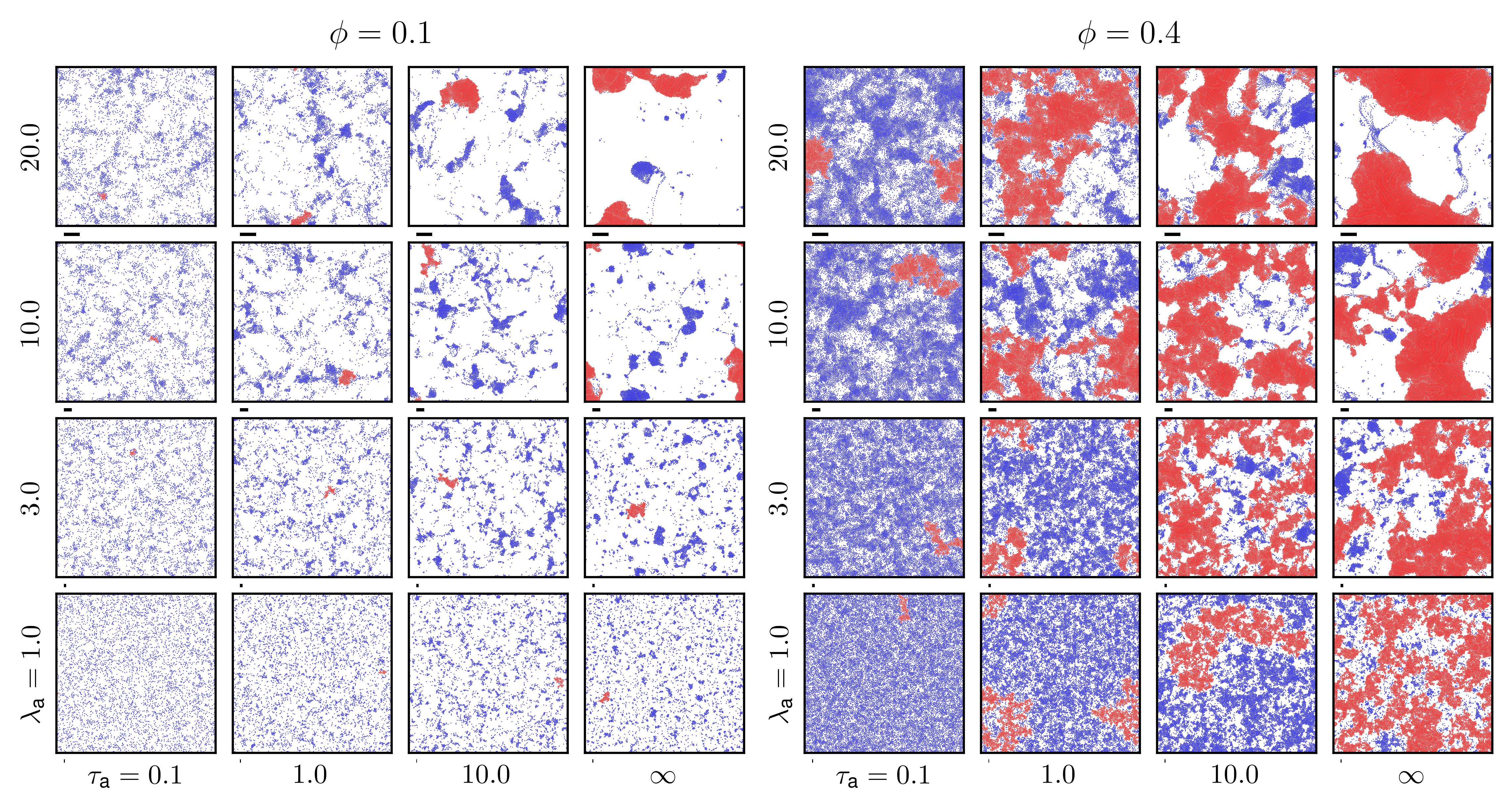}
    \caption{Representative snapshots of particles in an active noise field, for $\va=1$ and different values of $\la$ and $\taua$. Results are shown for two packing fractions: (left) $\phi=0.1$ ($N=5016$ particles) and (right)  $\phi=0.4$ ($N=20372$ particles).  The $\infty$ symbol denotes a ``quenched'' noise field which does not change with time. The black scale bar below each snapshot has length $\la$. Particles which make up the largest cluster are colored red (see Results and Methods); all other particles are colored blue. The size of the largest cluster, as measured by the radius of gyration $\rg=\sqrt{\sum_{j\neq i} |\br_i-\br_j|^2/(2N^2)}$, can significantly exceed the correlation length: $\rg\approx3.4\la$ for the snapshot with $\phi=0.1$, $\la=20$, $\tau=\infty$, and $\rg\approx4.2\la$ for the snapshot with $\phi=0.4$, $\la=20$, $\tau=\infty$. Except where mentioned otherwise, the dimensions of the simulation box are $200\times 200$ and the noise grid has spacing $0.5$ for all results reported in this article. All quantities in this article are presented in dimensionless units (see Results). Movies S1 and S2 show trajectories for all these values of $\phi$, $\la$, and $\taua$.}
    \label{fig:hard_sphere_multipanel}
\end{figure*}

In this article, we present a computational method that efficiently models spatiotemporally correlated active forces, enabling large-scale simulations of soft materials immersed in active fluids. We demonstrate the method on repulsive colloids, showing that the active forces drive the particles into dynamical assemblies whose size, structure, and dynamics can be tuned by the properties of the active fluid, ranging from small rotating droplets to system-spanning clusters with complex internal flows (Fig. \ref{fig:hard_sphere_multipanel}). Notably, the cluster scale can greatly exceed the active correlation length. These results elucidate how materials with complex structures and dynamics can emerge from a structureless suspension of active and passive components. 


Computer simulations are a powerful approach to elucidate mechanisms of assembly and organization, but have been limited by the wide spectrum of length and time scales associated with active matter. For example, simulating assembly driven by an active fluid, such as a bath of self-propelled particles \cite{Romanczuk2012,Fily2012,Czirok2000,Mishra2010,AditiSimha2002,Palacci2013,Palacci2014,Zottl2023a,Stenhammar2014,Levine2000,Redner2013,Czirok1997,Pedersen2024,Gokhale2022a,Batton2024,Omar2019,Grober2023}  or kinesin-powered microtubules \cite{Sanchez2012,DeCamp2015,Henkin2014,Lemma2019,Duclos2020,Berezney2022,Berezney2024}, requires computational effort proportional to the number of bath particles \cite{Frenkel2023}. Many computational techniques used to overcome these limits in traditional passive systems are not applicable to active materials, which lack separation of timescales and incur long-range boundary effects. For systems in which fluid flow is important, the cost is even steeper, since long-range forces must be computed \cite{Banchio2003,Martin-Gomez2019,Navarro2015,Ishikawa2008,Matas-Navarro2014}. Alternative methods that represent solvent dynamics explicitly, such as lattice Boltzmann \cite{deGraaf2016,Bardfalvy2019} or multiparticle collision dynamics \cite{Theers2018,Qi2022,Kozhukhov2022,Kozhukhov2024,Howard2019}, also incur high computational costs, and implementing appropriate boundary conditions in complex systems (such as dense nematics) is not straightforward. The immersed boundary method \cite{Peskin2002}, which couples straightforwardly to active nematic hydrodynamics \cite{Whitfield2016,Chandler2024}, scales too poorly with system size for many-particle simulations to be tractable. Furthermore, the nature of hydrodynamic interactions is system-specific and often poorly understood in dense active materials, making them challenging to describe by any of these methods. 

An alternative method is to represent the active bath implicitly. Based on the observation that the motion of tracer particles in an active bath can be modeled as a persistent random walk \cite{Wu2000,Sanchez2012,Maggi2014,Maggi2017,Koumakis2014,Ye2020,Hecht2024,Liebchen2022}, previous works have modeled active baths as time-correlated random forces that act independently on embedded particles \cite{Liao2021,Sprenger2023} (known as the active Ornstein-Uhlenbeck particle, or AOUP, model \cite{Szamel2014,Martin2021}). This approximation significantly reduces the computational cost of large-scale simulations, but omits the crucial feature discussed above that is common to many active matter systems regardless of their microscopic ingredients or symmetries --- spatial correlations among active forces.  These correlations give rise to emergent interparticle interactions that drive complex organizations of soft materials. 

The approach we present here seeks to overcome the aforementioned challenges while explicitly accounting for the correlations in active forces. 
We represent the bath as a \textit{spatiotemporally correlated} Gaussian noise field, which can describe the active forces or noise from a wide range of active materials, including self-propelled particles \cite{Palacci2013,Palacci2014,Redner2013,Fily2012,Stenhammar2014,Romanczuk2012,Zottl2023}, active nematics \cite{Doostmohammadi2018,Ramaswamy2003,Giomi2013,Giomi2014,DeCamp2015,Sanchez2012,Lemma2019,Duclos2020,Wu2017} and gels \cite{Sanchez2012,Henkin2014}, and vibrated granular materials \cite{Narayan2007,Deseigne2010,Deseigne2012,Weber2013,Kumar2014,Arora2024,Caprini2024}. To model so-called ``wet'' active matter systems \cite{Marchetti2013}, in which a dense liquid transmits active forces via flow fields, the noise field can be made incompressible. The model is computationally very tractable and straightforward to implement; we have implemented it in the GPU-enabled molecular dynamics package HOOMD-blue \cite{Anderson2020}. Further, the model allows independently varying the active speed and correlation length and time of the bath, enabling one to disentangle the effects of these parameters on emergent behaviors. Crucially, these parameters can be directly mapped to specific experimental systems (see Discussion).

To illustrate our method, we consider a system of repulsive particles in a compressible active fluid, such as vibrated granular objects \cite{Feitosa2002,Deseigne2010,Caprini2024,Narayan2007,Koumakis2016} or a 2D layer of a 3D fluid (to which heavy colloids may sediment \cite{Palacci2010,Dhar2024}). However, the method can also model incompressible fluids, which we will consider briefly here and explore in more detail in a subsequent work.
We first consider a single probe particle in the active noise field.  We show that its motion can be described as a persistent random walk, consistent with experiments and active bath models that only consider time-correlated noise. We measure how the noise field parameters map to an effective diffusivity and persistence time of the tracer. In contrast, we demonstrate that the behavior of many-particle systems in an active noise field can differ significantly from collections of independently-propelled particles, in particular forming the dynamical clusters shown in Fig. \ref{fig:hard_sphere_multipanel}. We quantitatively characterize how the length and time scales of these structures and their internal dynamics depend on the particle packing fractions and control parameters of the active fluid: the magnitude, and correlation length and time of the active flow. By analyzing the correlations between noise magnitude and particle density, we reveal the mechanism for clustering. Our method enables simulating systems on length and time scales much greater than previously possible, and thus is a powerful tool for studying the effects of active flows on diverse complex soft materials.

\section*{Results}

\subsection*{Active bath model}

Our active bath model is based on the observation that active flows are often chaotic \cite{Dombrowski2004,Doostmohammadi2018,Alert2022,Keta2024}, yet are correlated in space (over a typical length scale $\la$) and time (on a scale $\taua$), and have a typical speed ($\va$). We thus treat the active fluid velocity as a stochastic vector field $\bxi(\mathbf{r},t)$ for each point in space $\br$ and time $t$ with Gaussian statistics characterized by mean zero and variance:
\begin{equation}
    \langle \xi_{\mu}(\mathbf{r},t) \xi_{\nu}(\mathbf{r}',t') \rangle = \va^2 \delta_{\mu\nu}e^{-|t-t'|/\taua} e^{-|\mathbf{r}-\mathbf{r}'|/\la}, \label{eq:noise_statistics}
\end{equation}
where $\mu$ and $\nu$ denote Cartesian components. This approach has precedent in the fluid dynamics literature, where correlated Gaussian random fields have been used to model passive advection in turbulent flows \cite{Kraichnan1970,Kraichnan1994,Klyatskin1974,Klyatskin2003,Majda1999}. Most previous studies of inertial turbulence have focused on the ``inertial range'' (where velocity fluctuations are scale-free \cite{Kraichnan1994,Frisch1998}) or the short-correlation time limit (where analytic calculations are tractable \cite{Kraichnan1994,Ueda2016,Monnai2008,Dentz2003}). However, recent work has shown that adding spatiotemporally correlated noise to a Landau-Ginzburg model can alter behavior near a critical point \cite{Maggi2022} and promote phase separation \cite{Paoluzzi2024}. In our case, the active length and time scales $\la$ and $\taua$ will serve as important parameters for tuning the organization of active composite materials. 

\begin{figure}[htpb!]
    \centering
    \includegraphics[width=\linewidth]{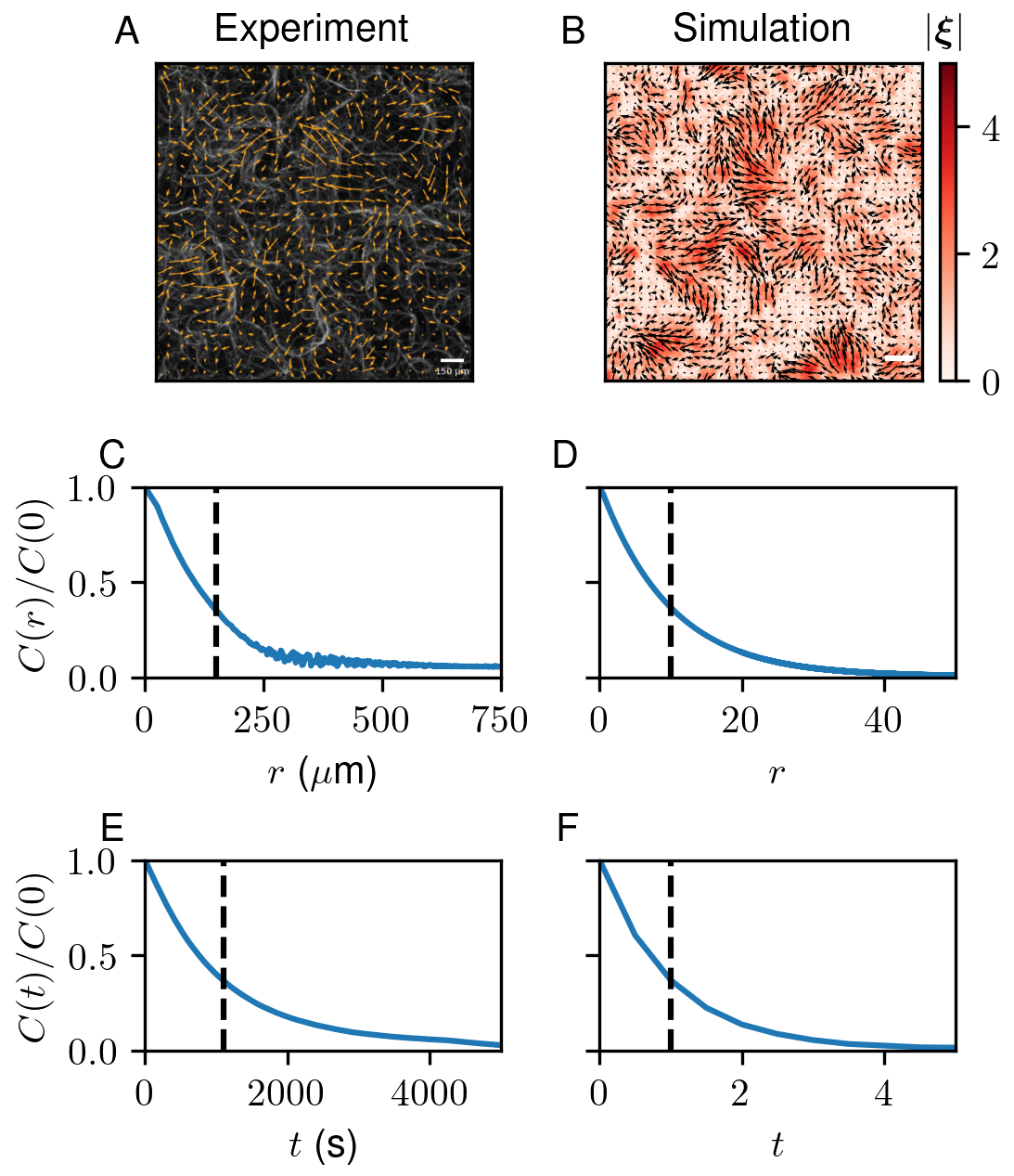}
    \caption{(A) Experimental image of an active gel (credit: Sattvic Ray). Orange arrows indicate the velocity field, and the white scale bar, representing the velocity correlation length, is 150$\mu$m. (B) Snapshot of a spatiotemporally correlated (``active'') noise field, $\boldsymbol{\xi}$ with $\la=10$, $\va=1$, and $\taua=1$ (in simulation units). The correlation length and field of view are chosen to correspond to those of the experiment in panel A, with the simulation unit length set to $\sigma=15\mu$m and unit time set to $t_0=1100$s. The magnitude of the noise field is indicated by red shading. Black arrows depict the noise vector field (where the length of an arrow is proportional to the magnitude of the noise at that point). Field points are subsampled every six grid spacings to show the vector field clearly. (C) Experimental and (D) simulated velocity spatial correlation functions. Black dashed lines represent the correlation length, equal to the value of $r$ at which $C(r)/C(0)=1/e$. (E) Experimental and (F) simulated velocity time correlation functions. Black dashed lines represent the correlation time, equal to the value of $t$ at which $C(t)/C(0)=1/e$.}
    \label{fig:active_noise_vs_expt}
\end{figure}

It is very straightforward to infer the active bath parameters from experiments. Fig. \ref{fig:active_noise_vs_expt}A shows an experimental image of an kinesin-microtubule active gel, with its velocity field (orange arrows) obtained from particle image velocimetry \cite{Sanchez2012}. From such a velocity field trajectory, one can compute a velocity autocorrelation function. We identify the time scale and length scale that characterize the decay of the autocorrelation function with the active noise parameters $\taua$ and $\la$, respectively. Additionally, we identify the root mean squared velocity with the magnitude of the active noise ($\va$). In Fig \ref{fig:active_noise_vs_expt}B we show a snapshot from an active noise simulation with a value of $\la$ comparable to the experiment. In both experiment and simulation, the velocity field is clearly chaotic, yet correlated over a length scale $\la$. In Fig. \ref{fig:active_noise_vs_expt}C we show the experimental spatial velocity correlation function, and in Fig. \ref{fig:active_noise_vs_expt}D we show the corresponding simulated active noise correlation function. Both correlation functions exhibit an exponential decay over a distance $\la$ ($\approx 150\mu$m in the experiment). Similarly, we show the experimental and simulated velocity/noise time correlation functions in Figs. \ref{fig:active_noise_vs_expt}E and F, respectively. Again, both correlation functions exhibit an exponential decay, here over a time $\taua$ ($\approx 1100$s in the experiment).

Details of the procedure we use to generate active noise trajectories are in Methods. In general, the noise field is compressible, $\nabla \cdot \bxi\neq 0$; it can be made incompressible via a projection operator method (see Methods). Particles are advected by the active noise field, and interact with each other via a purely repulsive potential, according to an overdamped Langevin equation (see Methods).

\subsection*{Reduced units}

Unless otherwise indicated, we report all quantities in terms of a unit length $\sigma$ (the particle diameter), a unit energy $\epsilon$ (the energy scale for particle repulsion), and a unit time $t_0=\gamma \sigma^2/\epsilon$, where $\gamma$ is the friction coefficient of the particles in the bath.

\subsection*{Tracer particle dynamics}

\begin{figure}[htpb!]
    \centering
    \includegraphics[width=\linewidth]{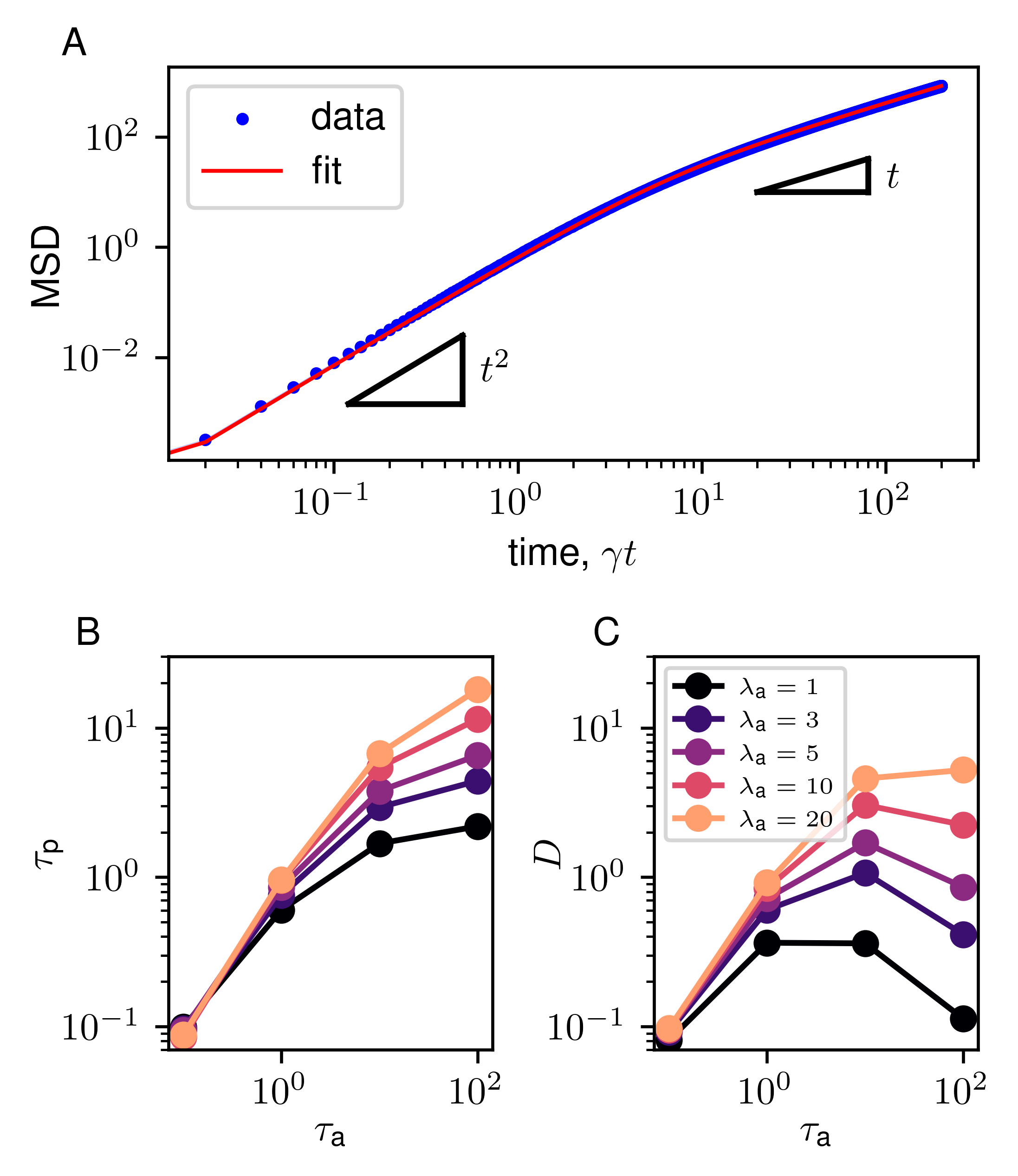}
    \caption{Motion of a tracer particle in an active noise field. Panel A shows the MSD of a tracer particle in an active noise field with $\la=3$, $\taua=10$, and $\va=1$. The MSD shows a crossover from ballistic ($\sim t^2$) motion at early times to diffusive ($\sim t$)  motion at late times. The simulation data (blue dots) are well-fit by the equation for the MSD of a persistent random walk, Eq. \ref{eq:persistent_random_walk} (red curve, obtained by least-squares regression using Scipy's optimize.curve\_fit function \cite{Virtanen2020}). Movie S3 shows a trajectory of a tracer particle, illustrating the persistent motion. Panels B and C show the fit parameters $\tau_{\text{p}}$ and $D$, respectively, versus $\taua$ for different values of $\la$ and $\va=1$.}
    \label{fig:msd}
\end{figure}

To illustrate the nature of motion in the active noise field, we have measured the mean squared displacement (MSD) of a single particle advected by the field. In Fig. \ref{fig:msd}A we show the tracer particle MSD for $\la=3$, $\taua=10$, and $\va=1$. Short-time ballistic motion gives way to long-time diffusive motion, which is consistent with experiments and simulations of tracer particles in active baths \cite{Wu2000,Ye2020,Sanchez2012} and motivates the AOUP model of active baths \cite{Seyforth2022,Liao2020}. In fact, we can make an explicit connection between the persistence time $\tau_{\text{p}}$ and diffusion constant $D$ of the tracer particle and the length and time scales of the active fluid. To do so, we assume a persistent-random-walk functional form for the MSD:
\begin{equation}
    \langle |\mathbf{r}_\text{p}(t)-\mathbf{r}_\text{p}(0)|^2 \rangle = 2dD\left(t+\tau_{\text{p}}(e^{-t/\tau_{\text{p}}}-1)\right), \label{eq:persistent_random_walk}
\end{equation} 
where $\mathbf{r}_{\text{p}}(t)$ is the position of the tracer particle at time $t$ and $d$ is the dimensionality, here equal to 2. We fit our MSD curves to Eq. \ref{eq:persistent_random_walk} (see Fig. \ref{supp_fig:msd_fit} for fits for all $\la$ and $\taua$) to extract values for $D$ and $\tau_{\text{p}}$, and we plot these quantities against $\la$ and $\taua$ in the bottom row of Fig. \ref{fig:msd}. We see that $\tau_{\text{p}}$ strictly increases with $\taua$ and, for all but the smallest $\taua$, with $\la$. This is reasonable -- the larger the correlated regions of the noise field and the longer they live, the longer the tracer particle will move persistently through those regions. Less intuitively, although $D$ increases with $\la$, it is a non-monotonic function of $\taua$. This can be understood by examining trajectories for $\taua=100$ (see Fig. \ref{supp_fig:single_particle_tau=100_trajectory} and Movie S3). The trajectories feature extended periods in which particles become trapped in low-mobility regions, and must wait for a fluctuation in the noise field (with timescale $\sim \taua$) to escape.

\subsection*{Active assembly simulations}

\begin{figure*}[ht]
    \centering
    \includegraphics[width=\linewidth]{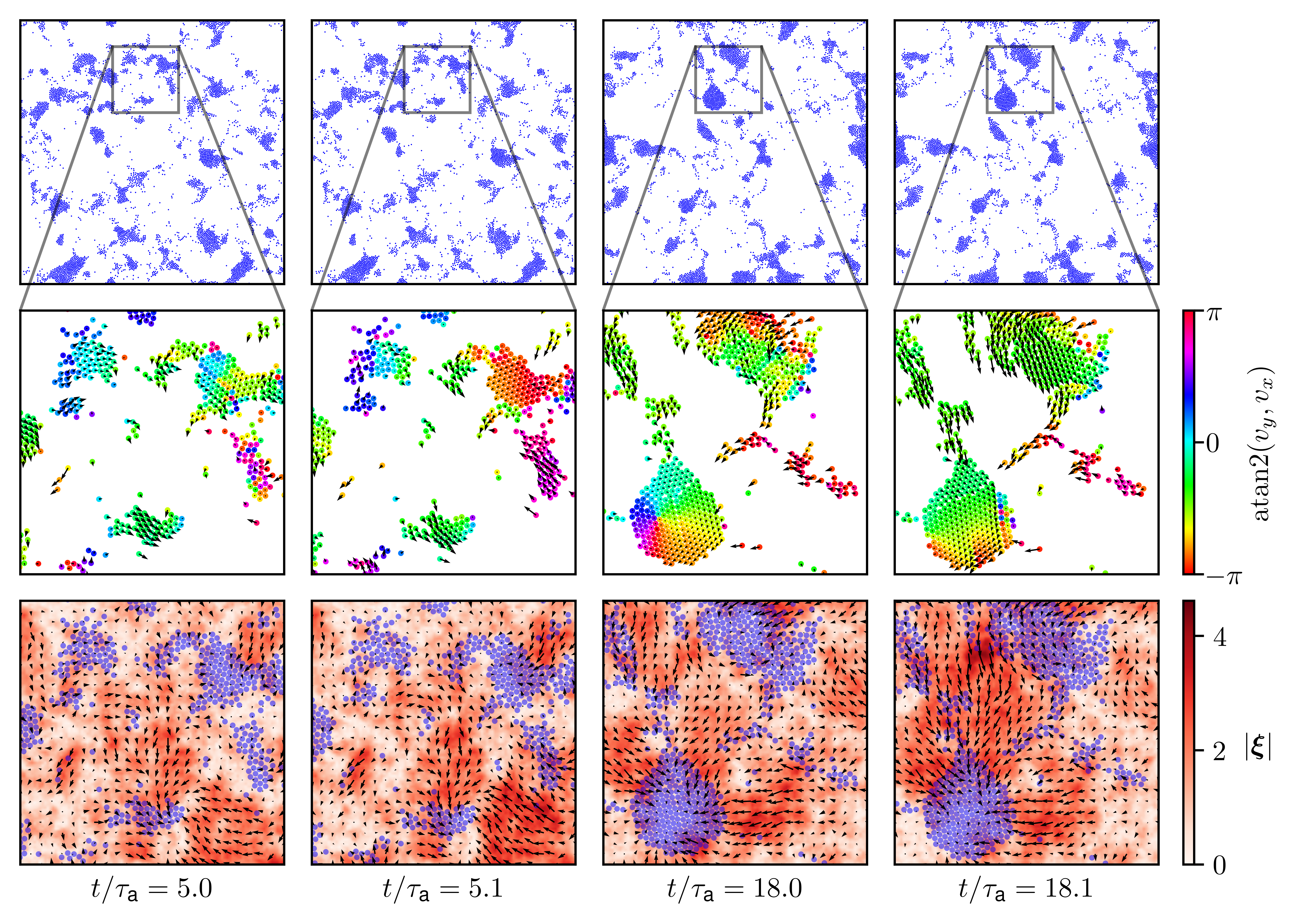}
    \caption{
    Snapshots from a trajectory of particles in an active noise field with $\va=1$, $\la=10$, and $\taua=10$, for a packing fraction of $\phi=0.1$ ($N=5016$). Snapshots in the left two columns are separated by a time interval much less than the correlation time, $\taua$; snapshots in the right two columns are similarly separated in time, but taken several correlation times later in the trajectory. The top row shows the entire simulation box, while the middle and bottom rows show a zoomed-in view focusing on one cluster. In the middle row, arrows indicate the magnitude and direction of the velocity of each particle, while colors indicate the orientation of the velocity vector. In the bottom row, both particles (translucent blue) and the active noise field (arrows and shades of red, as in Fig. \ref{fig:active_noise_vs_expt}) are shown. An animation corresponding to this trajectory is provided in Movie S4. 
    }
    \label{fig:hard_sphere_trajectory}
\end{figure*}

\begin{figure}[htpb!]
    \centering
    \includegraphics[width=\linewidth]{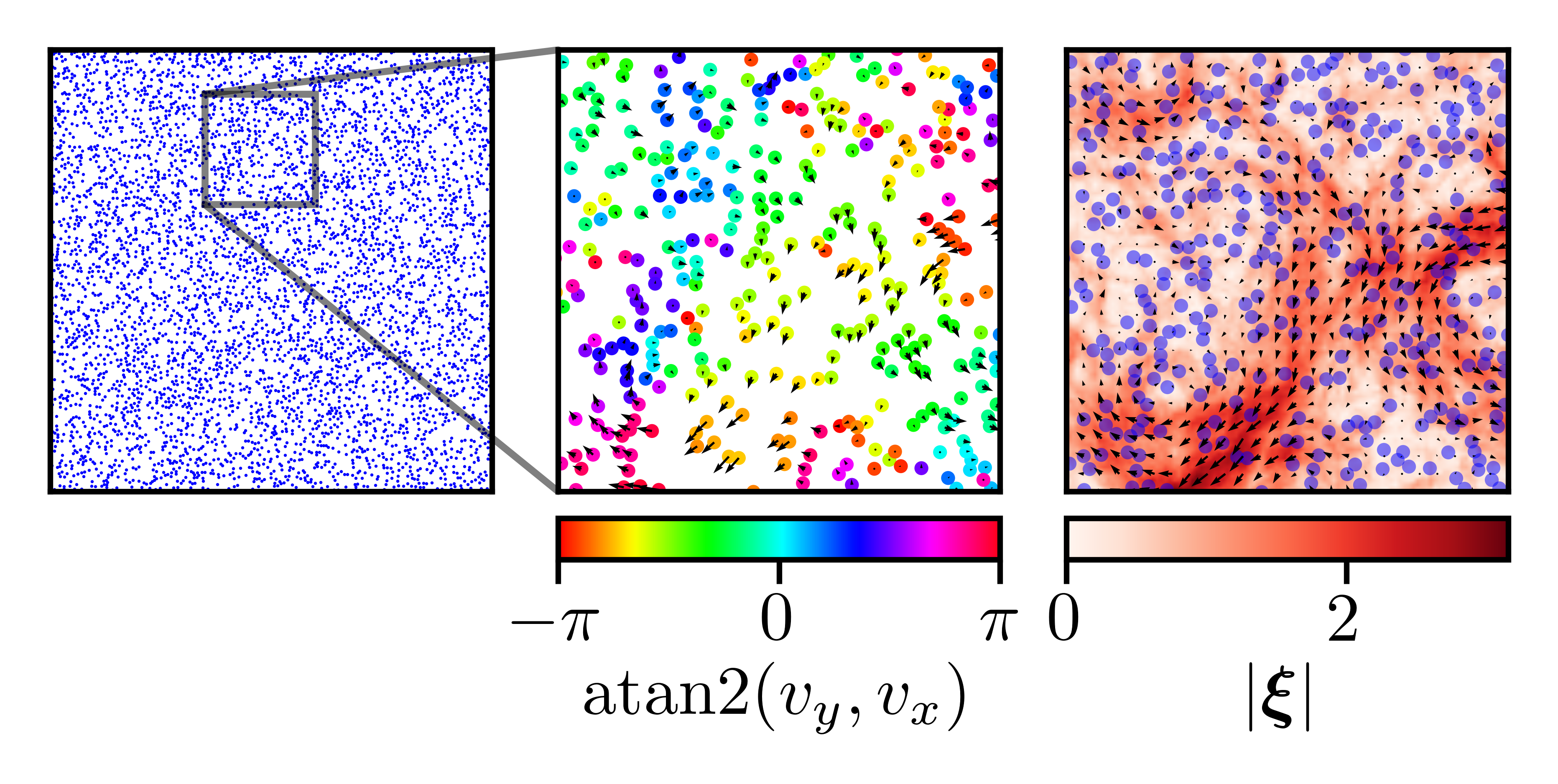}
    \caption{Snapshot from a trajectory of particles in an incompressible active noise field with $\va=1$, $\la=10$, and $\taua=10$, for a packing fraction of $\phi=0.1$ ($N=5016$). The left panel shows the entire simulation box, while the middle and right panels show a zoomed-in view. In the middle panel, arrows indicate the magnitude and direction of the velocity of each particle, while colors indicate the orientation of the velocity vector. In the right panel, both particles (translucent blue) and the active noise field (arrows and shades of red, as in Fig. \ref{fig:active_noise_vs_expt}) are shown.}
    \label{fig:incompressible}
\end{figure}

A single tracer particle in the active bath is well described as a persistent random walker, consistent with the above-mentioned simpler models of active baths that lack spatial correlations. In contrast, spatial correlations in the active bath dramatically affect multi-particle organization. To investigate assembly behavior, we performed the simulations over a wide range of $\taua$, $\la$, and $\va$. We show representative configurations from one such trajectory in Fig. \ref{fig:hard_sphere_trajectory} and snapshots for $\va=1$ and varying $\la$ and $\taua$ in Fig. \ref{fig:hard_sphere_multipanel}. For small $\la$ and $\taua$, the particles are dispersed uniformly throughout the simulation box and show no obvious spatial organization, as expected for a passive hard sphere fluid. However, as both $\la$ and $\taua$ increase, particles form dense clusters, with sizes that increase with $\la$ and $\taua$. Moreover, the particles form finite-sized clusters with correlated velocities that result in highly collective motion (such as cluster rotation, see Fig. \ref{fig:hard_sphere_trajectory}). 

This behavior is a marked contrast to motility-induced phase separation (MIPS), in which self-propelled particles form assemblies of unlimited size whose internal dynamics (at least in 2D) are slow and glassy \cite{Redner2013,Omar2021,Berthier2019,Wysocki2014}.
Indeed, if we start from a MIPS-like configuration in which the disks are closely packed rather than randomly dispersed, the crystal quickly breaks apart and we observe the same apparent steady state of dynamic finite-sized clusters (Fig. \ref{supp_fig:phase_sep_trajectory}). Only for $\phi=0.4$ and relatively large values of $\la$ and $\taua$ do we observe system-spanning clusters. These findings are consistent with simulations by Foffano et al. \cite{Foffano2019}, who observed dynamic clustering of colloids in a quasi-two-dimensional continuum active nematic. Importantly, the comparative simplicity of our approach, and our ability to tune $\la$ and $\taua$, allows us to determine how clustering depends individually on $\la$ and $\taua$, thus disentangling the influence of active length and time scales on emergent organization.

\textit{Clustering mechanism and the importance of compressibility.} The compressibility of the noise field is crucial for cluster formation. Indeed, simulations of repulsive disks in an incompressible active noise field exhibit no large-scale clusters (Fig. \ref{fig:incompressible}), although closely spaced disks exhibit spatiotemporally correlated velocities as in the case of compressible noise. As we will see in the following section, clustering arises from the tendency of disks to settle in low-noise regions, which cannot occur without ``sinks'' ($\nabla \cdot \bxi <0$). Our model thus predicts that purely repulsive colloids would not cluster, for example, within active nematics confined to a 2D interface \cite{Sanchez2012}. However, our model predicts that clustering can occur within a 2D slice of a 3D incompressible fluid (since the projected flow \textit{is} compressible in general), as observed in the simulations of Foffano et al. \cite{Foffano2019}.

\begin{figure*}[ht]
    \centering
    \includegraphics[width=\linewidth]{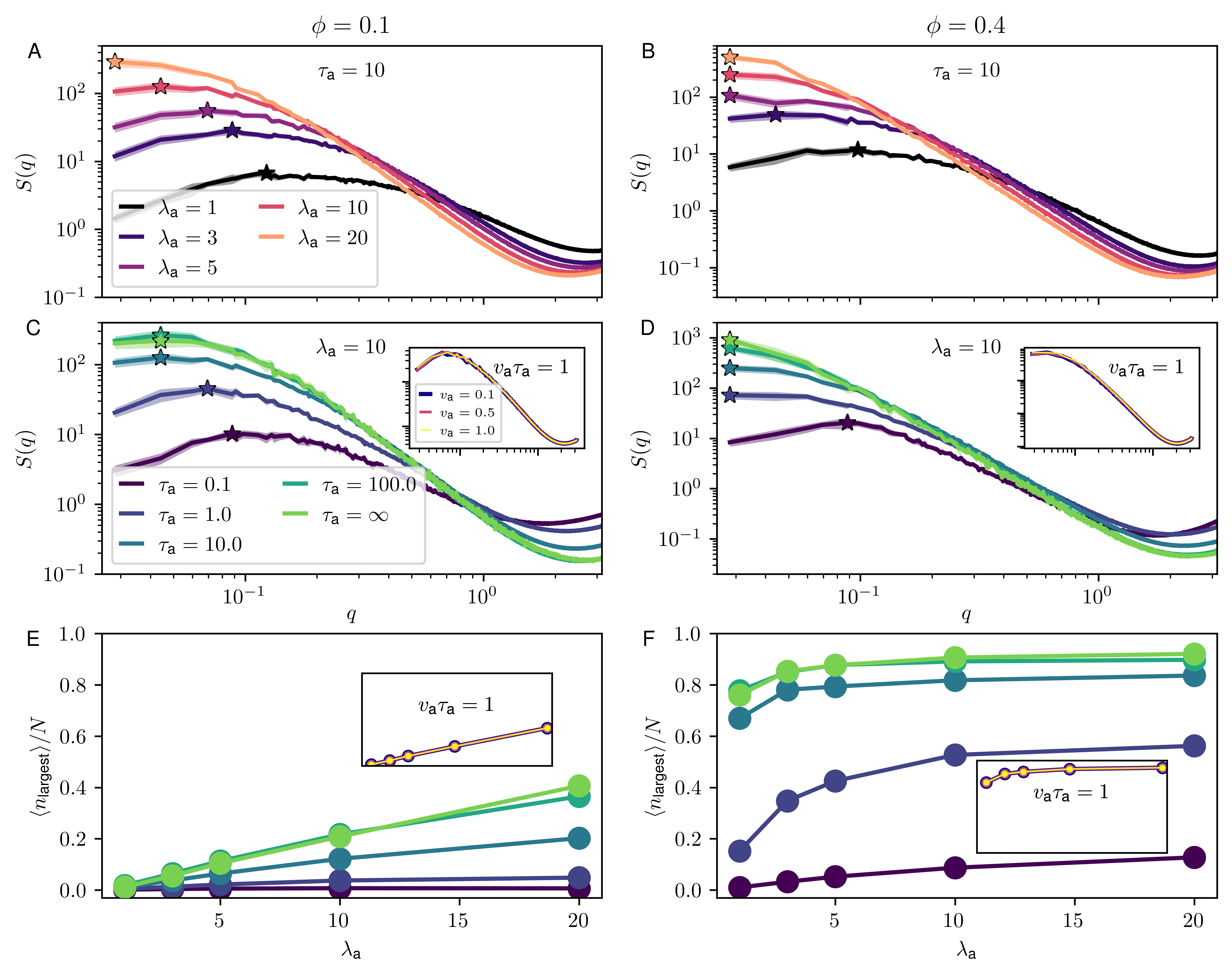}
    \caption {
    Measures of structure for particles in an active noise field. Panels A, C, E: $\phi=0.1$; Panels B, D, F: $\phi=0.4.$ Panels A,B: Structure factor $S(q)$ for fixed $\taua=10$ for different values of $\la$. Panels C,D: Structure factor $S(q)$ for fixed $\la=5$ for different values of $\taua$. In each case the star indicates the maximum value of $S(q)$. Note that this maximum value tends to increase with $\la$ and $\taua$. The insets show $S(q)$ versus $q$ for different $\va$, but fixed $\va \taua$. Panels E,F: Average size of the largest cluster, $\nLA$ for varying $\la$ as a function of $\taua$. The insets show $\nLA$ as a function of $\la$ for different $\va$, but fixed $\va \taua$.}
    \label{fig:sq_nlargest}
\end{figure*}

We quantify the structure of repulsive disk clusters in a compressible noise field via the static structure factor, $S(q)=\frac{1}{N}\left|\sum_je^{i\bq\cdot\mathbf{r}_j}\right|^2$ (where $N$ is the total number of particles and $\mathbf{r}_j$ is the position of particle $j$), which we show for selected $\la$ and $\taua$ in the top two rows of Fig. \ref{fig:sq_nlargest} (see SM Fig. \ref{supp_fig:sq_vary_va} for a wider range of $\la$, $\taua$, and $\va$). Unlike for a phase-separated system, in which $S(q)$ has a peak at the smallest nonzero wavevector $q$, here $S(q)$ exhibits a peak for a value of $q$ intermediate between those associated with the system size ($q=2\pi/L$) and the particle diameter ($q=2\pi$) for all but the largest $\la$ and $\taua$. This peak shifts to smaller $q$ (i.e., larger scales)  as both $\la$ and $\taua$ increase. The scaling of the peak locations with these two parameters is shown in Fig. \ref{supp_fig:sq_peak_scaling}. Notably, the length scale associated with the peak is not simply equal to the active noise correlation length $\la$, and is in fact much larger than $\la$. Note also that the quantity $\va \taua$ constitutes a second active length scale. The structure factor is sensitive only to $\la$ and to $\va \taua$; that is, if we change $\va$ and $\taua$ while keeping their product fixed, $S(q)$ is unchanged (see insets in the middle panels of Fig. \ref{fig:sq_nlargest} as well as Fig. \ref{supp_fig:sq_vary_va}). The only difference we have observed between trajectories with equal $\va \taua$ is that the smaller (larger) $\va$, the slower (faster) the particle motion (see Fig. \ref{supp_fig:vary_va_trajectories}).

Although the peak in $S(q)$ implies a characteristic length scale for density fluctuations, the distribution of cluster sizes $P(n)$ is broad, decaying as a power law with an exponential cutoff at large $n$ for most parameters (see Fig. \ref{supp_fig:csd}; for $\phi=0.4$ and large $\la$, $\taua$ there is a peak at large $n$, consistent with a single macroscopic cluster). Das and Barma \cite{Das2000,Das2001} observed similarly dynamic clusters with broadly-distributed sizes in a 1D lattice model of particles sliding on a fluctuating surface. This wide variation in cluster sizes makes it challenging to obtain good averages (especially at long wavelengths), which accounts for the noisiness of the $S(q)$ plots in Fig. \ref{fig:sq_nlargest}. However, although $P(n)$ (as well as its mass-weighted version, see Fig. \ref{supp_fig:csd_mw}) decays steadily with $n$ for all $\la$ and $\taua$ (when $\phi=0.1$), the distribution of {\it largest} cluster sizes, $P(\nL)$, has a well-defined peak which increases with $\la$ and $\taua$ (see Fig. \ref{supp_fig:largest_cluster_histograms}). 
In the bottom row of Fig. \ref{fig:sq_nlargest} we plot the average largest cluster size, $\nLA$, normalized by the total number of particles. While $\nLA$ tends to increase with both $\la$ and $\taua$, it only approaches $N$ for $\phi=0.4$. For $\phi=0.1$, $\nLA$ is significantly smaller than $N$, even when the peak of $S(q)$ occurs at the smallest allowed $q$. We attribute this to the highly anisotropic shape of the clusters (see Figs. \ref{fig:hard_sphere_multipanel} and \ref{fig:hard_sphere_trajectory}). Highly elongated clusters can span lengths comparable to that of the simulation box (thus contributing to $S(q=2\pi/L)$ even when they contain significantly fewer than $N$ particles (see Fig. \ref{supp_fig:elongated_clusters} for examples in addition to the top right panels of Fig. \ref{fig:hard_sphere_multipanel}). 

\subsection*{Fluctuating hydrodynamics}
Continuum theories \cite{Marchetti2013,Julicher2018}, such as fluctuating hydrodynamics \cite{Agranov2021,Han2021}, have provided valuable insights into the collective organization of active materials. In an attempt to better understand active noise-induced clustering, we derive (see SI Section \ref{section:particle_density}) a hydrodynamic equation for the density of repulsive disks, $\rho(\mathbf{r},t) = \sum_{i=1}^N\delta(\mathbf{r}-\mathbf{r}_i(t))$:
\begin{align*}
    \frac{\partial \rho(\mathbf{r},t)}{\partial t} = &-\nabla_{\mathbf{r}} \cdot \left[\rho(\mathbf{r},t)\bxi(\mathbf{r},t)\right] \\
    &+\nabla_{\mathbf{r}} \cdot \left[\rho(\mathbf{r},t)\int \diff{\mathbf{r}'}\rho(\mathbf{r}',t)\nabla_{\mathbf{r}}u(\mathbf{r}-\mathbf{r'}) \right], \numberthis \label{eq:continuum}
\end{align*}
where $u$ is the inter-particle potential (see Methods). Linear stability analysis of Eq. \ref{eq:continuum} predicts that a state with spatially uniform density is stable with respect to linear, noise-averaged density fluctuations (see SI Section \ref{section:linear_stability}). One must therefore go beyond this simple approach to understand clustering. However, the fact that the noise is temporally correlated (i.e. non-Markovian) makes this task highly challenging: it results in an infinite hierarchy of equations for statistical moments of the density, whose closure is unknown \cite{Klyatskin1974a,Schekochihin2001,Rowan2024}. Perturbation theory can be employed to study the short-correlation-time limit \cite{Schekochihin2001,Ueda2016}, but in our system clusters are most prominent for relatively long correlation times. Therefore, we investigate the clustering mechanism numerically.

\subsection*{Clustering mechanism}
\begin{figure}[ht]
    \centering
    \includegraphics[width=\linewidth]{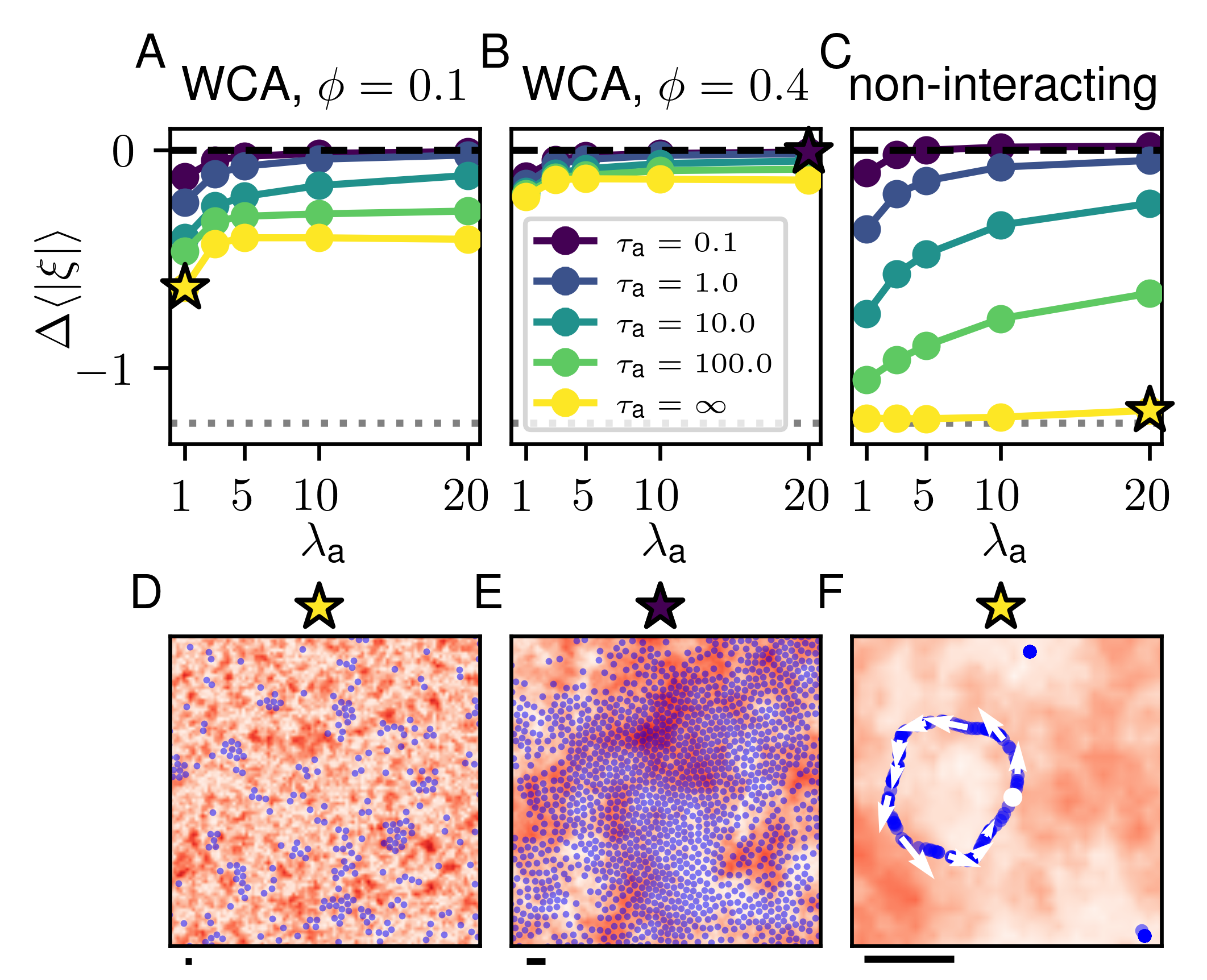}
    \caption{Panels A-C: difference between particle-averaged and space-averaged noise magnitude, $\xidiff$, as a function of $\la$ for varying $\taua$. The black dashed line at zero indicates no difference between the conditional average and the average over all space; the gray dotted line marks what the difference would be if every particle sat at a point with zero noise. Panels A, D:  $\phi=0.1$; Panels B, E: $\phi=0.4$; Panels C, F: non-interacting particles with $\phi=0.1$. Panels D-F: example configurations illustrating large negative values of $\xidiff$ (Panels D, F) and a value of $\xidiff$ close to zero (Panel E). Configurations correspond to conditions marked by stars in the top row. The snapshots are zoomed-in to better show details; the black lines indicate the value of $\la$. Panel F shows non-interacting particles trapped in a ``vortex'', with one particle shown in white and its velocity at different times shown as white arrows along the vortex. Other blue markers not in the vortex represent particles trapped in local minima.}
    \label{fig:particle_noise_corr}
\end{figure}

To elucidate the mechanism of cluster formation, we examine the correlation between noise and particle density.  Fig. \ref{fig:particle_noise_corr} shows the average magnitude of the noise field at positions where particles are present (the ``particle-averaged'' noise magnitude), $\xipart=\sum_{i=1}^N\langle |\bxi(\mathbf{r}_i(t),t)|\rangle/N$ minus the noise field magnitude averaged over all space (the ``space-averaged'' noise magnitude), $\xispace=\sum_{\mathbf{r}}\langle |\bxi(\mathbf{r},t)|\rangle/N_xN_y$, where $\mathbf{r}$ denotes points in the active noise grid and $N_x$ and $N_y$ are the number of grid points along $x$ and $y$. We denote the difference as $\xidiff \equiv \xipart  - \xispace $. If particles were distributed uniformly in space, then this quantity would be zero on average. We instead observe that $
\xidiff<0$ for a wide range of $\la$ and $\taua$, which largely coincides with the parameters for which we see clustering. The magnitude of $\xidiff$ increases uniformly with $\taua$, suggesting that particles settle in low-mobility regions of the noise field if given enough time (i.e. if the timescale of changes in the field is longer than the time it takes for particles to ``find'' low-noise regions). The configurations in the bottom row of Fig. \ref{fig:hard_sphere_trajectory} support this idea. If every particle sat at a position of zero noise, then $\xidiff$ would simply equal minus the space-averaged noise, $-\va\sqrt{\pi/2}$ in 2D. Volume exclusion among the disks prevents the difference between particle- and space-averaged noise from closely approaching this value, and explains why $\xidiff$ is significantly smaller for the higher packing fraction $\phi=0.4$ than for $\phi=0.1$. To support this point, we have also computed $\xidiff$ for collections of non-interacting particles, for which the limiting value of $-\va\sqrt{\pi/2}$ is approached much more closely (Fig. \ref{fig:particle_noise_corr}, rightmost column). The associated non-interacting configurations also suggest a reason why $\xidiff$ increases modestly with $\la$: at large $\la$, particles tend to become trapped in vortices in the noise field rather than at local minima (see Fig. \ref{fig:particle_noise_corr}F).

While the tendency of particles to settle in low-mobility regions of the noise field explains the existence of clusters, it does not explain why the cluster length scales (Fig. \ref{fig:sq_nlargest}) are much larger than the characteristic noise field scale $\la$. Examination of dynamical trajectories suggests a mechanism driving larger clusters. Because particles within a cluster tend to have correlated velocities over a timescale $\taua$, clusters exhibit persistent translational and rotational motions (see Fig. \ref{fig:hard_sphere_trajectory} and Movies S1, S2, and S4). Collisions between such clusters with each other and free particles then lead to further cluster growth as well as net rotation of clusters. While this mechanism is reminiscent of MIPS, note that the persistent dynamics and rotation of the clusters themselves is very different than in a MIPS system.

\section*{\label{sec:conclusion}Discussion}

\begin{figure}[ht]
    \centering
    \includegraphics[width=\linewidth]{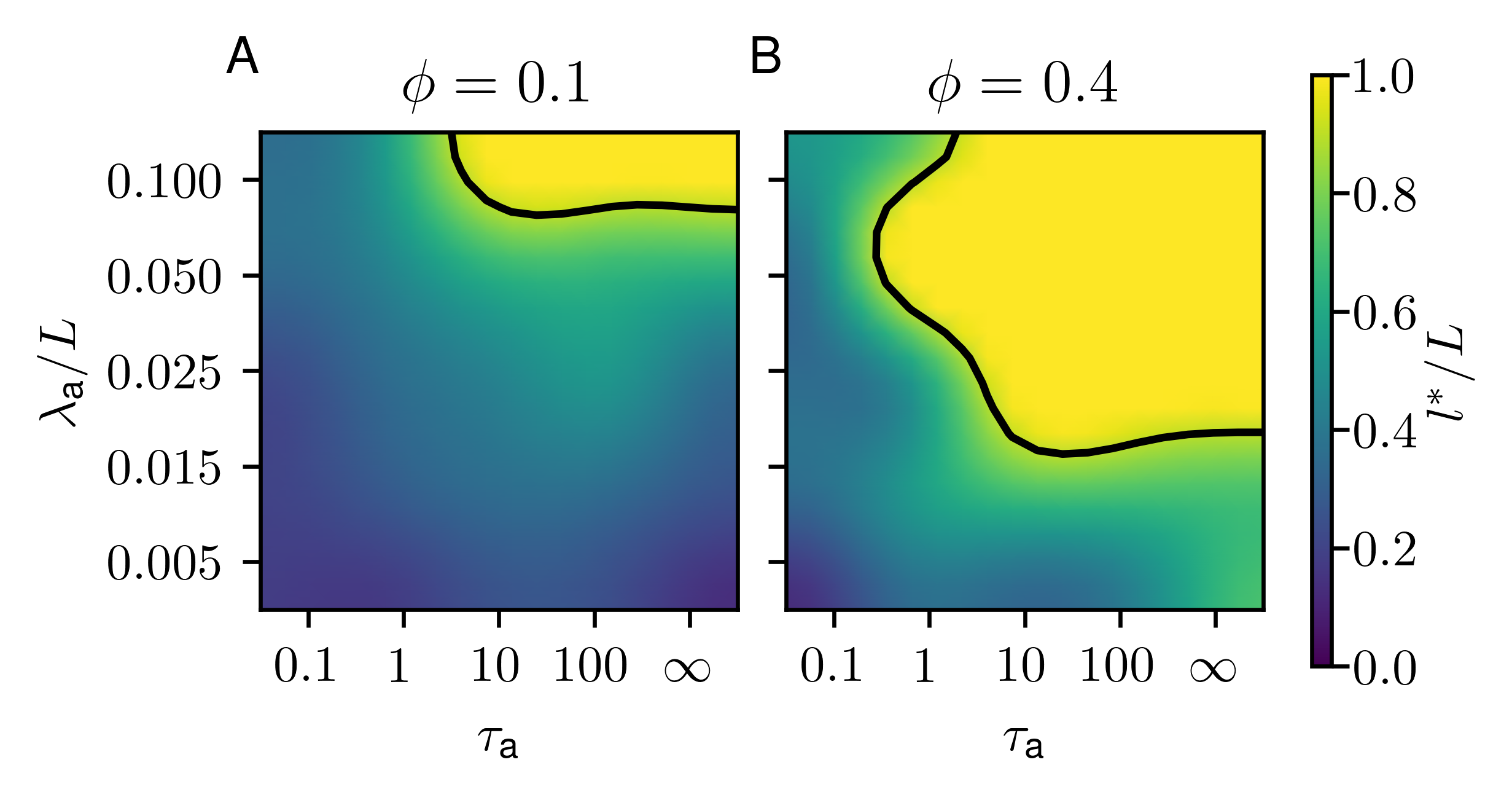}
    \caption {
    ``Phase diagram'' of repulsive disk clustering in an active noise field. Heatmaps show the length scale $l^*=2\pi/q^*$ associated with the peak of the structure factor, $S(q^*)$, as a function of $\la$ and $\taua$ for (A) $\phi=0.1$ and (B) $\phi=0.4$. Here, both $\la$ and $l^*$ are normalized by the linear system size $L=200$ to emphasize how system-spanning clusters can arise even for correlation lengths far below the system size. Black lines represent contours equal to 0.9, i.e. 90\% of the linear system size. Bicubic interpolation was used to smooth the heat maps.}
    \label{fig:sq_phase_diagram}
\end{figure}

We have shown how active flows, modeled by a spatiotemporally correlated noise field, can sculpt the organization of soft materials. For repulsive colloids, these flows lead to structures which have significantly different properties than those formed by the passive system or by self-propelled particles. The cluster dimensions can be tuned by the active noise correlation length and time, which can be experimentally controlled by changing parameters such as activity or channel height \cite{Lemma2019,Varghese2020,Chandrakar2020}. We summarize how cluster sizes depend on these parameters in Fig. \ref{fig:sq_phase_diagram}. The ability to tune the degree of clustering may in turn enable spatiotemporal control over chemical reactions among colloids. Locally enhancing the concentration of molecules grafted to or produced at colloidal surfaces could, for example, optimize reaction efficiencies \cite{Castellana2014} or promote chemical communication between a controlled number of colloids \cite{Goth2024,Huang2024}. 

While active phase separation can be arrested by direct interactions between the particles \cite{Theeyancheri2024a,Adorjani2024}, tuning the correlation length and time scales of the active fluid allows for controlled cluster sizes that are much larger than the scale of particle interactions.

Another important advantage of our model is that its parameters -- the mean active flow speed $\va$ and the active velocity correlation length $\la$ and time $\taua$ -- can be directly measured in experiments. With such measurements, the model can provide bespoke active bath models of specific experimental systems.

Our model extends trivially to 3D systems and to more complicated particle models. The latter could yield interesting collective behavior even in incompressible flows (in which repulsive disks fail to cluster). For example, elongated objects tend to flow-align \cite{Ericksen1959,Leslie1966,Helfrich1969}. Could active noise be exploited to manipulate the sizes and lifetimes of nematic domains? Similarly, our model could prove useful for investigating self-organization of crosslinked filament networks in active fluids \cite{Berezney2022,Berezney2024}. 

We have implemented our simulation code using GPUs to achieve very fast run times, interfaced it with the widely-used HOOMD-blue GPU-enabled simulation package \cite{Anderson2020}, and made it publicly available on GitHub (see Methods for URLs). Our computational approach is a simple, efficient, and versatile method to model the effect of active fluids, opening the door to fast and large-scale simulations of complex soft materials in active flows.

\section*{\label{sec:model}Methods}

\subsection*{Active bath model}

To generate realizations of the noise field, we follow a methodology described in Refs. \cite{Garcia-Ojalvo1999,Sagues2007}. Discretizing space on a regular Cartesian grid and assuming periodic boundary conditions, each allowed Fourier component $\tilde{\bxi}_{\bq}(t)$ of the noise field evolves according to:
\begin{equation}
    \partial_t \tilde{\bxi}_{\bq}(t) = -\tilde{\bxi}_{\bq}(t)/\taua + \va \sqrt{\frac{2\tilde{c}_{\bq}}{\taua}}\tilde{\boldsymbol{\eta}}_{\bq}(t), \label{eq:fourier_ou}
\end{equation}
where $\tilde{\boldsymbol{\eta}}_{\mathbf{q}}$ is a zero-mean white noise: 
\begin{equation}
    \langle \tilde{\eta}^{\mu}_{\bq}(t) \tilde{\eta}^{\nu}_{\bq'}(t') \rangle = N\delta_{\mu\nu} \delta_{\bq+\bq'}\delta(t-t'),
\end{equation} and $\tilde{c}_{\bq}$ is the Fourier transform of the desired spatial correlation function. For exponential spatial correlations, we have:
\begin{subequations}
\begin{align}
    \tilde{c}_{\bq} &= \frac{c_{\bq}}{\sum_{\bq}c_{\bq}} \\
        c_{\bq} &= \frac{1}{(1+\la^2|\bq|^2)^{(d+1)/2}},
\end{align}
\end{subequations}
The normalization of $\tilde{c}_{\bq}$ ensures that the magnitude of the noise variance is governed solely by $\va$. In our numerical implementation, we first divide our simulation box into a grid of size $n_x\times n_y$ in two dimensions ($n_x\times n_y \times n_z$ in three dimensions) with positions at grid centers labeled $\mathbf{R}$. We create an initial reciprocal-space noise field $\tilde{\bxi}_{\bq}(t=0)$ by generating independent Gaussian random numbers with variance 1 for each $\mathbf{R}$, taking the discrete Fourier transform (using NumPy's \cite{Harris2020} fast Fourier transform implementation), and multiplying the resulting field by $\va \sqrt{\tilde{c}_{\bq}}$. To advance the dynamics, we solve Eq. \ref{eq:fourier_ou} using an Euler-Maruyama scheme \cite{Kloeden1999}:
\begin{equation}
    \tilde{\bxi}_{\bq}(t+\Delta t) = (1-\Delta t/\taua)\tilde{\bxi}_{\bq}(t) + \va \sqrt{\frac{2\tilde{c}_{\bq}}{\taua}} \int_0^{\Delta t}\diff{s}\tilde{\boldsymbol{\eta}}_{\bq}(s),
\end{equation} 
where the noise increment $\int_0^{\Delta t}\diff{s}\tilde{\boldsymbol{\eta}}_{\bq}(s)$ is computed by generating a Gaussian random number of variance $\Delta t$ at each grid point $\mathbf{R}$ and then taking the discrete Fourier transform of this white noise field. The correlated noise field $\bxi(\mathbf{R},t)$ at each grid point $\mathbf{R}$ and time $t$ is then calculated by taking the discrete inverse Fourier transform of $\tilde{\bxi}_{\bq}(t)$:
\begin{equation}
    \bxi(\mathbf{R},t) = \frac{1}{n_xn_y}\sum_{\bq}e^{-i\bq\cdot\mathbf{R}} \tilde{\bxi}_{\bq}(t). \label{eq:noise_real_space}
\end{equation}
To compute values of the noise field at points in space $\mathbf{r}$ between the grid points $\mathbf{R}$, we use bilinear (in two dimensions) or trilinear (in three dimensions) interpolation (as implemented in SciPy's ``ndimage.map\_coordinates'' function \cite{Virtanen2020}) of each component of $\bxi(\mathbf{R},t)$. We have computed spatial and temporal correlation functions of our noise fields for varying $\la$ and $\taua$, confirming that they have the desired statistics (Figs. \ref{supp_fig:noise_tcorr}, \ref{supp_fig:noise_rcorr}).

To make the noise field incompressible, one can impose the condition $\nabla \cdot \bxi=0$ (equivalently, $\tilde{\bxi}_{\bq}\cdot \bq=0$) by projecting out components of $\tilde{\bxi}_{\bq}$ parallel to $\bq$ at each time step:
\begin{equation}
    \tilde{\bxi}_{\bq}(t) \rightarrow \left(\mathbf{I}-\hat{\bq}\hat{\bq}\right)\tilde{\bxi}_{\bq}(t),
\end{equation}
where $\mathbf{I}$ is the identity matrix and $\hat{\bq}$ is a unit vector with the same direction as $\mathbf{q}$. Following \cite{Saad2016}, we modify this procedure slightly to account for the discrete grid:
\begin{equation}
    \tilde{\bxi}_{\bq}(t) \rightarrow \left(\mathbf{I}-\tilde{\bq}\tilde{\bq}\right)\tilde{\bxi}_{\bq}(t),
\end{equation}
where, in 2D:
\begin{equation}
    \tilde{\bq} = \left(\frac{1}{L/n_x}\sin{(q_xL/n_x)}, \frac{1}{L/n_y}\sin{(q_yL/n_y)}\right),
\end{equation}
where $L$ is the linear system size and $q_x$, $q_y$ are Cartesian components of $\mathbf{q}$. We have confirmed that this procedure yields a divergence-free noise field (Fig. \ref{supp_fig:field_div}).

\subsection*{Tracer particle dynamics}

To illustrate the nature of motion in the active noise field, we have measured the mean squared displacement (MSD) of a single particle advected by the field. Specifically, this particle (with position $\mathbf{r}_{\text{p}}(t)$ at time $t$), obeys an overdamped Langevin equation:
\begin{equation}
    \frac{\diff{\mathbf{r}_{\text{p}}(t)}}{\diff{t}} = \bxi(\mathbf{r}_{\text{p}}(t),t).\label{eq:tracer_particle}
\end{equation}
We solve Eq. \ref{eq:tracer_particle} using standard Euler-Maruyama integration \cite{Kloeden1999} with a time step of $\Delta t=10^{-2}$ and compute the MSD $\langle |\mathbf{r}_{\text{p}}(t)-\mathbf{r}_{\text{p}}(0)|^2 \rangle$ by averaging over 100 independent trajectories of length 2000 time units.

\subsection*{Active assembly simulations}

We consider a collection of area-excluding particles in a 2D active noise field at moderate densities (packing fractions of $\phi=0.1, 0.4$). As for the tracer particle, the interacting particle dynamics obeys an overdamped Langevin equation:
\begin{equation}
    \frac{\diff{\mathbf{r}_i}}{\diff{t}} = -\nabla_{\mathbf{r}_i}U(\mathbf{r}_1,\dots,\mathbf{r}_N) + \bxi(\mathbf{r}_i,t), \label{eq:multiparticle_langevin}
\end{equation}
where $U = \sum_{i,j>i}u(|\mathbf{r}_i-\mathbf{r}_j|)$ is a sum of Weeks-Chandler-Andersen (WCA) pair potentials \cite{Weeks1971}:
\begin{equation}
    u_{\text{WCA}}(r) = \begin{cases}
    4\left[\left(\frac{1}{r}\right)^{12}-\left(\frac{1}{r}\right)^6\right] + 1, & r\leq 2^{1/6} \\
    0, & r>2^{1/6}.
    \end{cases}
\end{equation}
(Recall the reduced unit system described in Results.) For a given simulation, we initialized the particles in a random, non-overlapping configuration and then propagated the dynamics of Eq. \ref{eq:multiparticle_langevin} (using Euler-Maruyama integration \cite{Kloeden1999} with a time step $\Delta t=10^{-4}$) for $250$ unit times. We repeated this procedure 50 times for each $\phi$ over a range of $\la$, $\taua$, and $\va$ to obtain an ensemble of trajectories for each parameter set. We saved configurations every unit time. Except for the cluster size distributions (Fig. \ref{supp_fig:csd}, \ref{supp_fig:csd_mw}), which we obtained by binning data from all trajectories, reported quantities were obtained by averaging over the 50 independent trajectories. The error bars we report represent $\pm 2\times $SE, where SE is the standard error over trajectories. This approximately corresponds to a 95\% confidence interval. To identify distinct clusters and hence compute cluster sizes, we used a standard algorithm for identifying particle clusters \cite{Hoshen1976}, using a distance cutoff $r_c=1.2$ to identify connected particles.

\subsection*{Software implementation}
Our active noise generator is implemented in Python 3 and uses GPU acceleration via the CuPy \cite{Okuta2017} package to enable fast generation of the random numbers. Particle dynamics is performed using HOOMD-blue \cite{Anderson2020}, and we have created a Python package to interface the active noise generator with HOOMD-blue. A single, 2D, $2.5\times 10^6$-time step trajectory with $N=20372$ particles and a $400\times400$ active noise grid takes $\approx1.5$ hours to run on NERSC's Perlmutter system, using one AMD EPYC 7763 CPU and one NVIDIA A100 GPU. The code is available along with example usage and analysis scripts on Github:
\begin{itemize}
    \item Active noise generator:
    \begin{sloppypar}
        https://github.com/Layne28/ActiveNoise
    \end{sloppypar}
    \item Interface to HOOMD-blue:
    \begin{sloppypar}
        https://github.com/Layne28/ActiveNoiseHoomd
    \end{sloppypar}
    \item Example of usage:
    \begin{sloppypar}
        https://github.com/Layne28/active-assembly-hoomd
    \end{sloppypar}
    \item Analysis scripts: 
    \begin{sloppypar}
        https://github.com/Layne28/AnalysisTools
    \end{sloppypar}
\end{itemize}

\begin{acknowledgments}
We acknowledge inspiring conversations with John Berezney and Sattvic Ray. We also thank Sattvic Ray for providing the experimental image and velocity data in Fig. 2. This work was primarily supported by the Department of Energy (DOE) DE-SC0022291 (L.B.F. and M.F.H.). A.B. acknowledges support from the NSF through DMR-MRSEC 2011846 and DMR 2202353. Computing resources were provided the National Energy Research Scientific Computing Center (NERSC), a Department of Energy Office of Science User Facility using NERSC award BES-ERCAP0026774. Initial results were generated using the NSF ACCESS allocation TG-MCB090163 (Bridges-2) and the Brandeis HPCC which is partially supported by the NSF through DMR-MRSEC 2011846 and OAC-1920147.
\end{acknowledgments}



\bibliography{references}

\providecommand{\latin}[1]{#1}
\makeatletter
\providecommand{\doi}
  {\begingroup\let\do\@makeother\dospecials
  \catcode`\{=1 \catcode`\}=2 \doi@aux}
\providecommand{\doi@aux}[1]{\endgroup\texttt{#1}}
\makeatother
\providecommand*\mcitethebibliography{\thebibliography}
\csname @ifundefined\endcsname{endmcitethebibliography}
  {\let\endmcitethebibliography\endthebibliography}{}
\begin{mcitethebibliography}{120}
\providecommand*\natexlab[1]{#1}
\providecommand*\mciteSetBstSublistMode[1]{}
\providecommand*\mciteSetBstMaxWidthForm[2]{}
\providecommand*\mciteBstWouldAddEndPuncttrue
  {\def\EndOfBibitem{\unskip.}}
\providecommand*\mciteBstWouldAddEndPunctfalse
  {\let\EndOfBibitem\relax}
\providecommand*\mciteSetBstMidEndSepPunct[3]{}
\providecommand*\mciteSetBstSublistLabelBeginEnd[3]{}
\providecommand*\EndOfBibitem{}
\mciteSetBstSublistMode{f}
\mciteSetBstMaxWidthForm{subitem}{(\alph{mcitesubitemcount})}
\mciteSetBstSublistLabelBeginEnd
  {\mcitemaxwidthsubitemform\space}
  {\relax}
  {\relax}

\bibitem[Whitesides and Grzybowski(2002)Whitesides, and
  Grzybowski]{Whitesides2002}
Whitesides,~G.~M.; Grzybowski,~B. Self-{{Assembly}} at {{All Scales}}.
  \emph{Science} \textbf{2002}, \emph{295}, 2418--2421\relax
\mciteBstWouldAddEndPuncttrue
\mciteSetBstMidEndSepPunct{\mcitedefaultmidpunct}
{\mcitedefaultendpunct}{\mcitedefaultseppunct}\relax
\EndOfBibitem
\bibitem[Boles \latin{et~al.}(2016)Boles, Engel, and Talapin]{Boles2016}
Boles,~M.~A.; Engel,~M.; Talapin,~D.~V. Self-{{Assembly}} of {{Colloidal
  Nanocrystals}}: {{From Intricate Structures}} to {{Functional Materials}}.
  \emph{Chemical Reviews} \textbf{2016}, \emph{116}, 11220--11289\relax
\mciteBstWouldAddEndPuncttrue
\mciteSetBstMidEndSepPunct{\mcitedefaultmidpunct}
{\mcitedefaultendpunct}{\mcitedefaultseppunct}\relax
\EndOfBibitem
\bibitem[Hagan and Grason(2021)Hagan, and Grason]{Hagan2021}
Hagan,~M.~F.; Grason,~G.~M. Equilibrium Mechanisms of Self-Limiting Assembly.
  \emph{Reviews of Modern Physics} \textbf{2021}, \emph{93}, 025008\relax
\mciteBstWouldAddEndPuncttrue
\mciteSetBstMidEndSepPunct{\mcitedefaultmidpunct}
{\mcitedefaultendpunct}{\mcitedefaultseppunct}\relax
\EndOfBibitem
\bibitem[Grober \latin{et~al.}(2023)Grober, Palaia, U{\c c}ar, Hannezo, {\v
  S}ari{\'c}, and Palacci]{Grober2023}
Grober,~D.; Palaia,~I.; U{\c c}ar,~M.~C.; Hannezo,~E.; {\v S}ari{\'c},~A.;
  Palacci,~J. Unconventional Colloidal Aggregation in Chiral Bacterial Baths.
  \emph{Nature Physics} \textbf{2023}, \emph{19}, 1680--1688\relax
\mciteBstWouldAddEndPuncttrue
\mciteSetBstMidEndSepPunct{\mcitedefaultmidpunct}
{\mcitedefaultendpunct}{\mcitedefaultseppunct}\relax
\EndOfBibitem
\bibitem[Berezney \latin{et~al.}(2022)Berezney, Goode, Fraden, and
  Dogic]{Berezney2022}
Berezney,~J.; Goode,~B.~L.; Fraden,~S.; Dogic,~Z. Extensile to Contractile
  Transition in Active Microtubule--Actin Composites Generates Layered Asters
  with Programmable Lifetimes. \emph{Proceedings of the National Academy of
  Sciences} \textbf{2022}, \emph{119}, e2115895119\relax
\mciteBstWouldAddEndPuncttrue
\mciteSetBstMidEndSepPunct{\mcitedefaultmidpunct}
{\mcitedefaultendpunct}{\mcitedefaultseppunct}\relax
\EndOfBibitem
\bibitem[Berezney \latin{et~al.}(2024)Berezney, Ray, Kolvin, Bowick, Fraden,
  and Dogic]{Berezney2024}
Berezney,~J.; Ray,~S.; Kolvin,~I.; Bowick,~M.; Fraden,~S.; Dogic,~Z.
  Controlling Assembly and Oscillations of Elastic Membranes with an Active
  Fluid. https://arxiv.org/abs/2408.14699v1, 2024\relax
\mciteBstWouldAddEndPuncttrue
\mciteSetBstMidEndSepPunct{\mcitedefaultmidpunct}
{\mcitedefaultendpunct}{\mcitedefaultseppunct}\relax
\EndOfBibitem
\bibitem[Lemma \latin{et~al.}(2019)Lemma, DeCamp, You, Giomi, and
  Dogic]{Lemma2019}
Lemma,~L.~M.; DeCamp,~S.~J.; You,~Z.; Giomi,~L.; Dogic,~Z. Statistical
  Properties of Autonomous Flows in {{2D}} Active Nematics. \emph{Soft Matter}
  \textbf{2019}, \emph{15}, 3264--3272\relax
\mciteBstWouldAddEndPuncttrue
\mciteSetBstMidEndSepPunct{\mcitedefaultmidpunct}
{\mcitedefaultendpunct}{\mcitedefaultseppunct}\relax
\EndOfBibitem
\bibitem[Romanczuk \latin{et~al.}(2012)Romanczuk, B{\"a}r, Ebeling, Lindner,
  and {Schimansky-Geier}]{Romanczuk2012}
Romanczuk,~P.; B{\"a}r,~M.; Ebeling,~W.; Lindner,~B.; {Schimansky-Geier},~L.
  Active {{Brownian}} Particles. \emph{The European Physical Journal Special
  Topics} \textbf{2012}, \emph{202}, 1--162\relax
\mciteBstWouldAddEndPuncttrue
\mciteSetBstMidEndSepPunct{\mcitedefaultmidpunct}
{\mcitedefaultendpunct}{\mcitedefaultseppunct}\relax
\EndOfBibitem
\bibitem[Fily and Marchetti(2012)Fily, and Marchetti]{Fily2012}
Fily,~Y.; Marchetti,~M.~C. Athermal {{Phase Separation}} of {{Self-Propelled
  Particles}} with {{No Alignment}}. \emph{Physical Review Letters}
  \textbf{2012}, \emph{108}, 235702\relax
\mciteBstWouldAddEndPuncttrue
\mciteSetBstMidEndSepPunct{\mcitedefaultmidpunct}
{\mcitedefaultendpunct}{\mcitedefaultseppunct}\relax
\EndOfBibitem
\bibitem[Czir{\'o}k and Vicsek(2000)Czir{\'o}k, and Vicsek]{Czirok2000}
Czir{\'o}k,~A.; Vicsek,~T. Collective Behavior of Interacting Self-Propelled
  Particles. \emph{Physica A: Statistical Mechanics and its Applications}
  \textbf{2000}, \emph{281}, 17--29\relax
\mciteBstWouldAddEndPuncttrue
\mciteSetBstMidEndSepPunct{\mcitedefaultmidpunct}
{\mcitedefaultendpunct}{\mcitedefaultseppunct}\relax
\EndOfBibitem
\bibitem[Mishra \latin{et~al.}(2010)Mishra, Baskaran, and
  Marchetti]{Mishra2010}
Mishra,~S.; Baskaran,~A.; Marchetti,~M.~C. Fluctuations and Pattern Formation
  in Self-Propelled Particles. \emph{Physical Review E} \textbf{2010},
  \emph{81}, 061916\relax
\mciteBstWouldAddEndPuncttrue
\mciteSetBstMidEndSepPunct{\mcitedefaultmidpunct}
{\mcitedefaultendpunct}{\mcitedefaultseppunct}\relax
\EndOfBibitem
\bibitem[Aditi~Simha and Ramaswamy(2002)Aditi~Simha, and
  Ramaswamy]{AditiSimha2002}
Aditi~Simha,~R.; Ramaswamy,~S. Hydrodynamic {{Fluctuations}} and
  {{Instabilities}} in {{Ordered Suspensions}} of {{Self-Propelled Particles}}.
  \emph{Physical Review Letters} \textbf{2002}, \emph{89}, 058101\relax
\mciteBstWouldAddEndPuncttrue
\mciteSetBstMidEndSepPunct{\mcitedefaultmidpunct}
{\mcitedefaultendpunct}{\mcitedefaultseppunct}\relax
\EndOfBibitem
\bibitem[Palacci \latin{et~al.}(2013)Palacci, Sacanna, Steinberg, Pine, and
  Chaikin]{Palacci2013}
Palacci,~J.; Sacanna,~S.; Steinberg,~A.~P.; Pine,~D.~J.; Chaikin,~P.~M. Living
  {{Crystals}} of {{Light-Activated Colloidal Surfers}}. \emph{Science}
  \textbf{2013}, \emph{339}, 936--940\relax
\mciteBstWouldAddEndPuncttrue
\mciteSetBstMidEndSepPunct{\mcitedefaultmidpunct}
{\mcitedefaultendpunct}{\mcitedefaultseppunct}\relax
\EndOfBibitem
\bibitem[Palacci \latin{et~al.}(2014)Palacci, Sacanna, Kim, Yi, Pine, and
  Chaikin]{Palacci2014}
Palacci,~J.; Sacanna,~S.; Kim,~S.-H.; Yi,~G.-R.; Pine,~D.~J.; Chaikin,~P.~M.
  Light-Activated Self-Propelled Colloids. \emph{Philosophical Transactions of
  the Royal Society A: Mathematical, Physical and Engineering Sciences}
  \textbf{2014}, \emph{372}, 20130372\relax
\mciteBstWouldAddEndPuncttrue
\mciteSetBstMidEndSepPunct{\mcitedefaultmidpunct}
{\mcitedefaultendpunct}{\mcitedefaultseppunct}\relax
\EndOfBibitem
\bibitem[Z{\"o}ttl and Stark(2023)Z{\"o}ttl, and Stark]{Zottl2023a}
Z{\"o}ttl,~A.; Stark,~H. Modeling {{Active Colloids}}: {{From Active Brownian
  Particles}} to {{Hydrodynamic}} and {{Chemical Fields}}. \emph{Annual Review
  of Condensed Matter Physics} \textbf{2023}, \emph{14}, 109--127\relax
\mciteBstWouldAddEndPuncttrue
\mciteSetBstMidEndSepPunct{\mcitedefaultmidpunct}
{\mcitedefaultendpunct}{\mcitedefaultseppunct}\relax
\EndOfBibitem
\bibitem[Stenhammar \latin{et~al.}(2014)Stenhammar, Marenduzzo, J.~Allen, and
  E.~Cates]{Stenhammar2014}
Stenhammar,~J.; Marenduzzo,~D.; J.~Allen,~R.; E.~Cates,~M. Phase Behaviour of
  Active {{Brownian}} Particles: The Role of Dimensionality. \emph{Soft Matter}
  \textbf{2014}, \emph{10}, 1489--1499\relax
\mciteBstWouldAddEndPuncttrue
\mciteSetBstMidEndSepPunct{\mcitedefaultmidpunct}
{\mcitedefaultendpunct}{\mcitedefaultseppunct}\relax
\EndOfBibitem
\bibitem[Levine \latin{et~al.}(2000)Levine, Rappel, and Cohen]{Levine2000}
Levine,~H.; Rappel,~W.-J.; Cohen,~I. Self-Organization in Systems of
  Self-Propelled Particles. \emph{Physical Review E} \textbf{2000}, \emph{63},
  017101\relax
\mciteBstWouldAddEndPuncttrue
\mciteSetBstMidEndSepPunct{\mcitedefaultmidpunct}
{\mcitedefaultendpunct}{\mcitedefaultseppunct}\relax
\EndOfBibitem
\bibitem[Redner \latin{et~al.}(2013)Redner, Hagan, and Baskaran]{Redner2013}
Redner,~G.~S.; Hagan,~M.~F.; Baskaran,~A. Structure and {{Dynamics}} of a
  {{Phase-Separating Active Colloidal Fluid}}. \emph{Physical Review Letters}
  \textbf{2013}, \emph{110}, 055701\relax
\mciteBstWouldAddEndPuncttrue
\mciteSetBstMidEndSepPunct{\mcitedefaultmidpunct}
{\mcitedefaultendpunct}{\mcitedefaultseppunct}\relax
\EndOfBibitem
\bibitem[Czir{\'o}k \latin{et~al.}(1997)Czir{\'o}k, Stanley, and
  Vicsek]{Czirok1997}
Czir{\'o}k,~A.; Stanley,~H.~E.; Vicsek,~T. Spontaneously Ordered Motion of
  Self-Propelled Particles. \emph{Journal of Physics A: Mathematical and
  General} \textbf{1997}, \emph{30}, 1375\relax
\mciteBstWouldAddEndPuncttrue
\mciteSetBstMidEndSepPunct{\mcitedefaultmidpunct}
{\mcitedefaultendpunct}{\mcitedefaultseppunct}\relax
\EndOfBibitem
\bibitem[Pedersen \latin{et~al.}(2024)Pedersen, Mukherjee, Doostmohammadi,
  Mondal, and Thijssen]{Pedersen2024}
Pedersen,~M.~C.; Mukherjee,~S.; Doostmohammadi,~A.; Mondal,~C.; Thijssen,~K.
  Active Particles Knead Three-Dimensional Gels into Open Crumbs. 2024\relax
\mciteBstWouldAddEndPuncttrue
\mciteSetBstMidEndSepPunct{\mcitedefaultmidpunct}
{\mcitedefaultendpunct}{\mcitedefaultseppunct}\relax
\EndOfBibitem
\bibitem[Gokhale \latin{et~al.}(2022)Gokhale, Li, Solon, Gore, and
  Fakhri]{Gokhale2022a}
Gokhale,~S.; Li,~J.; Solon,~A.; Gore,~J.; Fakhri,~N. Dynamic Clustering of
  Passive Colloids in Dense Suspensions of Motile Bacteria. \emph{Physical
  Review E} \textbf{2022}, \emph{105}, 054605\relax
\mciteBstWouldAddEndPuncttrue
\mciteSetBstMidEndSepPunct{\mcitedefaultmidpunct}
{\mcitedefaultendpunct}{\mcitedefaultseppunct}\relax
\EndOfBibitem
\bibitem[Batton and Rotskoff(2024)Batton, and Rotskoff]{Batton2024}
Batton,~C.~H.; Rotskoff,~G.~M. Microscopic Origin of Tunable Assembly Forces in
  Chiral Active Environments. \emph{Soft Matter} \textbf{2024}, \emph{20},
  4111--4126\relax
\mciteBstWouldAddEndPuncttrue
\mciteSetBstMidEndSepPunct{\mcitedefaultmidpunct}
{\mcitedefaultendpunct}{\mcitedefaultseppunct}\relax
\EndOfBibitem
\bibitem[Omar \latin{et~al.}(2019)Omar, Wu, Wang, and Brady]{Omar2019}
Omar,~A.~K.; Wu,~Y.; Wang,~Z.-G.; Brady,~J.~F. Swimming to {{Stability}}:
  {{Structural}} and {{Dynamical Control}} {\emph{via}} {{Active Doping}}.
  \emph{ACS Nano} \textbf{2019}, \emph{13}, 560--572\relax
\mciteBstWouldAddEndPuncttrue
\mciteSetBstMidEndSepPunct{\mcitedefaultmidpunct}
{\mcitedefaultendpunct}{\mcitedefaultseppunct}\relax
\EndOfBibitem
\bibitem[Sanchez \latin{et~al.}(2012)Sanchez, Chen, DeCamp, Heymann, and
  Dogic]{Sanchez2012}
Sanchez,~T.; Chen,~D. T.~N.; DeCamp,~S.~J.; Heymann,~M.; Dogic,~Z. Spontaneous
  Motion in Hierarchically Assembled Active Matter. \emph{Nature}
  \textbf{2012}, \emph{491}, 431--434\relax
\mciteBstWouldAddEndPuncttrue
\mciteSetBstMidEndSepPunct{\mcitedefaultmidpunct}
{\mcitedefaultendpunct}{\mcitedefaultseppunct}\relax
\EndOfBibitem
\bibitem[DeCamp \latin{et~al.}(2015)DeCamp, Redner, Baskaran, Hagan, and
  Dogic]{DeCamp2015}
DeCamp,~S.~J.; Redner,~G.~S.; Baskaran,~A.; Hagan,~M.~F.; Dogic,~Z.
  Orientational {{Order}} of {{Motile Defects}} in {{Active Nematics}}.
  \emph{Nature materials} \textbf{2015}, \emph{14}, 1110--1115\relax
\mciteBstWouldAddEndPuncttrue
\mciteSetBstMidEndSepPunct{\mcitedefaultmidpunct}
{\mcitedefaultendpunct}{\mcitedefaultseppunct}\relax
\EndOfBibitem
\bibitem[Henkin \latin{et~al.}(2014)Henkin, DeCamp, Chen, Sanchez, and
  Dogic]{Henkin2014}
Henkin,~G.; DeCamp,~S.~J.; Chen,~D. T.~N.; Sanchez,~T.; Dogic,~Z. Tunable
  Dynamics of Microtubule-Based Active Isotropic Gels. \emph{Philosophical
  Transactions of the Royal Society A: Mathematical, Physical and Engineering
  Sciences} \textbf{2014}, \emph{372}, 20140142\relax
\mciteBstWouldAddEndPuncttrue
\mciteSetBstMidEndSepPunct{\mcitedefaultmidpunct}
{\mcitedefaultendpunct}{\mcitedefaultseppunct}\relax
\EndOfBibitem
\bibitem[Duclos \latin{et~al.}(2020)Duclos, Adkins, Banerjee, Peterson,
  Varghese, Kolvin, Baskaran, Pelcovits, Powers, Baskaran, Toschi, Hagan,
  Streichan, Vitelli, Beller, and Dogic]{Duclos2020}
Duclos,~G. \latin{et~al.}  Topological Structure and Dynamics of
  Three-Dimensional Active Nematics. \emph{Science} \textbf{2020}, \emph{367},
  1120--1124\relax
\mciteBstWouldAddEndPuncttrue
\mciteSetBstMidEndSepPunct{\mcitedefaultmidpunct}
{\mcitedefaultendpunct}{\mcitedefaultseppunct}\relax
\EndOfBibitem
\bibitem[Frenkel and Smit(2023)Frenkel, and Smit]{Frenkel2023}
Frenkel,~D.; Smit,~B. \emph{Understanding Molecular Simulation: From Algorithms
  to Applications}, third edition ed.; Academic Press, an imprint of Elsevier:
  London San Diego,CA Cambridge, MA Kidlington, 2023\relax
\mciteBstWouldAddEndPuncttrue
\mciteSetBstMidEndSepPunct{\mcitedefaultmidpunct}
{\mcitedefaultendpunct}{\mcitedefaultseppunct}\relax
\EndOfBibitem
\bibitem[Banchio and Brady(2003)Banchio, and Brady]{Banchio2003}
Banchio,~A.~J.; Brady,~J.~F. Accelerated {{Stokesian}} Dynamics: {{Brownian}}
  Motion. \emph{The Journal of Chemical Physics} \textbf{2003}, \emph{118},
  10323--10332\relax
\mciteBstWouldAddEndPuncttrue
\mciteSetBstMidEndSepPunct{\mcitedefaultmidpunct}
{\mcitedefaultendpunct}{\mcitedefaultseppunct}\relax
\EndOfBibitem
\bibitem[{Mart{\'i}n-G{\'o}mez} \latin{et~al.}(2019){Mart{\'i}n-G{\'o}mez},
  Eisenstecken, Gompper, and Winkler]{Martin-Gomez2019}
{Mart{\'i}n-G{\'o}mez},~A.; Eisenstecken,~T.; Gompper,~G.; Winkler,~R.~G.
  Active {{Brownian}} Filaments with Hydrodynamic Interactions: Conformations
  and Dynamics. \emph{Soft Matter} \textbf{2019}, \emph{15}, 3957--3969\relax
\mciteBstWouldAddEndPuncttrue
\mciteSetBstMidEndSepPunct{\mcitedefaultmidpunct}
{\mcitedefaultendpunct}{\mcitedefaultseppunct}\relax
\EndOfBibitem
\bibitem[Navarro and Fielding(2015)Navarro, and Fielding]{Navarro2015}
Navarro,~R.~M.; Fielding,~S.~M. Clustering and Phase Behaviour of Attractive
  Active Particles with Hydrodynamics. \emph{Soft Matter} \textbf{2015},
  \emph{11}, 7525--7546\relax
\mciteBstWouldAddEndPuncttrue
\mciteSetBstMidEndSepPunct{\mcitedefaultmidpunct}
{\mcitedefaultendpunct}{\mcitedefaultseppunct}\relax
\EndOfBibitem
\bibitem[Ishikawa \latin{et~al.}(2008)Ishikawa, Locsei, and
  Pedley]{Ishikawa2008}
Ishikawa,~T.; Locsei,~J.~T.; Pedley,~T.~J. Development of Coherent Structures
  in Concentrated Suspensions of Swimming Model Micro-Organisms. \emph{Journal
  of Fluid Mechanics} \textbf{2008}, \emph{615}, 401--431\relax
\mciteBstWouldAddEndPuncttrue
\mciteSetBstMidEndSepPunct{\mcitedefaultmidpunct}
{\mcitedefaultendpunct}{\mcitedefaultseppunct}\relax
\EndOfBibitem
\bibitem[{Matas-Navarro} \latin{et~al.}(2014){Matas-Navarro}, Golestanian,
  Liverpool, and Fielding]{Matas-Navarro2014}
{Matas-Navarro},~R.; Golestanian,~R.; Liverpool,~T.~B.; Fielding,~S.~M.
  Hydrodynamic Suppression of Phase Separation in Active Suspensions.
  \emph{Physical Review E} \textbf{2014}, \emph{90}, 032304\relax
\mciteBstWouldAddEndPuncttrue
\mciteSetBstMidEndSepPunct{\mcitedefaultmidpunct}
{\mcitedefaultendpunct}{\mcitedefaultseppunct}\relax
\EndOfBibitem
\bibitem[{de Graaf} \latin{et~al.}(2016){de Graaf}, Menke, Mathijssen,
  Fabritius, Holm, and Shendruk]{deGraaf2016}
{de Graaf},~J.; Menke,~H.; Mathijssen,~A. J. T.~M.; Fabritius,~M.; Holm,~C.;
  Shendruk,~T.~N. Lattice-{{Boltzmann}} Hydrodynamics of Anisotropic Active
  Matter. \emph{The Journal of Chemical Physics} \textbf{2016}, \emph{144},
  134106\relax
\mciteBstWouldAddEndPuncttrue
\mciteSetBstMidEndSepPunct{\mcitedefaultmidpunct}
{\mcitedefaultendpunct}{\mcitedefaultseppunct}\relax
\EndOfBibitem
\bibitem[B{\'a}rdfalvy \latin{et~al.}(2019)B{\'a}rdfalvy, Nordanger, Nardini,
  Morozov, and Stenhammar]{Bardfalvy2019}
B{\'a}rdfalvy,~D.; Nordanger,~H.; Nardini,~C.; Morozov,~A.; Stenhammar,~J.
  Particle-Resolved Lattice {{Boltzmann}} Simulations of 3-Dimensional Active
  Turbulence. \emph{Soft Matter} \textbf{2019}, \emph{15}, 7747--7756\relax
\mciteBstWouldAddEndPuncttrue
\mciteSetBstMidEndSepPunct{\mcitedefaultmidpunct}
{\mcitedefaultendpunct}{\mcitedefaultseppunct}\relax
\EndOfBibitem
\bibitem[Theers \latin{et~al.}(2018)Theers, Westphal, Qi, Winkler, and
  Gompper]{Theers2018}
Theers,~M.; Westphal,~E.; Qi,~K.; Winkler,~R.~G.; Gompper,~G. Clustering of
  Microswimmers: Interplay of Shape and Hydrodynamics. \emph{Soft Matter}
  \textbf{2018}, \emph{14}, 8590--8603\relax
\mciteBstWouldAddEndPuncttrue
\mciteSetBstMidEndSepPunct{\mcitedefaultmidpunct}
{\mcitedefaultendpunct}{\mcitedefaultseppunct}\relax
\EndOfBibitem
\bibitem[Qi \latin{et~al.}(2022)Qi, Westphal, Gompper, and Winkler]{Qi2022}
Qi,~K.; Westphal,~E.; Gompper,~G.; Winkler,~R.~G. Emergence of Active
  Turbulence in Microswimmer Suspensions Due to Active Hydrodynamic Stress and
  Volume Exclusion. \emph{Communications Physics} \textbf{2022}, \emph{5},
  49\relax
\mciteBstWouldAddEndPuncttrue
\mciteSetBstMidEndSepPunct{\mcitedefaultmidpunct}
{\mcitedefaultendpunct}{\mcitedefaultseppunct}\relax
\EndOfBibitem
\bibitem[Kozhukhov and Shendruk(2022)Kozhukhov, and Shendruk]{Kozhukhov2022}
Kozhukhov,~T.; Shendruk,~T.~N. Mesoscopic Simulations of Active Nematics.
  \emph{Science Advances} \textbf{2022}, \emph{8}, eabo5788\relax
\mciteBstWouldAddEndPuncttrue
\mciteSetBstMidEndSepPunct{\mcitedefaultmidpunct}
{\mcitedefaultendpunct}{\mcitedefaultseppunct}\relax
\EndOfBibitem
\bibitem[Kozhukhov \latin{et~al.}(2024)Kozhukhov, Loewe, and
  Shendruk]{Kozhukhov2024}
Kozhukhov,~T.; Loewe,~B.; Shendruk,~T.~N. Mitigating Density Fluctuations in
  Particle-Based Active Nematic Simulations. \emph{Communications Physics}
  \textbf{2024}, \emph{7}, 251\relax
\mciteBstWouldAddEndPuncttrue
\mciteSetBstMidEndSepPunct{\mcitedefaultmidpunct}
{\mcitedefaultendpunct}{\mcitedefaultseppunct}\relax
\EndOfBibitem
\bibitem[Howard \latin{et~al.}(2019)Howard, Nikoubashman, and
  Palmer]{Howard2019}
Howard,~M.~P.; Nikoubashman,~A.; Palmer,~J.~C. Modeling Hydrodynamic
  Interactions in Soft Materials with Multiparticle Collision Dynamics.
  \emph{Current Opinion in Chemical Engineering} \textbf{2019}, \emph{23},
  34--43\relax
\mciteBstWouldAddEndPuncttrue
\mciteSetBstMidEndSepPunct{\mcitedefaultmidpunct}
{\mcitedefaultendpunct}{\mcitedefaultseppunct}\relax
\EndOfBibitem
\bibitem[Peskin(2002)]{Peskin2002}
Peskin,~C.~S. The Immersed Boundary Method. \emph{Acta Numerica} \textbf{2002},
  \emph{11}, 479--517\relax
\mciteBstWouldAddEndPuncttrue
\mciteSetBstMidEndSepPunct{\mcitedefaultmidpunct}
{\mcitedefaultendpunct}{\mcitedefaultseppunct}\relax
\EndOfBibitem
\bibitem[Whitfield and Hawkins(2016)Whitfield, and Hawkins]{Whitfield2016}
Whitfield,~C.~A.; Hawkins,~R.~J. Immersed {{Boundary Simulations}} of {{Active
  Fluid Droplets}}. \emph{PLOS ONE} \textbf{2016}, \emph{11}, e0162474\relax
\mciteBstWouldAddEndPuncttrue
\mciteSetBstMidEndSepPunct{\mcitedefaultmidpunct}
{\mcitedefaultendpunct}{\mcitedefaultseppunct}\relax
\EndOfBibitem
\bibitem[Chandler and Spagnolie(2024)Chandler, and Spagnolie]{Chandler2024}
Chandler,~T. G.~J.; Spagnolie,~S.~E. Active Nematic Response to a Deformable
  Body or Boundary: Elastic Deformations and Anchoring-Induced Flow. 2024\relax
\mciteBstWouldAddEndPuncttrue
\mciteSetBstMidEndSepPunct{\mcitedefaultmidpunct}
{\mcitedefaultendpunct}{\mcitedefaultseppunct}\relax
\EndOfBibitem
\bibitem[Wu and Libchaber(2000)Wu, and Libchaber]{Wu2000}
Wu,~X.-L.; Libchaber,~A. Particle {{Diffusion}} in a {{Quasi-Two-Dimensional
  Bacterial Bath}}. \emph{Physical Review Letters} \textbf{2000}, \emph{84},
  3017--3020\relax
\mciteBstWouldAddEndPuncttrue
\mciteSetBstMidEndSepPunct{\mcitedefaultmidpunct}
{\mcitedefaultendpunct}{\mcitedefaultseppunct}\relax
\EndOfBibitem
\bibitem[Maggi \latin{et~al.}(2014)Maggi, Paoluzzi, Pellicciotta, Lepore,
  Angelani, and Di~Leonardo]{Maggi2014}
Maggi,~C.; Paoluzzi,~M.; Pellicciotta,~N.; Lepore,~A.; Angelani,~L.;
  Di~Leonardo,~R. Generalized {{Energy Equipartition}} in {{Harmonic
  Oscillators Driven}} by {{Active Baths}}. \emph{Physical Review Letters}
  \textbf{2014}, \emph{113}, 238303\relax
\mciteBstWouldAddEndPuncttrue
\mciteSetBstMidEndSepPunct{\mcitedefaultmidpunct}
{\mcitedefaultendpunct}{\mcitedefaultseppunct}\relax
\EndOfBibitem
\bibitem[Maggi \latin{et~al.}(2017)Maggi, Paoluzzi, Angelani, and
  Di~Leonardo]{Maggi2017}
Maggi,~C.; Paoluzzi,~M.; Angelani,~L.; Di~Leonardo,~R. Memory-Less Response and
  Violation of the Fluctuation-Dissipation Theorem in Colloids Suspended in an
  Active Bath. \emph{Scientific Reports} \textbf{2017}, \emph{7}, 17588\relax
\mciteBstWouldAddEndPuncttrue
\mciteSetBstMidEndSepPunct{\mcitedefaultmidpunct}
{\mcitedefaultendpunct}{\mcitedefaultseppunct}\relax
\EndOfBibitem
\bibitem[Koumakis \latin{et~al.}(2014)Koumakis, Maggi, and
  Leonardo]{Koumakis2014}
Koumakis,~N.; Maggi,~C.; Leonardo,~R.~D. Directed Transport of Active Particles
  over Asymmetric Energy Barriers. \emph{Soft Matter} \textbf{2014}, \emph{10},
  5695--5701\relax
\mciteBstWouldAddEndPuncttrue
\mciteSetBstMidEndSepPunct{\mcitedefaultmidpunct}
{\mcitedefaultendpunct}{\mcitedefaultseppunct}\relax
\EndOfBibitem
\bibitem[Ye \latin{et~al.}(2020)Ye, Liu, Ye, Chen, and Yang]{Ye2020}
Ye,~S.; Liu,~P.; Ye,~F.; Chen,~K.; Yang,~M. Active Noise Experienced by a
  Passive Particle Trapped in an Active Bath. \emph{Soft Matter} \textbf{2020},
  \emph{16}, 4655--4660\relax
\mciteBstWouldAddEndPuncttrue
\mciteSetBstMidEndSepPunct{\mcitedefaultmidpunct}
{\mcitedefaultendpunct}{\mcitedefaultseppunct}\relax
\EndOfBibitem
\bibitem[Hecht \latin{et~al.}(2024)Hecht, Caprini, L{\"o}wen, and
  Liebchen]{Hecht2024}
Hecht,~L.; Caprini,~L.; L{\"o}wen,~H.; Liebchen,~B. How to Define Temperature
  in Active Systems? \emph{The Journal of Chemical Physics} \textbf{2024},
  \emph{161}, 224904\relax
\mciteBstWouldAddEndPuncttrue
\mciteSetBstMidEndSepPunct{\mcitedefaultmidpunct}
{\mcitedefaultendpunct}{\mcitedefaultseppunct}\relax
\EndOfBibitem
\bibitem[Liebchen and Mukhopadhyay(2022)Liebchen, and
  Mukhopadhyay]{Liebchen2022}
Liebchen,~B.; Mukhopadhyay,~A.~K. Interactions in Active Colloids.
  \emph{Journal of Physics: Condensed Matter} \textbf{2022}, \emph{34},
  083002\relax
\mciteBstWouldAddEndPuncttrue
\mciteSetBstMidEndSepPunct{\mcitedefaultmidpunct}
{\mcitedefaultendpunct}{\mcitedefaultseppunct}\relax
\EndOfBibitem
\bibitem[Liao and Vaikuntanathan(2021)Liao, and Vaikuntanathan]{Liao2021}
Liao,~Z.; Vaikuntanathan,~S. Energy Rectification in Active Gyroscopic Networks
  under Time-Periodic Modulations. \emph{Physical Review E} \textbf{2021},
  \emph{104}, 014601\relax
\mciteBstWouldAddEndPuncttrue
\mciteSetBstMidEndSepPunct{\mcitedefaultmidpunct}
{\mcitedefaultendpunct}{\mcitedefaultseppunct}\relax
\EndOfBibitem
\bibitem[Sprenger \latin{et~al.}(2023)Sprenger, Caprini, L{\"o}wen, and
  Wittmann]{Sprenger2023}
Sprenger,~A.~R.; Caprini,~L.; L{\"o}wen,~H.; Wittmann,~R. Dynamics of Active
  Particles with Translational and Rotational Inertia. \emph{Journal of
  Physics: Condensed Matter} \textbf{2023}, \emph{35}, 305101\relax
\mciteBstWouldAddEndPuncttrue
\mciteSetBstMidEndSepPunct{\mcitedefaultmidpunct}
{\mcitedefaultendpunct}{\mcitedefaultseppunct}\relax
\EndOfBibitem
\bibitem[Szamel(2014)]{Szamel2014}
Szamel,~G. Self-Propelled Particle in an External Potential: {{Existence}} of
  an Effective Temperature. \emph{Physical Review E} \textbf{2014}, \emph{90},
  012111\relax
\mciteBstWouldAddEndPuncttrue
\mciteSetBstMidEndSepPunct{\mcitedefaultmidpunct}
{\mcitedefaultendpunct}{\mcitedefaultseppunct}\relax
\EndOfBibitem
\bibitem[Martin \latin{et~al.}(2021)Martin, O'Byrne, Cates, Fodor, Nardini,
  Tailleur, and Van~Wijland]{Martin2021}
Martin,~D.; O'Byrne,~J.; Cates,~M.~E.; Fodor,~{\'E}.; Nardini,~C.;
  Tailleur,~J.; Van~Wijland,~F. Statistical Mechanics of Active
  {{Ornstein-Uhlenbeck}} Particles. \emph{Physical Review E} \textbf{2021},
  \emph{103}, 032607\relax
\mciteBstWouldAddEndPuncttrue
\mciteSetBstMidEndSepPunct{\mcitedefaultmidpunct}
{\mcitedefaultendpunct}{\mcitedefaultseppunct}\relax
\EndOfBibitem
\bibitem[Z{\"o}ttl and Stark(2023)Z{\"o}ttl, and Stark]{Zottl2023}
Z{\"o}ttl,~A.; Stark,~H. Modeling {{Active Colloids}}: {{From Active Brownian
  Particles}} to {{Hydrodynamic}} and {{Chemical Fields}}. \emph{Annual Review
  of Condensed Matter Physics} \textbf{2023}, \emph{14}, 109--127\relax
\mciteBstWouldAddEndPuncttrue
\mciteSetBstMidEndSepPunct{\mcitedefaultmidpunct}
{\mcitedefaultendpunct}{\mcitedefaultseppunct}\relax
\EndOfBibitem
\bibitem[Doostmohammadi \latin{et~al.}(2018)Doostmohammadi, {Ign{\'e}s-Mullol},
  Yeomans, and Sagu{\'e}s]{Doostmohammadi2018}
Doostmohammadi,~A.; {Ign{\'e}s-Mullol},~J.; Yeomans,~J.~M.; Sagu{\'e}s,~F.
  Active Nematics. \emph{Nature Communications} \textbf{2018}, \emph{9},
  3246\relax
\mciteBstWouldAddEndPuncttrue
\mciteSetBstMidEndSepPunct{\mcitedefaultmidpunct}
{\mcitedefaultendpunct}{\mcitedefaultseppunct}\relax
\EndOfBibitem
\bibitem[Ramaswamy \latin{et~al.}(2003)Ramaswamy, Simha, and
  Toner]{Ramaswamy2003}
Ramaswamy,~S.; Simha,~R.~A.; Toner,~J. Active Nematics on a Substrate:
  {{Giantnumber}} Fluctuations and Long-Time Tails. \emph{Europhysics Letters}
  \textbf{2003}, \emph{62}, 196\relax
\mciteBstWouldAddEndPuncttrue
\mciteSetBstMidEndSepPunct{\mcitedefaultmidpunct}
{\mcitedefaultendpunct}{\mcitedefaultseppunct}\relax
\EndOfBibitem
\bibitem[Giomi \latin{et~al.}(2013)Giomi, Bowick, Ma, and Marchetti]{Giomi2013}
Giomi,~L.; Bowick,~M.~J.; Ma,~X.; Marchetti,~M.~C. Defect {{Annihilation}} and
  {{Proliferation}} in {{Active Nematics}}. \emph{Physical Review Letters}
  \textbf{2013}, \emph{110}, 228101\relax
\mciteBstWouldAddEndPuncttrue
\mciteSetBstMidEndSepPunct{\mcitedefaultmidpunct}
{\mcitedefaultendpunct}{\mcitedefaultseppunct}\relax
\EndOfBibitem
\bibitem[Giomi \latin{et~al.}(2014)Giomi, Bowick, Mishra, Sknepnek, and
  Cristina~Marchetti]{Giomi2014}
Giomi,~L.; Bowick,~M.~J.; Mishra,~P.; Sknepnek,~R.; Cristina~Marchetti,~M.
  Defect Dynamics in Active Nematics. \emph{Philosophical Transactions of the
  Royal Society A: Mathematical, Physical and Engineering Sciences}
  \textbf{2014}, \emph{372}, 20130365\relax
\mciteBstWouldAddEndPuncttrue
\mciteSetBstMidEndSepPunct{\mcitedefaultmidpunct}
{\mcitedefaultendpunct}{\mcitedefaultseppunct}\relax
\EndOfBibitem
\bibitem[Wu \latin{et~al.}(2017)Wu, Hishamunda, Chen, DeCamp, Chang,
  {Fern{\'a}ndez-Nieves}, Fraden, and Dogic]{Wu2017}
Wu,~K.-T.; Hishamunda,~J.~B.; Chen,~D. T.~N.; DeCamp,~S.~J.; Chang,~Y.-W.;
  {Fern{\'a}ndez-Nieves},~A.; Fraden,~S.; Dogic,~Z. Transition from Turbulent
  to Coherent Flows in Confined Three-Dimensional Active Fluids. \emph{Science}
  \textbf{2017}, \emph{355}, eaal1979\relax
\mciteBstWouldAddEndPuncttrue
\mciteSetBstMidEndSepPunct{\mcitedefaultmidpunct}
{\mcitedefaultendpunct}{\mcitedefaultseppunct}\relax
\EndOfBibitem
\bibitem[Narayan \latin{et~al.}(2007)Narayan, Ramaswamy, and
  Menon]{Narayan2007}
Narayan,~V.; Ramaswamy,~S.; Menon,~N. Long-{{Lived Giant Number Fluctuations}}
  in a {{Swarming Granular Nematic}}. \emph{Science} \textbf{2007}, \emph{317},
  105--108\relax
\mciteBstWouldAddEndPuncttrue
\mciteSetBstMidEndSepPunct{\mcitedefaultmidpunct}
{\mcitedefaultendpunct}{\mcitedefaultseppunct}\relax
\EndOfBibitem
\bibitem[Deseigne \latin{et~al.}(2010)Deseigne, Dauchot, and
  Chat{\'e}]{Deseigne2010}
Deseigne,~J.; Dauchot,~O.; Chat{\'e},~H. Collective {{Motion}} of {{Vibrated
  Polar Disks}}. \emph{Physical Review Letters} \textbf{2010}, \emph{105},
  098001\relax
\mciteBstWouldAddEndPuncttrue
\mciteSetBstMidEndSepPunct{\mcitedefaultmidpunct}
{\mcitedefaultendpunct}{\mcitedefaultseppunct}\relax
\EndOfBibitem
\bibitem[Deseigne \latin{et~al.}(2012)Deseigne, L{\'e}onard, Dauchot, and
  Chat{\'e}]{Deseigne2012}
Deseigne,~J.; L{\'e}onard,~S.; Dauchot,~O.; Chat{\'e},~H. Vibrated Polar Disks:
  Spontaneous Motion, Binary Collisions, and Collective Dynamics. \emph{Soft
  Matter} \textbf{2012}, \emph{8}, 5629\relax
\mciteBstWouldAddEndPuncttrue
\mciteSetBstMidEndSepPunct{\mcitedefaultmidpunct}
{\mcitedefaultendpunct}{\mcitedefaultseppunct}\relax
\EndOfBibitem
\bibitem[Weber \latin{et~al.}(2013)Weber, Hanke, Deseigne, L{\'e}onard,
  Dauchot, Frey, and Chat{\'e}]{Weber2013}
Weber,~C.~A.; Hanke,~T.; Deseigne,~J.; L{\'e}onard,~S.; Dauchot,~O.; Frey,~E.;
  Chat{\'e},~H. Long-{{Range Ordering}} of {{Vibrated Polar Disks}}.
  \emph{Physical Review Letters} \textbf{2013}, \emph{110}, 208001\relax
\mciteBstWouldAddEndPuncttrue
\mciteSetBstMidEndSepPunct{\mcitedefaultmidpunct}
{\mcitedefaultendpunct}{\mcitedefaultseppunct}\relax
\EndOfBibitem
\bibitem[Kumar \latin{et~al.}(2014)Kumar, Soni, Ramaswamy, and Sood]{Kumar2014}
Kumar,~N.; Soni,~H.; Ramaswamy,~S.; Sood,~A.~K. Flocking at a Distance in
  Active Granular Matter. \emph{Nature Communications} \textbf{2014}, \emph{5},
  4688\relax
\mciteBstWouldAddEndPuncttrue
\mciteSetBstMidEndSepPunct{\mcitedefaultmidpunct}
{\mcitedefaultendpunct}{\mcitedefaultseppunct}\relax
\EndOfBibitem
\bibitem[Arora \latin{et~al.}(2024)Arora, Sadhukhan, Nandi, Bi, Sood, and
  Ganapathy]{Arora2024}
Arora,~P.; Sadhukhan,~S.; Nandi,~S.~K.; Bi,~D.; Sood,~A.~K.; Ganapathy,~R. A
  Shape-Driven Reentrant Jamming Transition in Confluent Monolayers of
  Synthetic Cell-Mimics. \emph{Nature Communications} \textbf{2024}, \emph{15},
  5645\relax
\mciteBstWouldAddEndPuncttrue
\mciteSetBstMidEndSepPunct{\mcitedefaultmidpunct}
{\mcitedefaultendpunct}{\mcitedefaultseppunct}\relax
\EndOfBibitem
\bibitem[Caprini \latin{et~al.}(2024)Caprini, Ldov, Gupta, Ellenberg, Wittmann,
  L{\"o}wen, and Scholz]{Caprini2024}
Caprini,~L.; Ldov,~A.; Gupta,~R.~K.; Ellenberg,~H.; Wittmann,~R.;
  L{\"o}wen,~H.; Scholz,~C. Emergent Memory from Tapping Collisions in Active
  Granular Matter. \emph{Communications Physics} \textbf{2024}, \emph{7},
  52\relax
\mciteBstWouldAddEndPuncttrue
\mciteSetBstMidEndSepPunct{\mcitedefaultmidpunct}
{\mcitedefaultendpunct}{\mcitedefaultseppunct}\relax
\EndOfBibitem
\bibitem[Marchetti \latin{et~al.}(2013)Marchetti, Joanny, Ramaswamy, Liverpool,
  Prost, Rao, and Simha]{Marchetti2013}
Marchetti,~M.~C.; Joanny,~J.~F.; Ramaswamy,~S.; Liverpool,~T.~B.; Prost,~J.;
  Rao,~M.; Simha,~R.~A. Hydrodynamics of Soft Active Matter. \emph{Reviews of
  Modern Physics} \textbf{2013}, \emph{85}, 1143--1189\relax
\mciteBstWouldAddEndPuncttrue
\mciteSetBstMidEndSepPunct{\mcitedefaultmidpunct}
{\mcitedefaultendpunct}{\mcitedefaultseppunct}\relax
\EndOfBibitem
\bibitem[Anderson \latin{et~al.}(2020)Anderson, Glaser, and
  Glotzer]{Anderson2020}
Anderson,~J.~A.; Glaser,~J.; Glotzer,~S.~C. {{HOOMD-blue}}: {{A Python}}
  Package for High-Performance Molecular Dynamics and Hard Particle {{Monte
  Carlo}} Simulations. \emph{Computational Materials Science} \textbf{2020},
  \emph{173}, 109363\relax
\mciteBstWouldAddEndPuncttrue
\mciteSetBstMidEndSepPunct{\mcitedefaultmidpunct}
{\mcitedefaultendpunct}{\mcitedefaultseppunct}\relax
\EndOfBibitem
\bibitem[Feitosa and Menon(2002)Feitosa, and Menon]{Feitosa2002}
Feitosa,~K.; Menon,~N. Breakdown of {{Energy Equipartition}} in a {{2D Binary
  Vibrated Granular Gas}}. \emph{Physical Review Letters} \textbf{2002},
  \emph{88}, 198301\relax
\mciteBstWouldAddEndPuncttrue
\mciteSetBstMidEndSepPunct{\mcitedefaultmidpunct}
{\mcitedefaultendpunct}{\mcitedefaultseppunct}\relax
\EndOfBibitem
\bibitem[Koumakis \latin{et~al.}(2016)Koumakis, Gnoli, Maggi, Puglisi, and
  Leonardo]{Koumakis2016}
Koumakis,~N.; Gnoli,~A.; Maggi,~C.; Puglisi,~A.; Leonardo,~R.~D. Mechanism of
  Self-Propulsion in {{3D-printed}} Active Granular Particles. \emph{New
  Journal of Physics} \textbf{2016}, \emph{18}, 113046\relax
\mciteBstWouldAddEndPuncttrue
\mciteSetBstMidEndSepPunct{\mcitedefaultmidpunct}
{\mcitedefaultendpunct}{\mcitedefaultseppunct}\relax
\EndOfBibitem
\bibitem[Palacci \latin{et~al.}(2010)Palacci, {Cottin-Bizonne}, Ybert, and
  Bocquet]{Palacci2010}
Palacci,~J.; {Cottin-Bizonne},~C.; Ybert,~C.; Bocquet,~L. Sedimentation and
  {{Effective Temperature}} of {{Active Colloidal Suspensions}}. \emph{Physical
  Review Letters} \textbf{2010}, \emph{105}, 088304\relax
\mciteBstWouldAddEndPuncttrue
\mciteSetBstMidEndSepPunct{\mcitedefaultmidpunct}
{\mcitedefaultendpunct}{\mcitedefaultseppunct}\relax
\EndOfBibitem
\bibitem[Dhar and Saintillan(2024)Dhar, and Saintillan]{Dhar2024}
Dhar,~T.; Saintillan,~D. Active Transport of a Passive Colloid in a Bath of
  Run-and-Tumble Particles. \emph{Scientific Reports} \textbf{2024}, \emph{14},
  11844\relax
\mciteBstWouldAddEndPuncttrue
\mciteSetBstMidEndSepPunct{\mcitedefaultmidpunct}
{\mcitedefaultendpunct}{\mcitedefaultseppunct}\relax
\EndOfBibitem
\bibitem[Dombrowski \latin{et~al.}(2004)Dombrowski, Cisneros, Chatkaew,
  Goldstein, and Kessler]{Dombrowski2004}
Dombrowski,~C.; Cisneros,~L.; Chatkaew,~S.; Goldstein,~R.~E.; Kessler,~J.~O.
  Self-{{Concentration}} and {{Large-Scale Coherence}} in {{Bacterial
  Dynamics}}. \emph{Physical Review Letters} \textbf{2004}, \emph{93},
  098103\relax
\mciteBstWouldAddEndPuncttrue
\mciteSetBstMidEndSepPunct{\mcitedefaultmidpunct}
{\mcitedefaultendpunct}{\mcitedefaultseppunct}\relax
\EndOfBibitem
\bibitem[Alert \latin{et~al.}(2022)Alert, Casademunt, and Joanny]{Alert2022}
Alert,~R.; Casademunt,~J.; Joanny,~J.-F. Active {{Turbulence}}. \emph{Annual
  Review of Condensed Matter Physics} \textbf{2022}, \emph{13}, 143--170\relax
\mciteBstWouldAddEndPuncttrue
\mciteSetBstMidEndSepPunct{\mcitedefaultmidpunct}
{\mcitedefaultendpunct}{\mcitedefaultseppunct}\relax
\EndOfBibitem
\bibitem[Keta \latin{et~al.}(2024)Keta, Klamser, Jack, and Berthier]{Keta2024}
Keta,~Y.-E.; Klamser,~J.~U.; Jack,~R.~L.; Berthier,~L. Emerging {{Mesoscale
  Flows}} and {{Chaotic Advection}} in {{Dense Active Matter}}. \emph{Physical
  Review Letters} \textbf{2024}, \emph{132}, 218301\relax
\mciteBstWouldAddEndPuncttrue
\mciteSetBstMidEndSepPunct{\mcitedefaultmidpunct}
{\mcitedefaultendpunct}{\mcitedefaultseppunct}\relax
\EndOfBibitem
\bibitem[Kraichnan(1970)]{Kraichnan1970}
Kraichnan,~R.~H. Diffusion by a {{Random Velocity Field}}. \emph{The Physics of
  Fluids} \textbf{1970}, \emph{13}, 22--31\relax
\mciteBstWouldAddEndPuncttrue
\mciteSetBstMidEndSepPunct{\mcitedefaultmidpunct}
{\mcitedefaultendpunct}{\mcitedefaultseppunct}\relax
\EndOfBibitem
\bibitem[Kraichnan(1994)]{Kraichnan1994}
Kraichnan,~R.~H. Anomalous Scaling of a Randomly Advected Passive Scalar.
  \emph{Physical Review Letters} \textbf{1994}, \emph{72}, 1016--1019\relax
\mciteBstWouldAddEndPuncttrue
\mciteSetBstMidEndSepPunct{\mcitedefaultmidpunct}
{\mcitedefaultendpunct}{\mcitedefaultseppunct}\relax
\EndOfBibitem
\bibitem[Klyatskin and Tatarski{\u \i}(1974)Klyatskin, and Tatarski{\u
  \i}]{Klyatskin1974}
Klyatskin,~V.~I.; Tatarski{\u \i},~V.~I. Diffusive Random Process Approximation
  in Certain Nonstationary Statistical Problems of Physics. \emph{Soviet
  Physics Uspekhi} \textbf{1974}, \emph{16}, 494--511\relax
\mciteBstWouldAddEndPuncttrue
\mciteSetBstMidEndSepPunct{\mcitedefaultmidpunct}
{\mcitedefaultendpunct}{\mcitedefaultseppunct}\relax
\EndOfBibitem
\bibitem[Klyatskin(2003)]{Klyatskin2003}
Klyatskin,~V.~I. Clustering and Diffusion of Particles and Passive Tracer
  Density in Random Hydrodynamic Flows. \emph{Physics-Uspekhi} \textbf{2003},
  \emph{46}, 667--688\relax
\mciteBstWouldAddEndPuncttrue
\mciteSetBstMidEndSepPunct{\mcitedefaultmidpunct}
{\mcitedefaultendpunct}{\mcitedefaultseppunct}\relax
\EndOfBibitem
\bibitem[Majda and Kramer(1999)Majda, and Kramer]{Majda1999}
Majda,~A.~J.; Kramer,~P.~R. Simplified Models for Turbulent Diffusion:
  {{Theory}}, Numerical Modelling, and Physical Phenomena. \emph{Physics
  Reports} \textbf{1999}, \emph{314}, 237--574\relax
\mciteBstWouldAddEndPuncttrue
\mciteSetBstMidEndSepPunct{\mcitedefaultmidpunct}
{\mcitedefaultendpunct}{\mcitedefaultseppunct}\relax
\EndOfBibitem
\bibitem[Frisch \latin{et~al.}(1998)Frisch, Mazzino, and
  Vergassola]{Frisch1998}
Frisch,~U.; Mazzino,~A.; Vergassola,~M. Intermittency in {{Passive Scalar
  Advection}}. \emph{Physical Review Letters} \textbf{1998}, \emph{80},
  5532--5535\relax
\mciteBstWouldAddEndPuncttrue
\mciteSetBstMidEndSepPunct{\mcitedefaultmidpunct}
{\mcitedefaultendpunct}{\mcitedefaultseppunct}\relax
\EndOfBibitem
\bibitem[Ueda(2016)]{Ueda2016}
Ueda,~M. Suppression of Thermal Noise in a Non-{{Markovian}} Random Velocity
  Field. \emph{Journal of Statistical Mechanics: Theory and Experiment}
  \textbf{2016}, \emph{2016}, 023206\relax
\mciteBstWouldAddEndPuncttrue
\mciteSetBstMidEndSepPunct{\mcitedefaultmidpunct}
{\mcitedefaultendpunct}{\mcitedefaultseppunct}\relax
\EndOfBibitem
\bibitem[Monnai \latin{et~al.}(2008)Monnai, Sugita, and Nakamura]{Monnai2008}
Monnai,~T.; Sugita,~A.; Nakamura,~K. Diffusion in the {{Markovian}} Limit of
  the Spatio-Temporal Colored Noise. \emph{EPL (Europhysics Letters)}
  \textbf{2008}, \emph{84}, 20005\relax
\mciteBstWouldAddEndPuncttrue
\mciteSetBstMidEndSepPunct{\mcitedefaultmidpunct}
{\mcitedefaultendpunct}{\mcitedefaultseppunct}\relax
\EndOfBibitem
\bibitem[Dentz \latin{et~al.}(2003)Dentz, Kinzelbach, Attinger, and
  Kinzelbach]{Dentz2003}
Dentz,~M.; Kinzelbach,~H.; Attinger,~S.; Kinzelbach,~W. Numerical Studies of
  the Transport Behavior of a Passive Solute in a Two-Dimensional
  Incompressible Random Flow Field. \emph{Physical Review E} \textbf{2003},
  \emph{67}, 046306\relax
\mciteBstWouldAddEndPuncttrue
\mciteSetBstMidEndSepPunct{\mcitedefaultmidpunct}
{\mcitedefaultendpunct}{\mcitedefaultseppunct}\relax
\EndOfBibitem
\bibitem[Maggi \latin{et~al.}(2022)Maggi, Gnan, Paoluzzi, Zaccarelli, and
  Crisanti]{Maggi2022}
Maggi,~C.; Gnan,~N.; Paoluzzi,~M.; Zaccarelli,~E.; Crisanti,~A. Critical Active
  Dynamics Is Captured by a Colored-Noise Driven Field Theory.
  \emph{Communications Physics} \textbf{2022}, \emph{5}, 55\relax
\mciteBstWouldAddEndPuncttrue
\mciteSetBstMidEndSepPunct{\mcitedefaultmidpunct}
{\mcitedefaultendpunct}{\mcitedefaultseppunct}\relax
\EndOfBibitem
\bibitem[Paoluzzi \latin{et~al.}(2024)Paoluzzi, Levis, Crisanti, and
  Pagonabarraga]{Paoluzzi2024}
Paoluzzi,~M.; Levis,~D.; Crisanti,~A.; Pagonabarraga,~I. Noise-{{Induced Phase
  Separation}} and {{Time Reversal Symmetry Breaking}} in {{Active Field
  Theories Driven}} by {{Persistent Noise}}. \emph{Physical Review Letters}
  \textbf{2024}, \emph{133}, 118301\relax
\mciteBstWouldAddEndPuncttrue
\mciteSetBstMidEndSepPunct{\mcitedefaultmidpunct}
{\mcitedefaultendpunct}{\mcitedefaultseppunct}\relax
\EndOfBibitem
\bibitem[Seyforth \latin{et~al.}(2022)Seyforth, Gomez, Rogers, Ross, and
  Ahmed]{Seyforth2022}
Seyforth,~H.; Gomez,~M.; Rogers,~W.~B.; Ross,~J.~L.; Ahmed,~W.~W.
  Nonequilibrium Fluctuations and Nonlinear Response of an Active Bath.
  \emph{Physical Review Research} \textbf{2022}, \emph{4}, 023043\relax
\mciteBstWouldAddEndPuncttrue
\mciteSetBstMidEndSepPunct{\mcitedefaultmidpunct}
{\mcitedefaultendpunct}{\mcitedefaultseppunct}\relax
\EndOfBibitem
\bibitem[Liao \latin{et~al.}(2020)Liao, Irvine, and Vaikuntanathan]{Liao2020}
Liao,~Z.; Irvine,~W. T.~M.; Vaikuntanathan,~S. Rectification in
  {{Nonequilibrium Parity Violating Metamaterials}}. \emph{Physical Review X}
  \textbf{2020}, \emph{10}, 021036\relax
\mciteBstWouldAddEndPuncttrue
\mciteSetBstMidEndSepPunct{\mcitedefaultmidpunct}
{\mcitedefaultendpunct}{\mcitedefaultseppunct}\relax
\EndOfBibitem
\bibitem[Virtanen \latin{et~al.}(2020)Virtanen, Gommers, Oliphant, Haberland,
  Reddy, Cournapeau, Burovski, Peterson, Weckesser, Bright, {van der Walt},
  Brett, Wilson, Millman, Mayorov, Nelson, Jones, Kern, Larson, Carey, Polat,
  Feng, Moore, VanderPlas, Laxalde, Perktold, Cimrman, Henriksen, Quintero,
  Harris, Archibald, Ribeiro, Pedregosa, {van Mulbregt}, and {SciPy 1.0
  Contributors}]{Virtanen2020}
Virtanen,~P. \latin{et~al.}  {{SciPy}} 1.0: {{Fundamental Algorithms}} for
  {{Scientific Computing}} in {{Python}}. \emph{Nature Methods} \textbf{2020},
  \emph{17}, 261--272\relax
\mciteBstWouldAddEndPuncttrue
\mciteSetBstMidEndSepPunct{\mcitedefaultmidpunct}
{\mcitedefaultendpunct}{\mcitedefaultseppunct}\relax
\EndOfBibitem
\bibitem[Omar \latin{et~al.}(2021)Omar, Klymko, GrandPre, and
  Geissler]{Omar2021}
Omar,~A.~K.; Klymko,~K.; GrandPre,~T.; Geissler,~P.~L. Phase {{Diagram}} of
  {{Active Brownian Spheres}}: {{Crystallization}} and the {{Metastability}} of
  {{Motility-Induced Phase Separation}}. \emph{Physical Review Letters}
  \textbf{2021}, \relax
\mciteBstWouldAddEndPunctfalse
\mciteSetBstMidEndSepPunct{\mcitedefaultmidpunct}
{}{\mcitedefaultseppunct}\relax
\EndOfBibitem
\bibitem[Berthier \latin{et~al.}(2019)Berthier, Flenner, and
  Szamel]{Berthier2019}
Berthier,~L.; Flenner,~E.; Szamel,~G. Glassy Dynamics in Dense Systems of
  Active Particles. \emph{The Journal of Chemical Physics} \textbf{2019},
  \emph{150}, 200901\relax
\mciteBstWouldAddEndPuncttrue
\mciteSetBstMidEndSepPunct{\mcitedefaultmidpunct}
{\mcitedefaultendpunct}{\mcitedefaultseppunct}\relax
\EndOfBibitem
\bibitem[Wysocki \latin{et~al.}(2014)Wysocki, Winkler, and
  Gompper]{Wysocki2014}
Wysocki,~A.; Winkler,~R.~G.; Gompper,~G. Cooperative Motion of Active
  {{Brownian}} Spheres in Three-Dimensional Dense Suspensions. \emph{EPL
  (Europhysics Letters)} \textbf{2014}, \emph{105}, 48004\relax
\mciteBstWouldAddEndPuncttrue
\mciteSetBstMidEndSepPunct{\mcitedefaultmidpunct}
{\mcitedefaultendpunct}{\mcitedefaultseppunct}\relax
\EndOfBibitem
\bibitem[Foffano \latin{et~al.}(2019)Foffano, Lintuvuori, Stratford, Cates, and
  Marenduzzo]{Foffano2019}
Foffano,~G.; Lintuvuori,~J.~S.; Stratford,~K.; Cates,~M.~E.; Marenduzzo,~D.
  Dynamic Clustering and Re-Dispersion in Concentrated Colloid-Active Gel
  Composites. \emph{Soft Matter} \textbf{2019}, \emph{15}, 6896--6902\relax
\mciteBstWouldAddEndPuncttrue
\mciteSetBstMidEndSepPunct{\mcitedefaultmidpunct}
{\mcitedefaultendpunct}{\mcitedefaultseppunct}\relax
\EndOfBibitem
\bibitem[Das and Barma(2000)Das, and Barma]{Das2000}
Das,~D.; Barma,~M. Particles {{Sliding}} on a {{Fluctuating Surface}}: {{Phase
  Separation}} and {{Power Laws}}. \emph{Physical Review Letters}
  \textbf{2000}, \emph{85}, 1602--1605\relax
\mciteBstWouldAddEndPuncttrue
\mciteSetBstMidEndSepPunct{\mcitedefaultmidpunct}
{\mcitedefaultendpunct}{\mcitedefaultseppunct}\relax
\EndOfBibitem
\bibitem[Das \latin{et~al.}(2001)Das, Barma, and Majumdar]{Das2001}
Das,~D.; Barma,~M.; Majumdar,~S.~N. Fluctuation-Dominated Phase Ordering Driven
  by Stochastically Evolving Surfaces: {{Depth}} Models and Sliding Particles.
  \emph{Physical Review E} \textbf{2001}, \emph{64}, 046126\relax
\mciteBstWouldAddEndPuncttrue
\mciteSetBstMidEndSepPunct{\mcitedefaultmidpunct}
{\mcitedefaultendpunct}{\mcitedefaultseppunct}\relax
\EndOfBibitem
\bibitem[J{\"u}licher \latin{et~al.}(2018)J{\"u}licher, Grill, and
  Salbreux]{Julicher2018}
J{\"u}licher,~F.; Grill,~S.~W.; Salbreux,~G. Hydrodynamic Theory of Active
  Matter. \emph{Reports on Progress in Physics} \textbf{2018}, \emph{81},
  076601\relax
\mciteBstWouldAddEndPuncttrue
\mciteSetBstMidEndSepPunct{\mcitedefaultmidpunct}
{\mcitedefaultendpunct}{\mcitedefaultseppunct}\relax
\EndOfBibitem
\bibitem[Agranov \latin{et~al.}(2021)Agranov, Ro, Kafri, and
  Lecomte]{Agranov2021}
Agranov,~T.; Ro,~S.; Kafri,~Y.; Lecomte,~V. Exact Fluctuating Hydrodynamics of
  Active Lattice Gases---Typical Fluctuations. \emph{Journal of Statistical
  Mechanics: Theory and Experiment} \textbf{2021}, \emph{2021}, 083208\relax
\mciteBstWouldAddEndPuncttrue
\mciteSetBstMidEndSepPunct{\mcitedefaultmidpunct}
{\mcitedefaultendpunct}{\mcitedefaultseppunct}\relax
\EndOfBibitem
\bibitem[Han \latin{et~al.}(2021)Han, Fruchart, Scheibner, Vaikuntanathan,
  De~Pablo, and Vitelli]{Han2021}
Han,~M.; Fruchart,~M.; Scheibner,~C.; Vaikuntanathan,~S.; De~Pablo,~J.~J.;
  Vitelli,~V. Fluctuating Hydrodynamics of Chiral Active Fluids. \emph{Nature
  Physics} \textbf{2021}, \emph{17}, 1260--1269\relax
\mciteBstWouldAddEndPuncttrue
\mciteSetBstMidEndSepPunct{\mcitedefaultmidpunct}
{\mcitedefaultendpunct}{\mcitedefaultseppunct}\relax
\EndOfBibitem
\bibitem[Klyatskin and Tatarski{\u \i}(1974)Klyatskin, and Tatarski{\u
  \i}]{Klyatskin1974a}
Klyatskin,~V.~I.; Tatarski{\u \i},~V.~I. Diffusive Random Process Approximation
  in Certainnonstationary Statistical Problems of Physics. \emph{Soviet Physics
  Uspekhi} \textbf{1974}, \emph{16}, 494\relax
\mciteBstWouldAddEndPuncttrue
\mciteSetBstMidEndSepPunct{\mcitedefaultmidpunct}
{\mcitedefaultendpunct}{\mcitedefaultseppunct}\relax
\EndOfBibitem
\bibitem[Schekochihin and Kulsrud(2001)Schekochihin, and
  Kulsrud]{Schekochihin2001}
Schekochihin,~A.~A.; Kulsrud,~R.~M. Finite-Correlation-Time Effects in the
  Kinematic Dynamo Problem. \emph{Physics of Plasmas} \textbf{2001}, \emph{8},
  4937--4953\relax
\mciteBstWouldAddEndPuncttrue
\mciteSetBstMidEndSepPunct{\mcitedefaultmidpunct}
{\mcitedefaultendpunct}{\mcitedefaultseppunct}\relax
\EndOfBibitem
\bibitem[Rowan(2024)]{Rowan2024}
Rowan,~K. On {{Anomalous Diffusion}} in the {{Kraichnan Model}} and
  {{Correlated-in-Time Variants}}. \emph{Archive for Rational Mechanics and
  Analysis} \textbf{2024}, \emph{248}, 93\relax
\mciteBstWouldAddEndPuncttrue
\mciteSetBstMidEndSepPunct{\mcitedefaultmidpunct}
{\mcitedefaultendpunct}{\mcitedefaultseppunct}\relax
\EndOfBibitem
\bibitem[Varghese \latin{et~al.}(2020)Varghese, Baskaran, Hagan, and
  Baskaran]{Varghese2020}
Varghese,~M.; Baskaran,~A.; Hagan,~M.~F.; Baskaran,~A. Confinement-{{Induced
  Self-Pumping}} in {{3D Active Fluids}}. \emph{Physical Review Letters}
  \textbf{2020}, \emph{125}, 268003\relax
\mciteBstWouldAddEndPuncttrue
\mciteSetBstMidEndSepPunct{\mcitedefaultmidpunct}
{\mcitedefaultendpunct}{\mcitedefaultseppunct}\relax
\EndOfBibitem
\bibitem[Chandrakar \latin{et~al.}(2020)Chandrakar, Varghese, Aghvami,
  Baskaran, Dogic, and Duclos]{Chandrakar2020}
Chandrakar,~P.; Varghese,~M.; Aghvami,~S.; Baskaran,~A.; Dogic,~Z.; Duclos,~G.
  Confinement {{Controls}} the {{Bend Instability}} of {{Three-Dimensional
  Active Liquid Crystals}}. \emph{Physical Review Letters} \textbf{2020},
  \emph{125}, 257801\relax
\mciteBstWouldAddEndPuncttrue
\mciteSetBstMidEndSepPunct{\mcitedefaultmidpunct}
{\mcitedefaultendpunct}{\mcitedefaultseppunct}\relax
\EndOfBibitem
\bibitem[Castellana \latin{et~al.}(2014)Castellana, Wilson, Xu, Joshi, Cristea,
  Rabinowitz, Gitai, and Wingreen]{Castellana2014}
Castellana,~M.; Wilson,~M.~Z.; Xu,~Y.; Joshi,~P.; Cristea,~I.~M.;
  Rabinowitz,~J.~D.; Gitai,~Z.; Wingreen,~N.~S. Enzyme Clustering Accelerates
  Processing of Intermediates through Metabolic Channeling. \emph{Nature
  Biotechnology} \textbf{2014}, \emph{32}, 1011--1018\relax
\mciteBstWouldAddEndPuncttrue
\mciteSetBstMidEndSepPunct{\mcitedefaultmidpunct}
{\mcitedefaultendpunct}{\mcitedefaultseppunct}\relax
\EndOfBibitem
\bibitem[G{\"o}th and Dzubiella(2024)G{\"o}th, and Dzubiella]{Goth2024}
G{\"o}th,~N.; Dzubiella,~J. Collective Dynamics and Elasto-Chemical Cluster
  Waves in Communicating Colloids with Explicit Size Response. 2024\relax
\mciteBstWouldAddEndPuncttrue
\mciteSetBstMidEndSepPunct{\mcitedefaultmidpunct}
{\mcitedefaultendpunct}{\mcitedefaultseppunct}\relax
\EndOfBibitem
\bibitem[Huang \latin{et~al.}(2024)Huang, Lin, and He]{Huang2024}
Huang,~Y.; Lin,~Z.; He,~Q. Directional and {{Reconfigurable Assembly}} of
  {{Active Colloidal Motors Triggered}} by {{Chemical Communications}}.
  \emph{Advanced Functional Materials} \textbf{2024}, \emph{34}, 2311136\relax
\mciteBstWouldAddEndPuncttrue
\mciteSetBstMidEndSepPunct{\mcitedefaultmidpunct}
{\mcitedefaultendpunct}{\mcitedefaultseppunct}\relax
\EndOfBibitem
\bibitem[Theeyancheri \latin{et~al.}(2024)Theeyancheri, Chaki, Bhattacharjee,
  and Chakrabarti]{Theeyancheri2024a}
Theeyancheri,~L.; Chaki,~S.; Bhattacharjee,~T.; Chakrabarti,~R. Dynamic
  Clustering of Active Rings. \emph{Physical Review Research} \textbf{2024},
  \emph{6}, L012038\relax
\mciteBstWouldAddEndPuncttrue
\mciteSetBstMidEndSepPunct{\mcitedefaultmidpunct}
{\mcitedefaultendpunct}{\mcitedefaultseppunct}\relax
\EndOfBibitem
\bibitem[Adorj{\'a}ni \latin{et~al.}(2024)Adorj{\'a}ni, Lib{\'a}l, Reichhardt,
  and Reichhardt]{Adorjani2024}
Adorj{\'a}ni,~B.; Lib{\'a}l,~A.; Reichhardt,~C.; Reichhardt,~C. J.~O.
  Motility-Induced Phase Separation and Frustration in Active Matter
  Swarmalators. \emph{Physical Review E} \textbf{2024}, \emph{109},
  024607\relax
\mciteBstWouldAddEndPuncttrue
\mciteSetBstMidEndSepPunct{\mcitedefaultmidpunct}
{\mcitedefaultendpunct}{\mcitedefaultseppunct}\relax
\EndOfBibitem
\bibitem[Ericksen(1959)]{Ericksen1959}
Ericksen,~J.~L. Anisotropic Fluids. \emph{Archive for Rational Mechanics and
  Analysis} \textbf{1959}, \emph{4}, 231--237\relax
\mciteBstWouldAddEndPuncttrue
\mciteSetBstMidEndSepPunct{\mcitedefaultmidpunct}
{\mcitedefaultendpunct}{\mcitedefaultseppunct}\relax
\EndOfBibitem
\bibitem[Leslie(1966)]{Leslie1966}
Leslie,~F.~M. {{SOME CONSTITUTIVE EQUATIONS FOR ANISOTROPIC FLUIDS}}. \emph{The
  Quarterly Journal of Mechanics and Applied Mathematics} \textbf{1966},
  \emph{19}, 357--370\relax
\mciteBstWouldAddEndPuncttrue
\mciteSetBstMidEndSepPunct{\mcitedefaultmidpunct}
{\mcitedefaultendpunct}{\mcitedefaultseppunct}\relax
\EndOfBibitem
\bibitem[Helfrich(1969)]{Helfrich1969}
Helfrich,~W. Molecular {{Theory}} of {{Flow Alignment}} of {{Nematic Liquid
  Crystals}}. \emph{The Journal of Chemical Physics} \textbf{1969}, \emph{50},
  100--106\relax
\mciteBstWouldAddEndPuncttrue
\mciteSetBstMidEndSepPunct{\mcitedefaultmidpunct}
{\mcitedefaultendpunct}{\mcitedefaultseppunct}\relax
\EndOfBibitem
\bibitem[{Garc{\'i}a-Ojalvo} and Sancho(1999){Garc{\'i}a-Ojalvo}, and
  Sancho]{Garcia-Ojalvo1999}
{Garc{\'i}a-Ojalvo},~J.; Sancho,~J.~M. \emph{Noise in Spatially Extended
  Systems}; Institute for Nonlinear Science; Springer: New York, 1999\relax
\mciteBstWouldAddEndPuncttrue
\mciteSetBstMidEndSepPunct{\mcitedefaultmidpunct}
{\mcitedefaultendpunct}{\mcitedefaultseppunct}\relax
\EndOfBibitem
\bibitem[Sagu{\'e}s \latin{et~al.}(2007)Sagu{\'e}s, Sancho, and
  {Garc{\'i}a-Ojalvo}]{Sagues2007}
Sagu{\'e}s,~F.; Sancho,~J.~M.; {Garc{\'i}a-Ojalvo},~J. Spatiotemporal Order out
  of Noise. \emph{Reviews of Modern Physics} \textbf{2007}, \emph{79},
  829--882\relax
\mciteBstWouldAddEndPuncttrue
\mciteSetBstMidEndSepPunct{\mcitedefaultmidpunct}
{\mcitedefaultendpunct}{\mcitedefaultseppunct}\relax
\EndOfBibitem
\bibitem[Harris \latin{et~al.}(2020)Harris, Millman, Van Der~Walt, Gommers,
  Virtanen, Cournapeau, Wieser, Taylor, Berg, Smith, Kern, Picus, Hoyer,
  Van~Kerkwijk, Brett, Haldane, Del~R{\'i}o, Wiebe, Peterson,
  {G{\'e}rard-Marchant}, Sheppard, Reddy, Weckesser, Abbasi, Gohlke, and
  Oliphant]{Harris2020}
Harris,~C.~R. \latin{et~al.}  Array Programming with {{NumPy}}. \emph{Nature}
  \textbf{2020}, \emph{585}, 357--362\relax
\mciteBstWouldAddEndPuncttrue
\mciteSetBstMidEndSepPunct{\mcitedefaultmidpunct}
{\mcitedefaultendpunct}{\mcitedefaultseppunct}\relax
\EndOfBibitem
\bibitem[Kloeden and Platen(1999)Kloeden, and Platen]{Kloeden1999}
Kloeden,~P.~E.; Platen,~E. \emph{Numerical Solution of Stochastic Differential
  Equations}, corrected third printing ed.; Applications of Mathematics 23;
  Springer: Berlin Heidelberg, 1999\relax
\mciteBstWouldAddEndPuncttrue
\mciteSetBstMidEndSepPunct{\mcitedefaultmidpunct}
{\mcitedefaultendpunct}{\mcitedefaultseppunct}\relax
\EndOfBibitem
\bibitem[Saad and Sutherland(2016)Saad, and Sutherland]{Saad2016}
Saad,~T.; Sutherland,~J.~C. Comment on ``{{Diffusion}} by a Random Velocity
  Field'' [{{Phys}}. {{Fluids}} 13, 22 (1970)]. \emph{Physics of Fluids}
  \textbf{2016}, \emph{28}, 119101\relax
\mciteBstWouldAddEndPuncttrue
\mciteSetBstMidEndSepPunct{\mcitedefaultmidpunct}
{\mcitedefaultendpunct}{\mcitedefaultseppunct}\relax
\EndOfBibitem
\bibitem[Weeks \latin{et~al.}(1971)Weeks, Chandler, and Andersen]{Weeks1971}
Weeks,~J.~D.; Chandler,~D.; Andersen,~H.~C. Role of {{Repulsive Forces}} in
  {{Determining}} the {{Equilibrium Structure}} of {{Simple Liquids}}.
  \emph{The Journal of Chemical Physics} \textbf{1971}, \emph{54},
  5237--5247\relax
\mciteBstWouldAddEndPuncttrue
\mciteSetBstMidEndSepPunct{\mcitedefaultmidpunct}
{\mcitedefaultendpunct}{\mcitedefaultseppunct}\relax
\EndOfBibitem
\bibitem[Hoshen and Kopelman(1976)Hoshen, and Kopelman]{Hoshen1976}
Hoshen,~J.; Kopelman,~R. Percolation and Cluster Distribution. {{I}}.
  {{Cluster}} Multiple Labeling Technique and Critical Concentration Algorithm.
  \emph{Physical Review B} \textbf{1976}, \emph{14}, 3438--3445\relax
\mciteBstWouldAddEndPuncttrue
\mciteSetBstMidEndSepPunct{\mcitedefaultmidpunct}
{\mcitedefaultendpunct}{\mcitedefaultseppunct}\relax
\EndOfBibitem
\bibitem[Okuta \latin{et~al.}(2017)Okuta, Unno, Nishino, Hido, and
  Loomis]{Okuta2017}
Okuta,~R.; Unno,~Y.; Nishino,~D.; Hido,~S.; Loomis,~C. {{CuPy}}: A
  {{NumPy-compatible}} Library for {{NVIDIA GPU}} Calculations. Proceedings of
  Workshop on Machine Learning Systems ({{LearningSys}}) in the Thirty-First
  Annual Conference on Neural Information Processing Systems ({{NIPS}}).
  2017\relax
\mciteBstWouldAddEndPuncttrue
\mciteSetBstMidEndSepPunct{\mcitedefaultmidpunct}
{\mcitedefaultendpunct}{\mcitedefaultseppunct}\relax
\EndOfBibitem
\end{mcitethebibliography}


\begin{thebibliography}{2}%
\makeatletter
\providecommand \@ifxundefined [1]{%
 \@ifx{#1\undefined}
}%
\providecommand \@ifnum [1]{%
 \ifnum #1\expandafter \@firstoftwo
 \else \expandafter \@secondoftwo
 \fi
}%
\providecommand \@ifx [1]{%
 \ifx #1\expandafter \@firstoftwo
 \else \expandafter \@secondoftwo
 \fi
}%
\providecommand \natexlab [1]{#1}%
\providecommand \enquote  [1]{``#1''}%
\providecommand \bibnamefont  [1]{#1}%
\providecommand \bibfnamefont [1]{#1}%
\providecommand \citenamefont [1]{#1}%
\providecommand \href@noop [0]{\@secondoftwo}%
\providecommand \href [0]{\begingroup \@sanitize@url \@href}%
\providecommand \@href[1]{\@@startlink{#1}\@@href}%
\providecommand \@@href[1]{\endgroup#1\@@endlink}%
\providecommand \@sanitize@url [0]{\catcode `\\12\catcode `\$12\catcode
  `\&12\catcode `\#12\catcode `\^12\catcode `\_12\catcode `\%12\relax}%
\providecommand \@@startlink[1]{}%
\providecommand \@@endlink[0]{}%
\providecommand \url  [0]{\begingroup\@sanitize@url \@url }%
\providecommand \@url [1]{\endgroup\@href {#1}{\urlprefix }}%
\providecommand \urlprefix  [0]{URL }%
\providecommand \Eprint [0]{\href }%
\providecommand \doibase [0]{https://doi.org/}%
\providecommand \selectlanguage [0]{\@gobble}%
\providecommand \bibinfo  [0]{\@secondoftwo}%
\providecommand \bibfield  [0]{\@secondoftwo}%
\providecommand \translation [1]{[#1]}%
\providecommand \BibitemOpen [0]{}%
\providecommand \bibitemStop [0]{}%
\providecommand \bibitemNoStop [0]{.\EOS\space}%
\providecommand \EOS [0]{\spacefactor3000\relax}%
\providecommand \BibitemShut  [1]{\csname bibitem#1\endcsname}%
\let\auto@bib@innerbib\@empty
\bibitem [{\citenamefont {Dean}(1996)}]{Dean1996}%
  \BibitemOpen
  \bibfield  {author} {\bibinfo {author} {\bibfnamefont {D.~S.}\ \bibnamefont
  {Dean}},\ }\bibfield  {title} {\bibinfo {title} {Langevin equation for the
  density of a system of interacting {{Langevin}} processes},\ }\href
  {https://doi.org/10.1088/0305-4470/29/24/001} {\bibfield  {journal} {\bibinfo
   {journal} {Journal of Physics A: Mathematical and General}\ }\textbf
  {\bibinfo {volume} {29}},\ \bibinfo {pages} {L613} (\bibinfo {year}
  {1996})}\BibitemShut {NoStop}%
\bibitem [{\citenamefont {Virtanen}\ \emph {et~al.}(2020)\citenamefont
  {Virtanen}, \citenamefont {Gommers}, \citenamefont {Oliphant}, \citenamefont
  {Haberland}, \citenamefont {Reddy}, \citenamefont {Cournapeau}, \citenamefont
  {Burovski}, \citenamefont {Peterson}, \citenamefont {Weckesser},
  \citenamefont {Bright}, \citenamefont {{van der Walt}}, \citenamefont
  {Brett}, \citenamefont {Wilson}, \citenamefont {Millman}, \citenamefont
  {Mayorov}, \citenamefont {Nelson}, \citenamefont {Jones}, \citenamefont
  {Kern}, \citenamefont {Larson}, \citenamefont {Carey}, \citenamefont {Polat},
  \citenamefont {Feng}, \citenamefont {Moore}, \citenamefont {VanderPlas},
  \citenamefont {Laxalde}, \citenamefont {Perktold}, \citenamefont {Cimrman},
  \citenamefont {Henriksen}, \citenamefont {Quintero}, \citenamefont {Harris},
  \citenamefont {Archibald}, \citenamefont {Ribeiro}, \citenamefont
  {Pedregosa}, \citenamefont {{van Mulbregt}},\ and\ \citenamefont {{SciPy 1.0
  Contributors}}}]{Virtanen2020}%
  \BibitemOpen
  \bibfield  {author} {\bibinfo {author} {\bibfnamefont {P.}~\bibnamefont
  {Virtanen}}, \bibinfo {author} {\bibfnamefont {R.}~\bibnamefont {Gommers}},
  \bibinfo {author} {\bibfnamefont {T.~E.}\ \bibnamefont {Oliphant}}, \bibinfo
  {author} {\bibfnamefont {M.}~\bibnamefont {Haberland}}, \bibinfo {author}
  {\bibfnamefont {T.}~\bibnamefont {Reddy}}, \bibinfo {author} {\bibfnamefont
  {D.}~\bibnamefont {Cournapeau}}, \bibinfo {author} {\bibfnamefont
  {E.}~\bibnamefont {Burovski}}, \bibinfo {author} {\bibfnamefont
  {P.}~\bibnamefont {Peterson}}, \bibinfo {author} {\bibfnamefont
  {W.}~\bibnamefont {Weckesser}}, \bibinfo {author} {\bibfnamefont
  {J.}~\bibnamefont {Bright}}, \bibinfo {author} {\bibfnamefont {S.~J.}\
  \bibnamefont {{van der Walt}}}, \bibinfo {author} {\bibfnamefont
  {M.}~\bibnamefont {Brett}}, \bibinfo {author} {\bibfnamefont
  {J.}~\bibnamefont {Wilson}}, \bibinfo {author} {\bibfnamefont {K.~J.}\
  \bibnamefont {Millman}}, \bibinfo {author} {\bibfnamefont {N.}~\bibnamefont
  {Mayorov}}, \bibinfo {author} {\bibfnamefont {A.~R.~J.}\ \bibnamefont
  {Nelson}}, \bibinfo {author} {\bibfnamefont {E.}~\bibnamefont {Jones}},
  \bibinfo {author} {\bibfnamefont {R.}~\bibnamefont {Kern}}, \bibinfo {author}
  {\bibfnamefont {E.}~\bibnamefont {Larson}}, \bibinfo {author} {\bibfnamefont
  {C.~J.}\ \bibnamefont {Carey}}, \bibinfo {author} {\bibfnamefont
  {{\.I}.}~\bibnamefont {Polat}}, \bibinfo {author} {\bibfnamefont
  {Y.}~\bibnamefont {Feng}}, \bibinfo {author} {\bibfnamefont {E.~W.}\
  \bibnamefont {Moore}}, \bibinfo {author} {\bibfnamefont {J.}~\bibnamefont
  {VanderPlas}}, \bibinfo {author} {\bibfnamefont {D.}~\bibnamefont {Laxalde}},
  \bibinfo {author} {\bibfnamefont {J.}~\bibnamefont {Perktold}}, \bibinfo
  {author} {\bibfnamefont {R.}~\bibnamefont {Cimrman}}, \bibinfo {author}
  {\bibfnamefont {I.}~\bibnamefont {Henriksen}}, \bibinfo {author}
  {\bibfnamefont {E.~A.}\ \bibnamefont {Quintero}}, \bibinfo {author}
  {\bibfnamefont {C.~R.}\ \bibnamefont {Harris}}, \bibinfo {author}
  {\bibfnamefont {A.~M.}\ \bibnamefont {Archibald}}, \bibinfo {author}
  {\bibfnamefont {A.~H.}\ \bibnamefont {Ribeiro}}, \bibinfo {author}
  {\bibfnamefont {F.}~\bibnamefont {Pedregosa}}, \bibinfo {author}
  {\bibfnamefont {P.}~\bibnamefont {{van Mulbregt}}},\ and\ \bibinfo {author}
  {\bibnamefont {{SciPy 1.0 Contributors}}},\ }\bibfield  {title} {\bibinfo
  {title} {{{SciPy}} 1.0: {{Fundamental Algorithms}} for {{Scientific
  Computing}} in {{Python}}},\ }\href
  {https://doi.org/10.1038/s41592-019-0686-2} {\bibfield  {journal} {\bibinfo
  {journal} {Nature Methods}\ }\textbf {\bibinfo {volume} {17}},\ \bibinfo
  {pages} {261} (\bibinfo {year} {2020})}\BibitemShut {NoStop}%
\end{thebibliography}%

\makeatletter\@input{yy.tex}\makeatother
\end{document}




\title{Supplementary Material: Active noise\textendash induced dynamic clustering of passive colloids}

\author{Layne B. Frechette}
\email{laynefrechette@brandeis.edu}
\author{Aparna Baskaran}
 \email{aparna@brandeis.edu}
 \author{Michael F. Hagan}%
\email{hagan@brandeis.edu}
\affiliation{%
 Martin Fisher School of Physics, Brandeis University, Waltham, Massachusetts 02453, USA
}%

\date{\today}

\maketitle

\onecolumngrid

\section{Derivation of particle density equation} \label{section:particle_density}

Consider a collection of $N$ particles interacting via a pair potential $u$ embedded in an active noise field $\bxi(\mathbf{r},t)$. Each particle $i$ obeys an overdamped Langevin equation:
\begin{equation}
    \frac{\diff{\mathbf{r}_i}}{\diff{t}} = -\sum_j\nabla u(\mathbf{r}_i-\mathbf{r}_j) + \bxi(\mathbf{r}_i(t),t), \label{eq:langevin}
\end{equation}
where the Gaussian noise field has mean and covariance:
\begin{subequations}
\begin{align}
    \langle \xi_{\mu} \rangle &= 0 \\
    \langle \xi_{\mu}(\mathbf{r},t) \xi_{\nu}(\mathbf{r}',t') \rangle &= \delta_{\mu\nu}\va^2 e^{-|t-t'|/\taua} e^{-|\mathbf{r}-\mathbf{r}'|/\la}.
\end{align}
\end{subequations}
We define the single-particle density field $\rho_i$ as:
\begin{equation}
    \rho_i(\mathbf{r},t) = \delta(\mathbf{r}_i(t)-\mathbf{r}),
\end{equation}
and the full density field as:
\begin{equation}
    \rho(\mathbf{r},t) = \sum_{i=1}^N \rho_i(\mathbf{r},t). \label{eq:full_density}
\end{equation}
Now, following Dean \cite{Dean1996}, we write an arbitrary function $f(\mathbf{r})$ as :
\begin{equation}
    f(\mathbf{r}_i(t)) = \int \diff{\mathbf{r}} f(\mathbf{r}) \rho_i(\mathbf{r},t),
\end{equation}
Then we invoke Ito calculus, Taylor expanding $f$ to second order:
\begin{equation}
    \diff{f(\mathbf{r}_i)} = \int \diff{\mathbf{r}} \rho_i(\mathbf{r},t) \left[ \nabla f(\mathbf{r}) \cdot \diff{\mathbf{r}} + \frac{1}{2} \nabla^2f(\mathbf{r})(\diff{\mathbf{r}})^2\right].
\end{equation}
The reason is that if $\bxi$ where a white noise, then $(\diff{\mathbf{r}})^2$ would be of order $\diff{t}$ since $\bxi \diff{t} \sim (\diff{t})^{1/2}$. In more mathematical language, the quantity $\bxi \diff{t}$ would be a Wiener process $\diff{W}$. However, our noise is colored in time and hence itself results from an Ornstein-Uhlenbeck process, $\diff{\bxi_q}/\diff{t} = -\bxi_q/\taua + (c_q)^{1/2} \boldsymbol{\eta}_q$, with $\boldsymbol{\eta}_q \diff{t}$ a Wiener process. In solving Eq. \ref{eq:langevin} numerically, we update $\mathbf{r}_i$ by multiplying the right hand side at time $t$ by an increment $\Delta t$, then we update $\bxi$ according to the Ornstein-Uhlenbeck process. That is to say, our $\bxi(\mathbf{r}_i(t),t)$ is \textit{not} the time derivative of a Wiener increment, hence our $\bxi(\mathbf{r}_i(t),t) \diff{t} \sim \diff{t}$ rather than $(\diff{t})^{1/2}$. Therefore, for our colored noise field, $(\diff{\mathbf{r}})^2$ is of order $(\diff{t})^2$ and can be safely neglected. Hence we write:
\begin{align*}
    \diff{f(\mathbf{r}_i)} &= \int \diff{\mathbf{r}} \rho_i(\mathbf{r},t) \nabla f(\mathbf{r}) \cdot \diff{\mathbf{r}} \\
    &= \int \diff{\mathbf{r}} \rho_i(\mathbf{r},t) \nabla f(\mathbf{r}) \cdot \left[-\sum_j\nabla u(\mathbf{r}-\mathbf{r}_j) \diff{t} + \bxi(\mathbf{r},t) \diff{t} \right] \\
    &= \int \diff{\mathbf{r}} f(\mathbf{r}) \diff{t} \left[ -\nabla \cdot \left(\rho_i(\mathbf{r},t) \bxi(\mathbf{r},t) \right) + \nabla \cdot \left(\rho_i(\mathbf{r},t) \sum_j \nabla u(\mathbf{r}-\mathbf{r}_j) \right)\right].
\end{align*}
Since we can also write:
\begin{equation}
\frac{\diff{f(\mathbf{r}_i(t))}}{\diff{t}} = \int \diff{\mathbf{r}} f(\mathbf{r}) \frac{\partial \rho_i(\mathbf{r},t)}{\partial t}
\end{equation}
and since $f(\mathbf{r})$ is an arbitrary function, we see that:
\begin{equation}
    \frac{\partial \rho_i(\mathbf{r},t)}{\partial t} = -\nabla \cdot \left(\rho_i(\mathbf{r},t) \bxi(\mathbf{r},t) \right) + \nabla \cdot \left(\rho_i(\mathbf{r},t) \sum_j \nabla u(\mathbf{r}-\mathbf{r}_j) \right).
\end{equation}
Recalling the definition of the total density (Eq. \ref{eq:full_density}) to rewrite the sum over $j$ as an integral and summing over all particles $i$, we arrive at an equation for the total density field:
\begin{equation}
\boxed{
    \frac{\partial \rho(\mathbf{r},t)}{\partial t} = -\nabla \cdot \left(\rho(\mathbf{r},t) \bxi(\mathbf{r},t)\right) + \nabla \cdot \left( \rho(\mathbf{r},t) \int \diff{\mathbf{r}'} \rho(\mathbf{r}',t) \nabla u(\mathbf{r}-\mathbf{r}')\right). \label{eq:density}
    }
\end{equation}
Note that since we don't have to worry about doing Ito calculus, we could have obtained this result much more directly simply by writing a continuity equation for the density:
\begin{equation}
    \frac{\partial \rho(\mathbf{r},t)}{\partial t} = -\nabla \cdot \mathbf{J}(\mathbf{r},t), \label{eq:continuity}
\end{equation}
where the mass current $\mathbf{J}$ is:
\begin{align*}
    \mathbf{J}(\mathbf{r},t) &= \sum_i \dot{\mathbf{r}}_i(t) \delta(\mathbf{r}_i(t)-\mathbf{r}) \numberthis \\
    &= \sum_i \left(-\sum_j\nabla u(\mathbf{r}_i-\mathbf{r}_j) + \bxi(\mathbf{r}_i(t),t)\right) \delta(\mathbf{r}_i(t)-\mathbf{r}) \\
    &= -\sum_{i,j} \nabla u(\mathbf{r}-\mathbf{r}_j) \rho_i(\mathbf{r},t) + \sum_i \rho_i(\mathbf{r},t) \bxi(\mathbf{r},t) \\
    &= -\rho(\mathbf{r},t) \int \diff{\mathbf{r}'} \rho(\mathbf{r}',t) \nabla u(\mathbf{r}-\mathbf{r}') + \rho(\mathbf{r},t)\bxi(\mathbf{r},t). \numberthis
\end{align*}
Plugging this back into Eq. \ref{eq:continuity} yields our previous result, Eq. \ref{eq:density}.

\section{Linear stability analysis} \label{section:linear_stability}
Let us consider the linear stability with respect to a homogeneous density by writing:
\begin{equation}
    \rho(\mathbf{r},t) = \overline{\rho} + \delta \rho(\mathbf{r},t).
\end{equation}
The linearized form of Eq. \ref{eq:density} reads:
\begin{equation}
    \frac{\partial \delta \rho(\mathbf{r},t)}{\partial t} = -\overline{\rho}\nabla \cdot \bxi(\mathbf{r},t) + \overline{\rho} \nabla \cdot \left(\int \diff{\mathbf{r}'} \delta \rho(\mathbf{r}',t) \nabla u(\mathbf{r}-\mathbf{r}')\right).
\end{equation}
(Recall that $\langle \bxi \rangle=0$.) Taking the spatial Fourier transform, we have:
\begin{equation}
    \frac{\partial \delta \tilde{\rho}(\mathbf{q},t)}{\partial t} = -i \overline{\rho} \mathbf{q}\cdot \tilde{\bxi}(\mathbf{q},t) - \overline{\rho} q^2 \tilde{u}(\mathbf{q}) \delta \tilde{\rho}(\mathbf{q},t). \label{eq:linear_fourier}
\end{equation}
If we take the noise average of Eq. \ref{eq:linear_fourier}, since the noise has zero mean, we get:
\begin{equation}
    \frac{\partial \langle \delta \tilde{\rho}(\mathbf{q},t)\rangle}{\partial t} = - \overline{\rho} q^2 \tilde{u}(\mathbf{q}) \langle \delta \tilde{\rho}(\mathbf{q},t) \rangle.
\end{equation}
For a purely repulsive pair potential, $\tilde{u}(\mathbf{q})$ should be positive, hence the homogeneous state is linearly stable with respect to noise-averaged density fluctuations. 

\section{Supplementary figures}

\begin{figure}[h!]
    \centering
    \includegraphics[width=\linewidth]{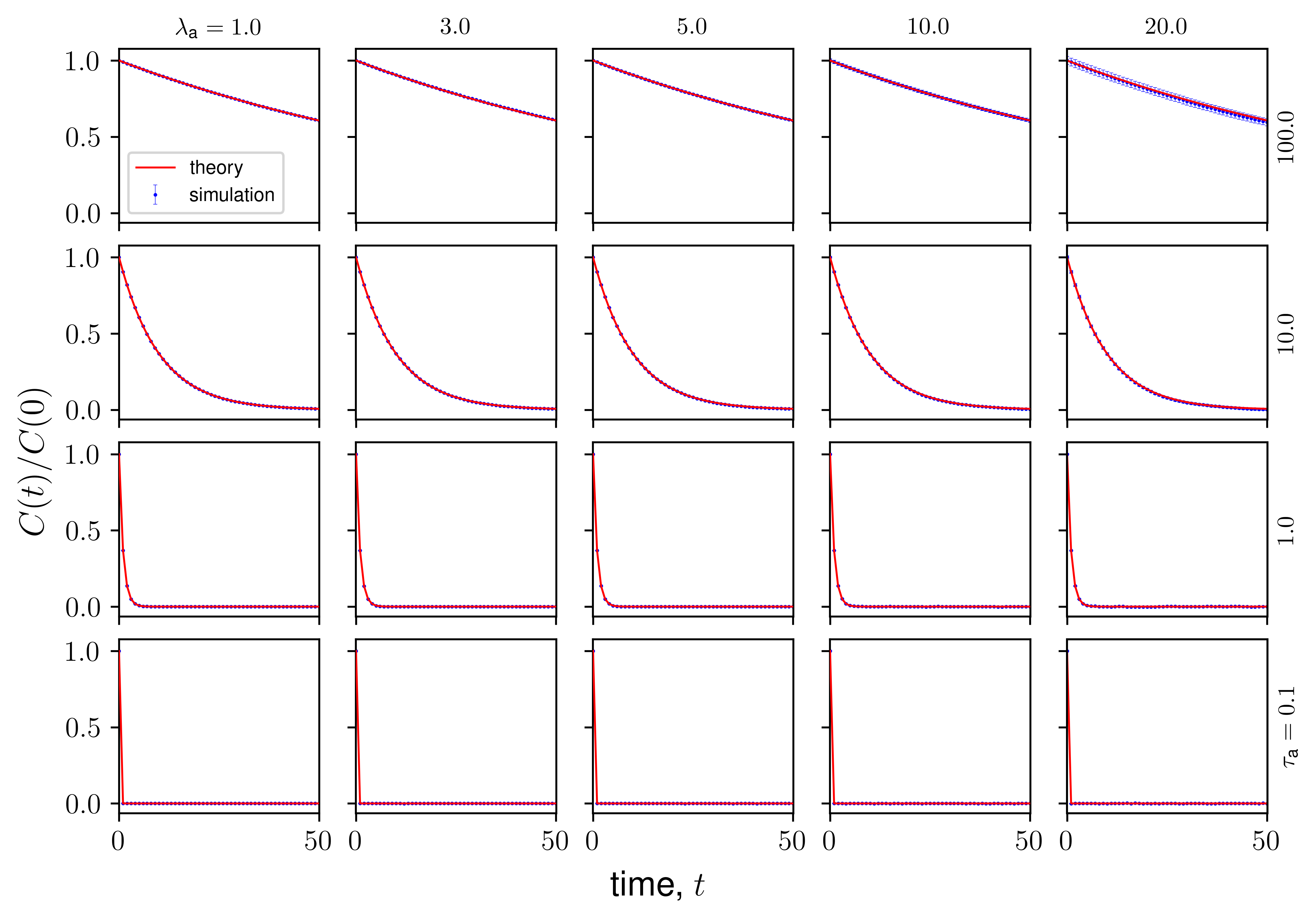}
    \caption{Noise time correlation for different $\la$ and $\taua$. Error bars, where visible, represent $\pm$ twice the standard error of the mean over 50 independent trajectories. ``Theory'' denotes the function $\exp{(-t/\taua)}$.}
    \label{supp_fig:noise_tcorr}
\end{figure}

\begin{figure}[h!]
    \centering
    \includegraphics[width=\linewidth]{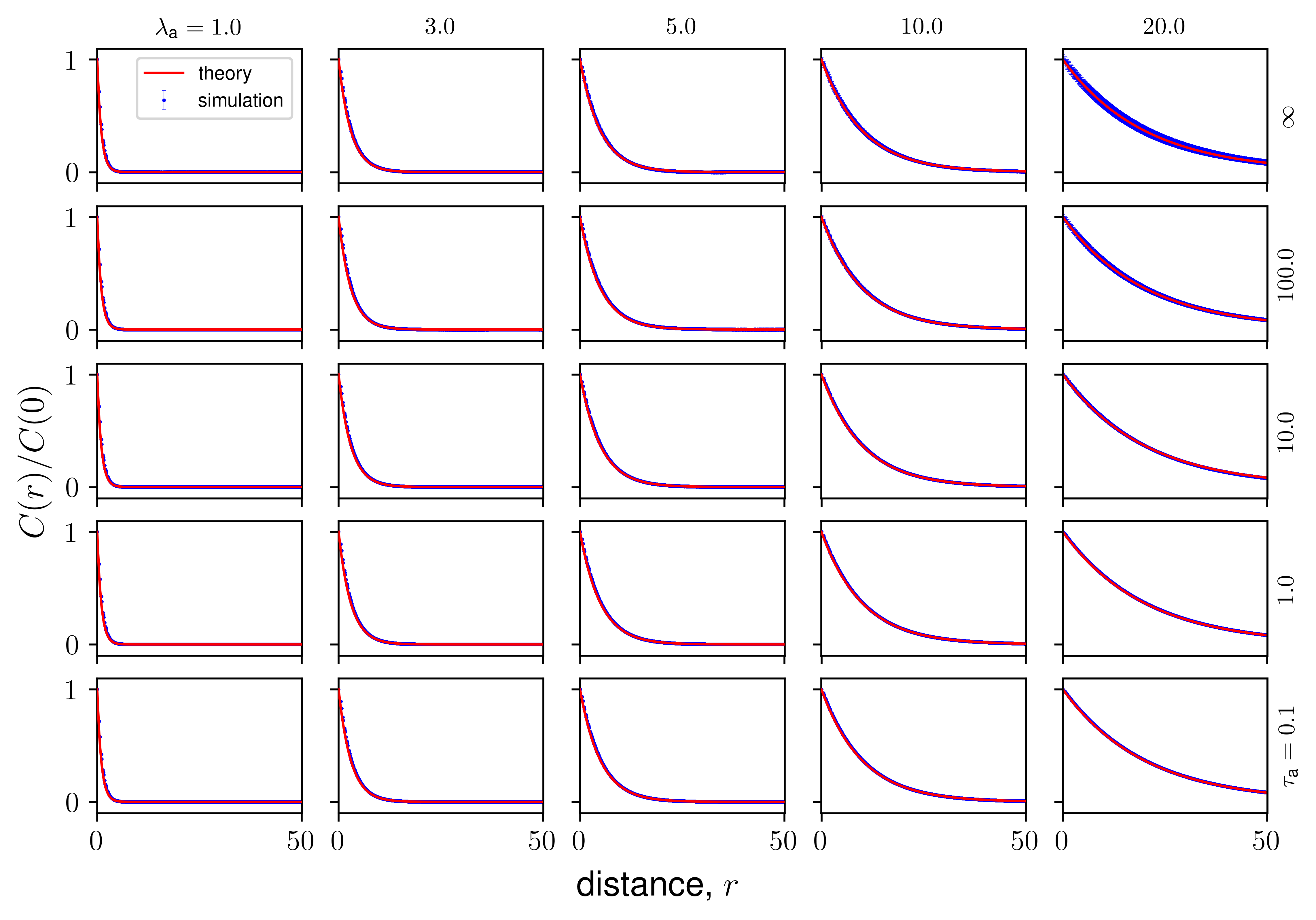}
    \caption{Noise spatial correlation for different $\la$ and $\taua$. Error bars, where visible, represent $\pm$ twice the standard error of the mean over 50 independent trajectories. ``Theory'' denotes the function $\exp{(-r/\la)}$.}
    \label{supp_fig:noise_rcorr}
\end{figure}

\begin{figure}[h!]
    \centering
    \includegraphics[width=\linewidth]{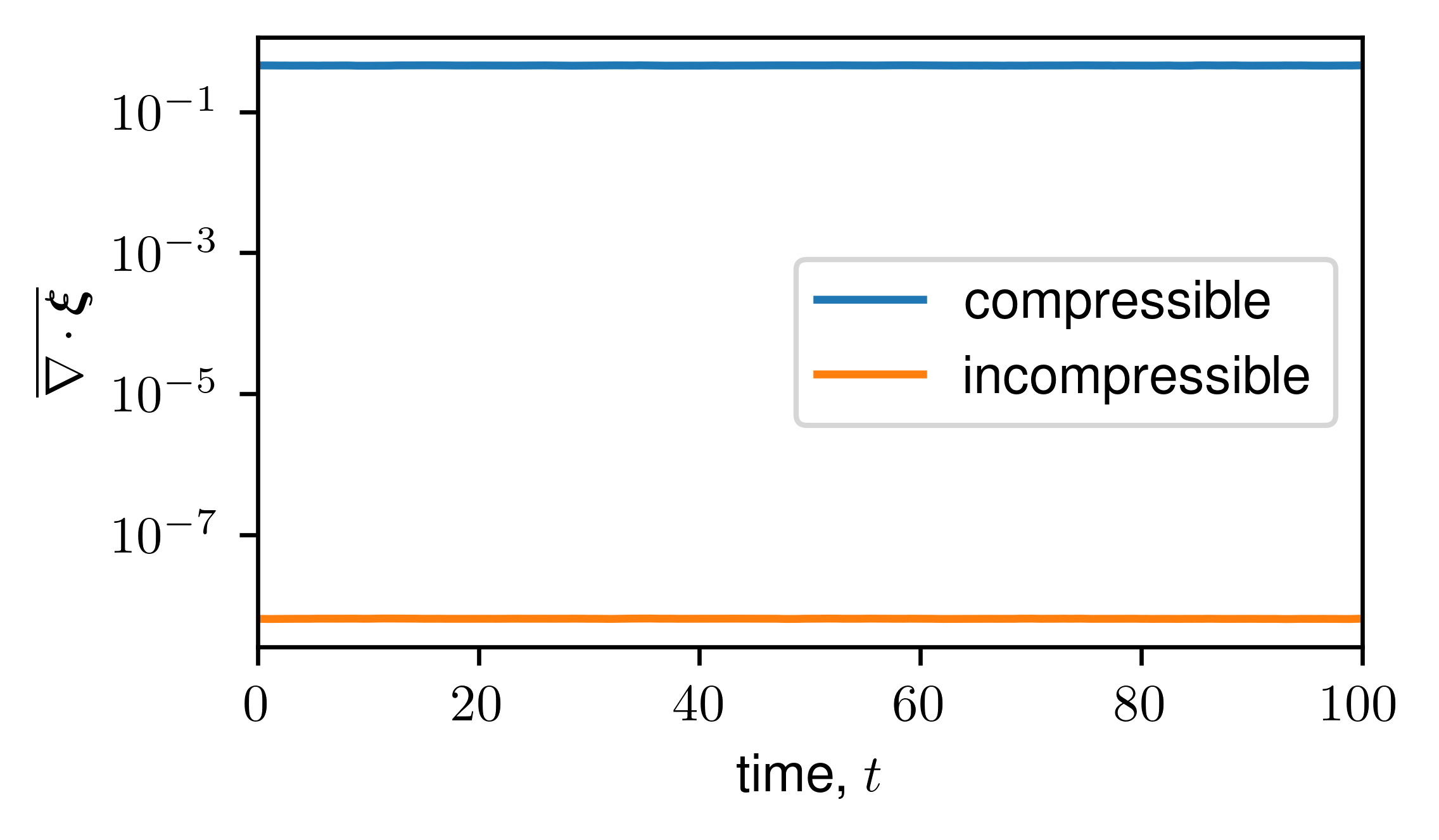}
    \caption{Divergence of the noise field, averaged over all grid points, as a function of time for compressible and incompressible noise.}
    \label{supp_fig:field_div}
\end{figure}

\begin{figure}[h!]
    \centering
    \includegraphics[width=\linewidth]{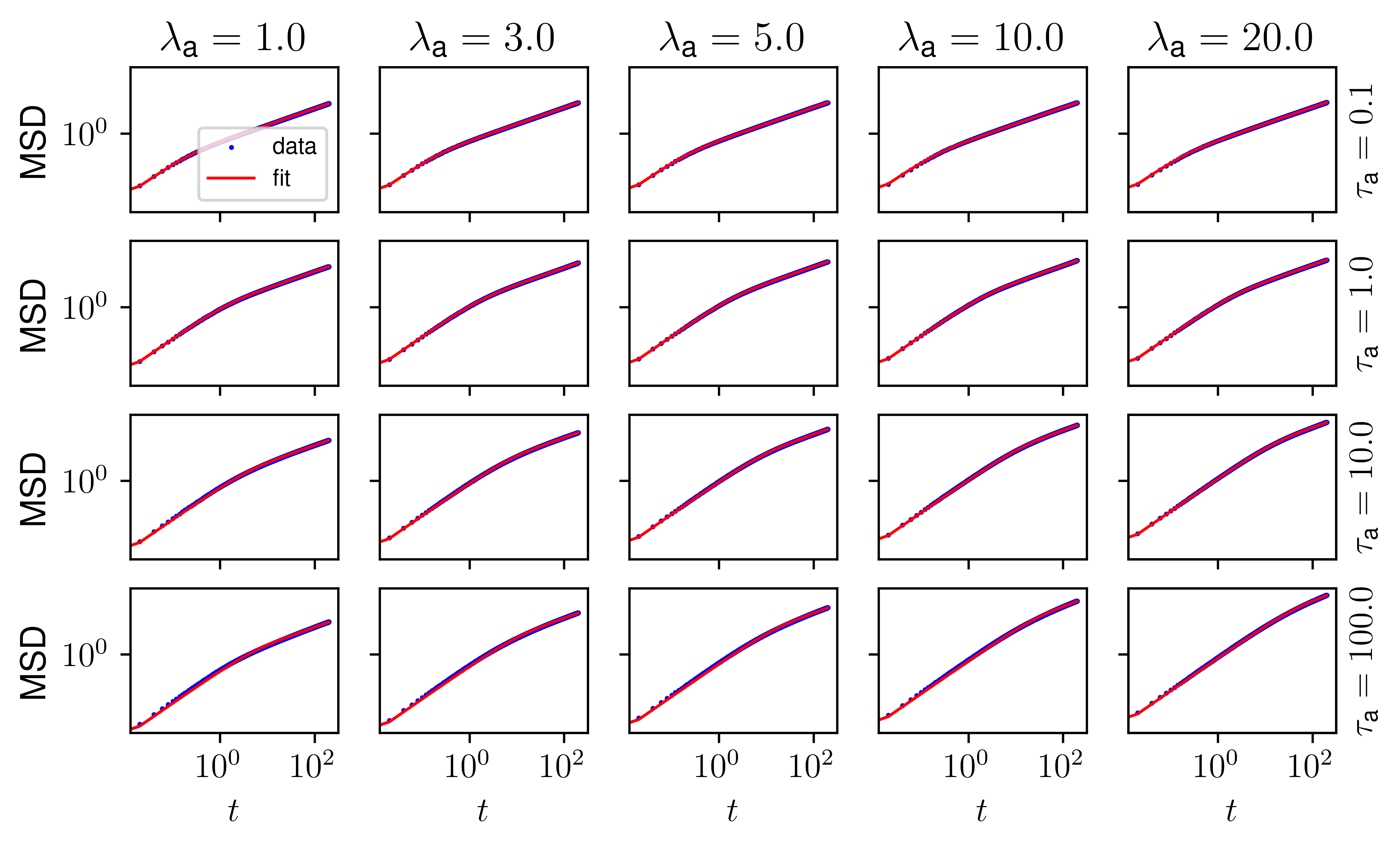}
    \caption{MSD versus time for a tracer particle in active noise fields with varying $\la$, $\taua$ and with $\va=1$. Fits (red curve) were obtained using least-squares regression using Scipy's optimize.curve\_fit function \cite{Virtanen2020}. We assumed a fitting form of $\langle |\mathbf{r}_p(t)-\mathbf{r}_p(0)|^2 \rangle = 2dD\left(t+\tau_{\text{p}}(e^{-t/\tau_{\text{p}}}-1)\right)$ (Eq. \ref{eq:persistent_random_walk} in the Main Text) with $d=2$.}
    \label{supp_fig:msd_fit}
\end{figure}

\begin{figure}[h!]
    \centering
    \includegraphics[width=\linewidth]{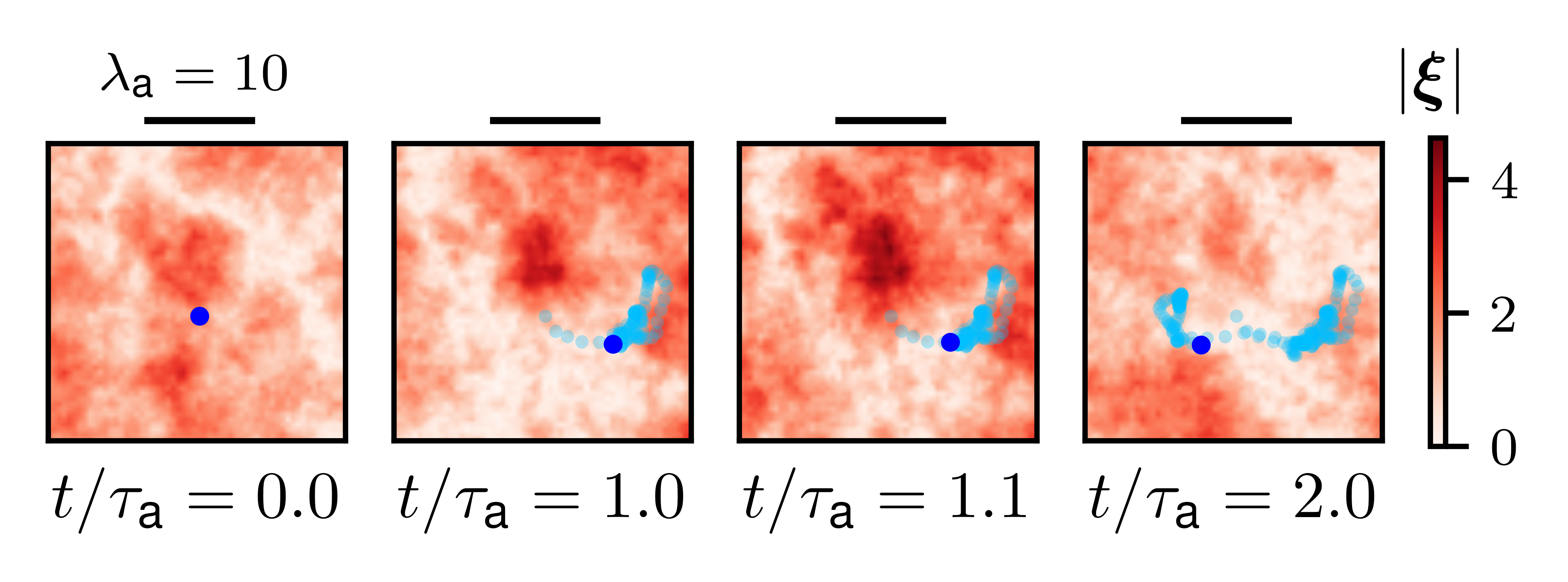}
    \caption{Trajectory of a single particle in an active noise field with $\la=10$, $\taua=100$, $\va=1$. The particle position at time $t$ is denoted with a dark blue marker, while positions at previous time points are denoted with light blue markers. The high density of light blue markers in particular regions indicates ``trapping'' of the particle in a low-noise region.}
    \label{supp_fig:single_particle_tau=100_trajectory}
\end{figure}


\begin{figure}[h!]
    \centering
    \includegraphics[width=\linewidth]{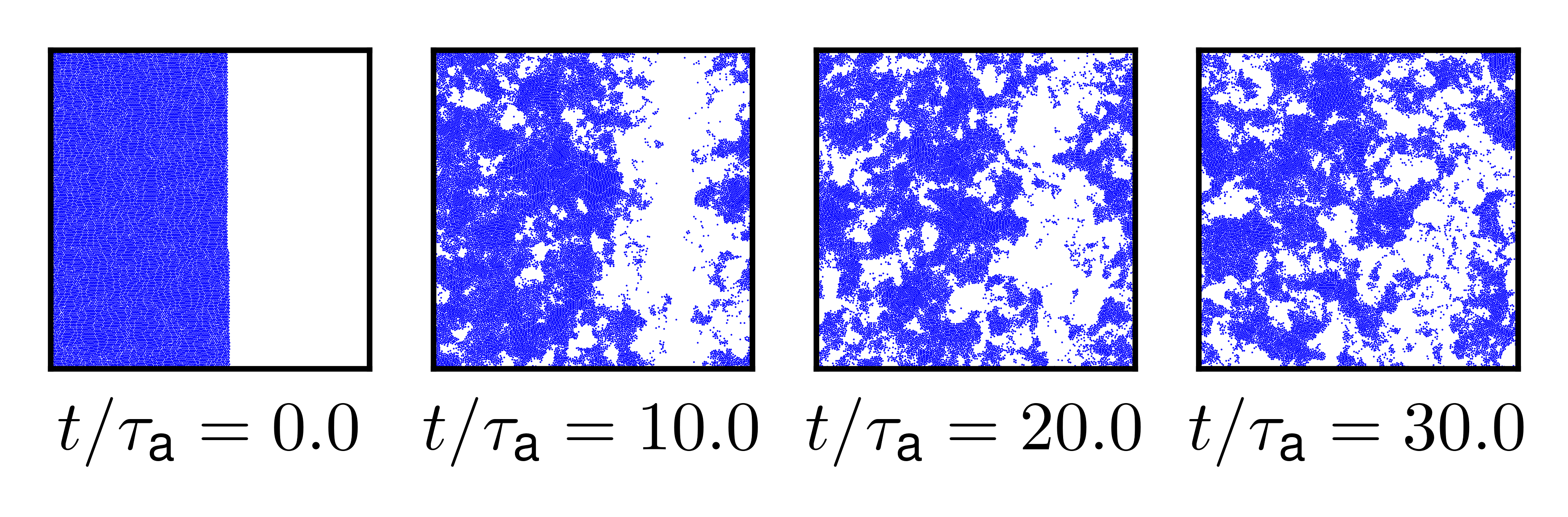}
    \caption{Trajectory of particles with a packing fraction of $\phi=0.4$ in an active noise field with $\taua=10$, $\la=3$, $\va=1$, initialized in a close-packed configuration. Rather than remaining a single cluster, the disks break up into many small clusters.}
    \label{supp_fig:phase_sep_trajectory}
\end{figure}

\begin{figure}[h!]
    \centering
    \includegraphics[width=\linewidth]{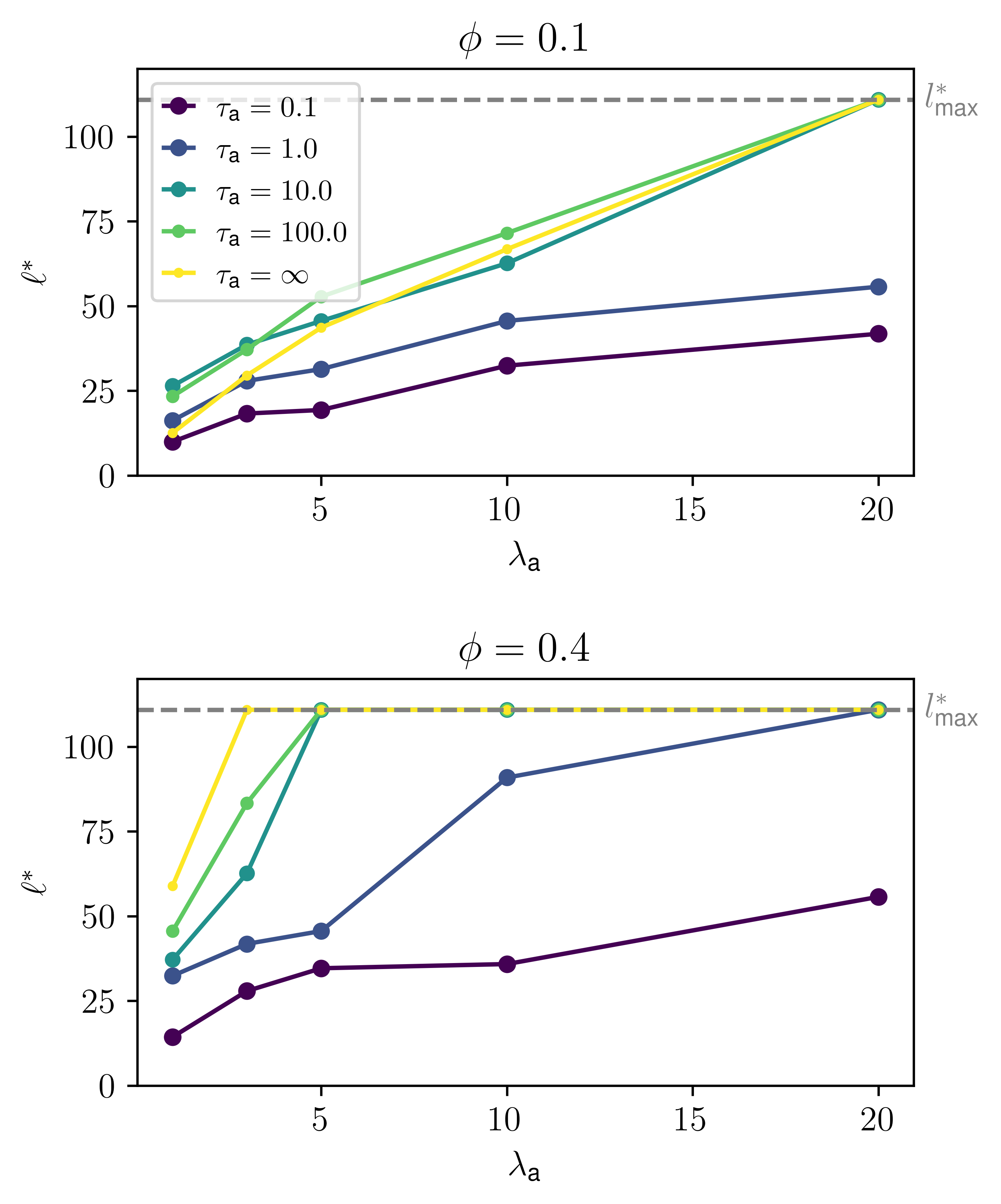}
    \caption{Length scale $\ell^*=\pi/q^*$ associated with $S(q)$ peak location $q^*$ for varying $\la$ and $\taua$. The quantity $\ell^*$ generally increases with $\la$ and $\taua$, although for $\taua=\infty$ and $\phi=0.1$ the noise field is less effective at promoting cluster motion, particularly at low $\la$, causing $\ell^*$ to be smaller at these conditions. Here, $\ell^*_{\text{max}}=\pi/q_{\text{min}}$ is the length scale associated with the smallest wavevector $q_{\text{min}}$ allowed in the periodic simulation box, i.e., it is the wavevector associated with the system size.}
    \label{supp_fig:sq_peak_scaling}
\end{figure}

\begin{figure}[h!]
    \centering
    \includegraphics[width=\linewidth]{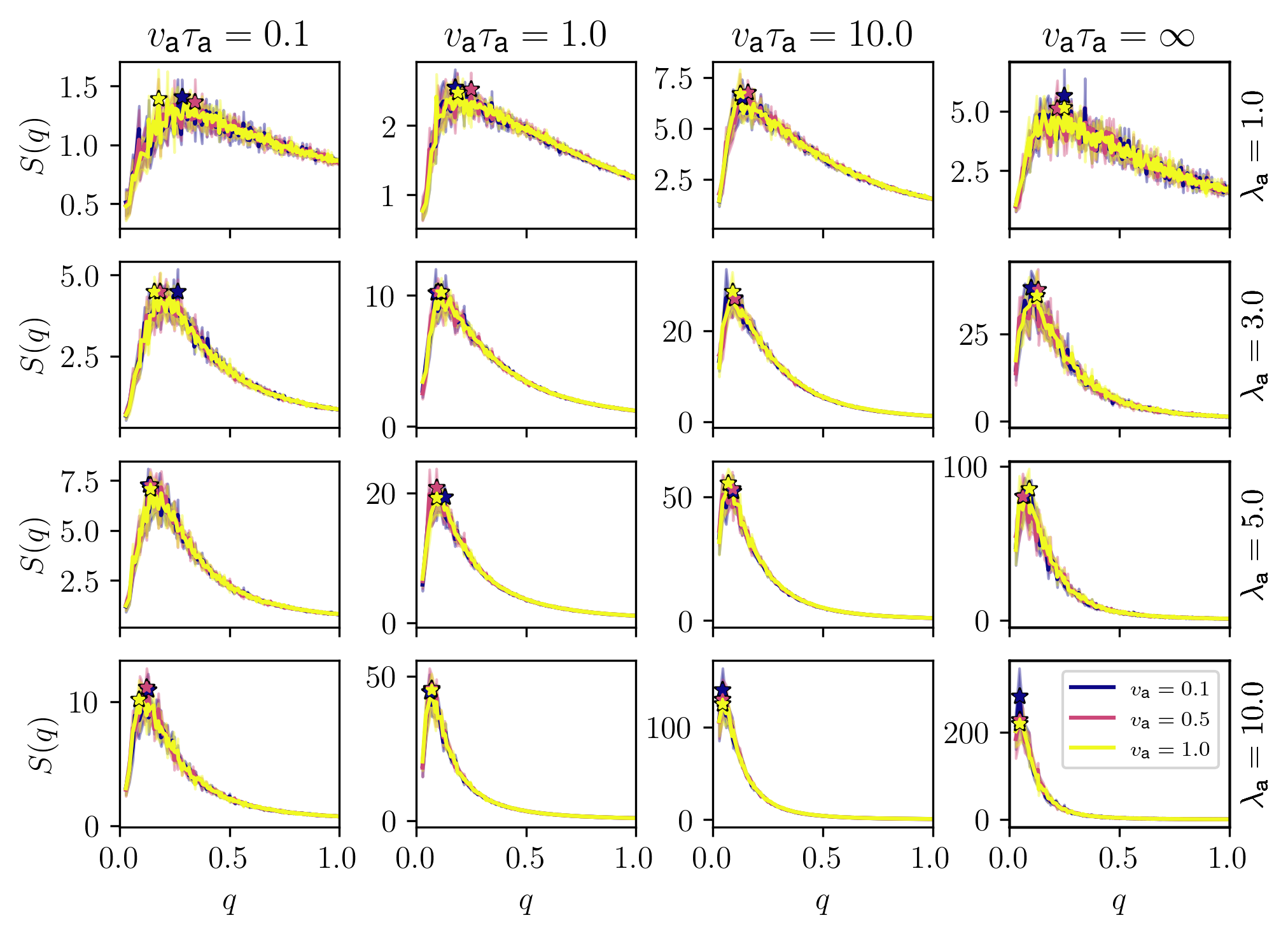}
    \caption{Structure factor for particles at packing fraction $\phi=0.1$ for varying $\va$. Systems with the same $\la$ and $\va \taua$ (i.e. the different colors within each panel, which have different $\va$ but the same $\va\taua$) have statistically indistinguishable structure factors.}
    \label{supp_fig:sq_vary_va}
\end{figure}

\begin{figure}[h!]
    \centering
    \includegraphics[width=\linewidth]{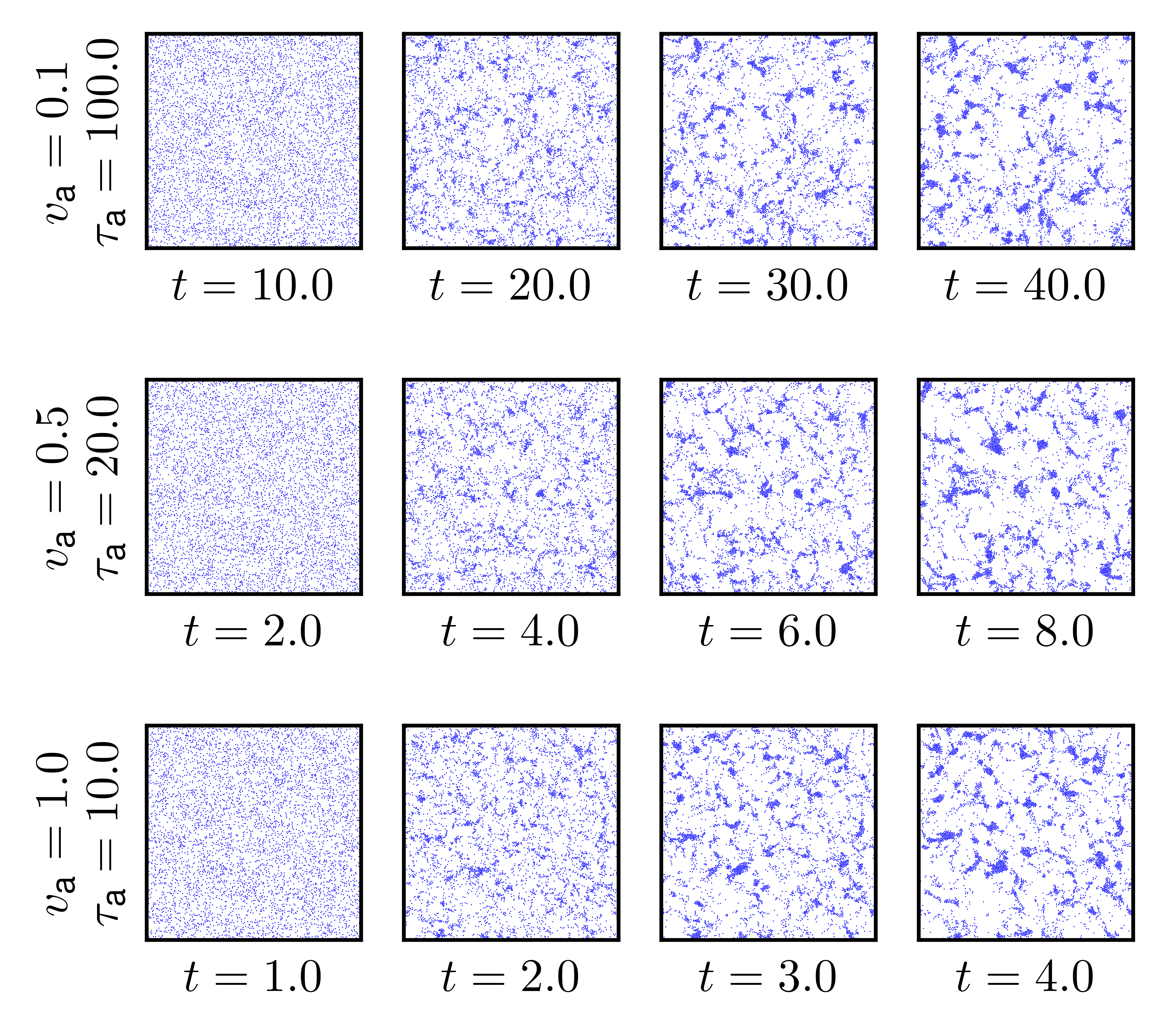}
    \caption{Hard sphere trajectories for $\phi=0.1$, $\la=10.0$, and $\va\taua=10$ for varying $\va$ (and $\taua$.) Starting from an initial random distribution of particles, trajectories with different $\va$ achieve comparably-sized clusters after different times, but trajectories with lower $\va$ take longer to form clusters than those with higher $\va$.}
    \label{supp_fig:vary_va_trajectories}
\end{figure}

\begin{figure}[h!]
    \centering
    \includegraphics[width=\linewidth]{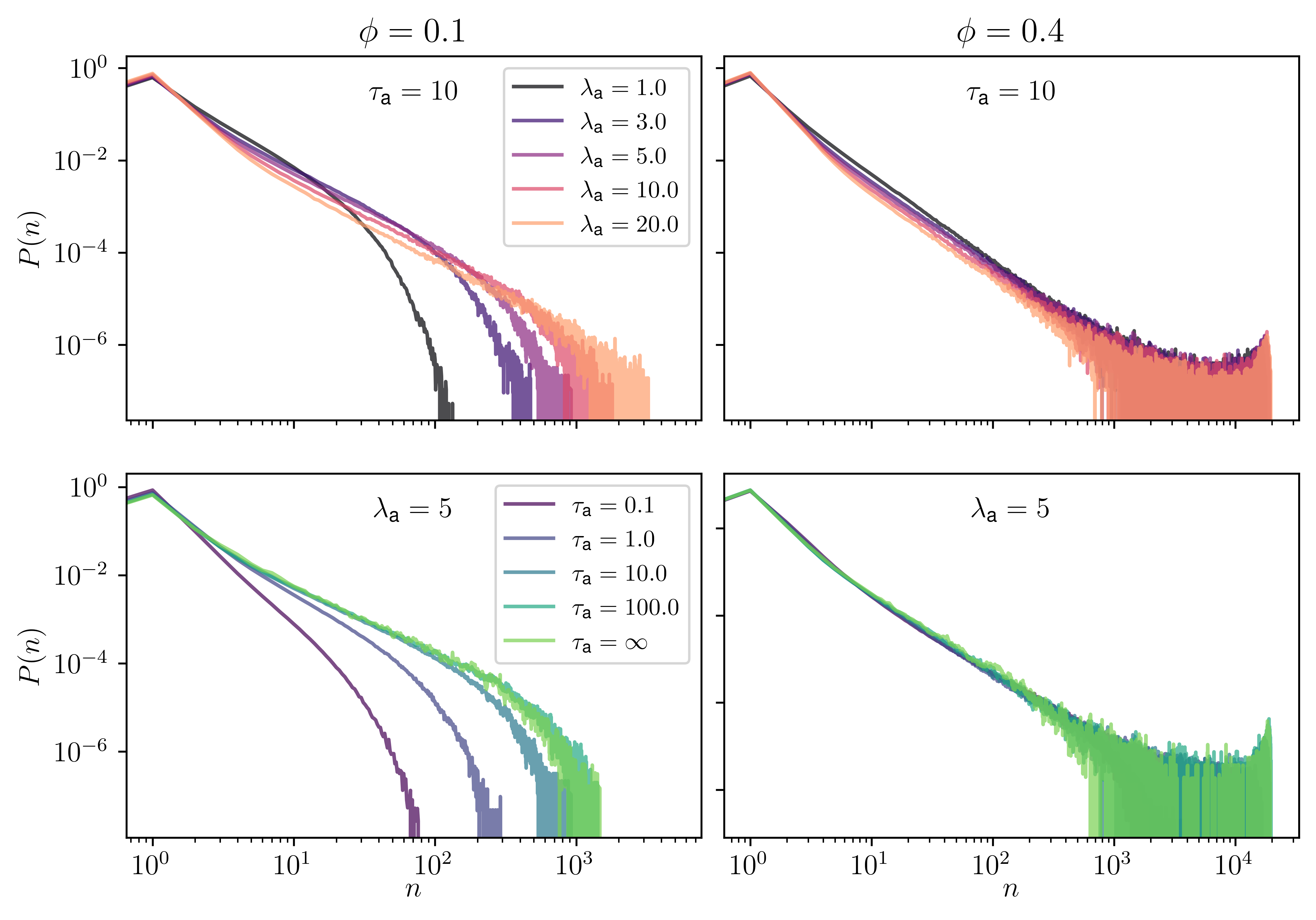}
    \caption{Cluster size distributions $P(n)$ for varying $\phi$, $\la$, and $\taua$.}
    \label{supp_fig:csd}
\end{figure}

\begin{figure}[h!]
    \centering
    \includegraphics[width=\linewidth]{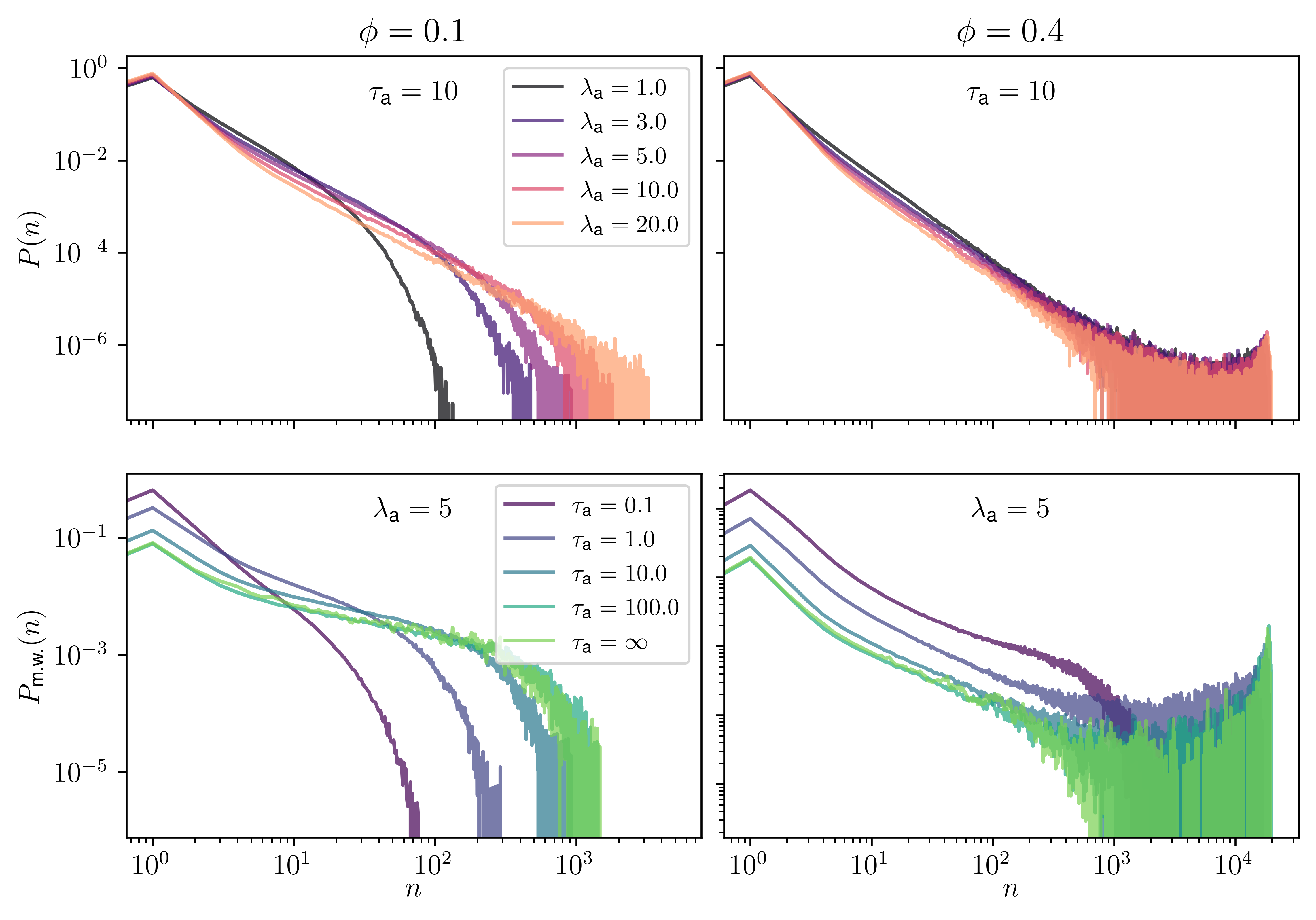}
    \caption{Mass-weighted cluster size distributions, $P_{\text{m.w.}}(n)=nP(n)/\langle n \rangle$, for varying $\phi$, $\la$, and $\taua$.}
    \label{supp_fig:csd_mw}
\end{figure}

\begin{figure}[h!]
    \centering
    \includegraphics[width=\linewidth]{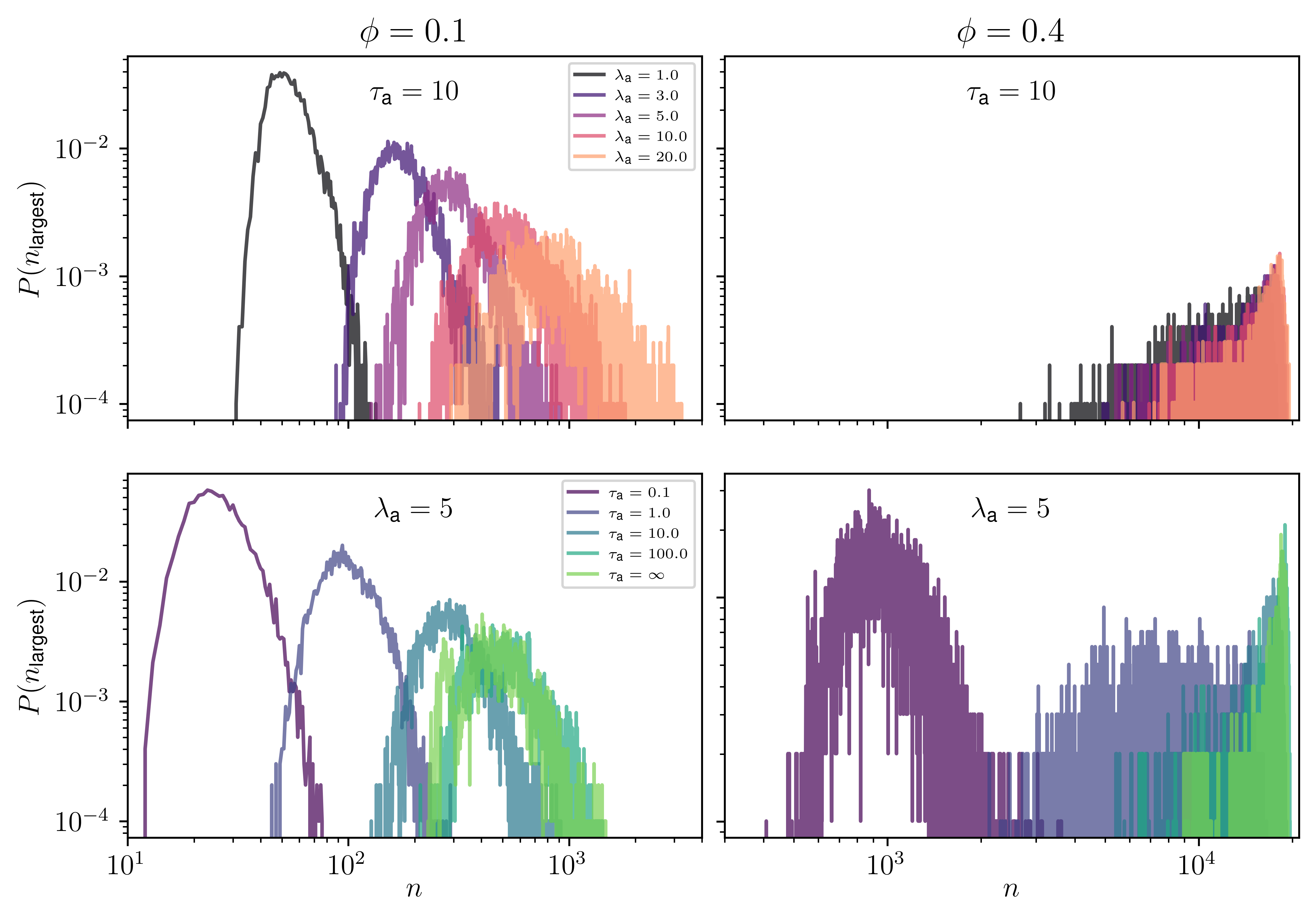}
    \caption{Histograms of the largest cluster size, $\nLA$, for varying $\phi$, $\la$, and $\taua$.}
    \label{supp_fig:largest_cluster_histograms}
\end{figure}

\begin{figure}[h!]
    \centering
    \includegraphics[width=\linewidth]{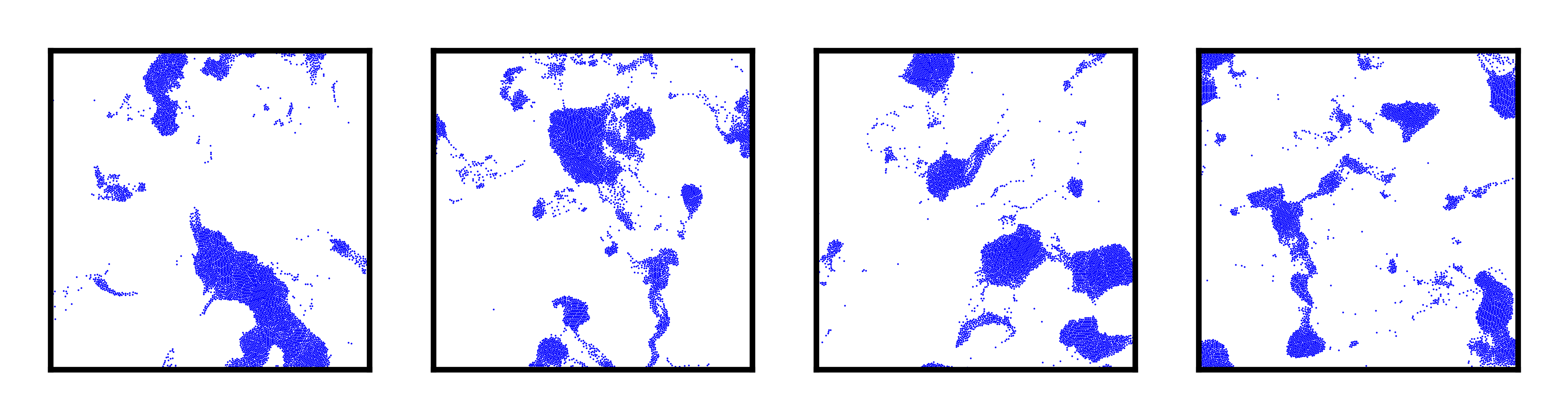}
    \caption{Final configurations from different trajectories (initialized with different random seeds) with $\phi=0.1$, $\la=20$, $\tau=100$, and $\va=1$, illustrating elongated, anisotropic clusters.}
    \label{supp_fig:elongated_clusters}
\end{figure}

\clearpage

\section*{Movie descriptions}
\begin{itemize}
    \item \textbf{Movie S1}: 2D particle trajectories for $\phi=0.1$, $\va=1$ and varying $\la$, $\taua$ (Fig. \ref{fig:hard_sphere_multipanel} main text). Each trajectory has length $250$ time units, the box dimensions are $200\times200$, and each box is subject to periodic boundary conditions.
    \item \textbf{Movie S2}: 2D particle trajectories for $\phi=0.4$, $\va=1$ and varying $\la$, $\taua$ (Fig. \ref{fig:hard_sphere_multipanel} main text). Each trajectory has length $250$ time units, the box dimensions are $200\times200$, and each box is subject to periodic boundary conditions.
    \item \textbf{Movie S3}: Trajectory of a 2D tracer particle in an active noise field with $\va=1$, $\la=10$, $\taua=100$. The scale bar has length $\la=10$. Movie shows a zoomed-in view of a $200\times200$ simulation box. Dark blue marker indicates the current particle position, while light blue markers indicate previous positions along the trajectory. Color indicates the magnitude of the active noise field (see Fig. \ref{fig:active_noise_vs_expt} main text).
    \item \textbf{Movie S4}: 2D particle trajectory for $\phi=0.1$, $\va=1$, $\la=10$, $\taua=10$. Top panel shows particles in a $200\times200$ simulation box; middle and bottom panels show a small portion of this box to better illustrate detailed particle motion. In the middle panel, color indicates the direction of each particle's velocity vector, while arrows indicate both the direction and magnitude of the velocity. In the bottom panel, shades of red represent the magnitude of the noise field and arrows represent the magnitude and direction of the noise field.
\end{itemize}

\bibliography{references}

\makeatletter\@input{xx.tex}\makeatother